\documentclass[aps,prx,citeautoscript,reprint,showpacs,superscriptaddress,twocolumn]{revtex4-2}

\usepackage[english]{babel}
\usepackage{amsmath,bm}
\usepackage{amsmath,amssymb,bbm,graphicx,comment,dsfont,braket,mathtools,units}
\usepackage{hyperref}
\hypersetup{colorlinks=true,citecolor=blue,linkcolor=blue,urlcolor=blue,pdfstartview=FitH}

\usepackage{color}
\definecolor{dkgreen}{rgb}{0,0.5,0}

\newcommand{\abs}[1]{\left\lvert #1 \right\rvert}
\newcommand\dif{\mathop{}\!\mathrm{d}}
\newcommand{\be}{\begin{equation}}
\newcommand{\ee}{\end{equation}}

\newcommand{\RePart}{\operatorname{Re}}

\DeclareMathOperator{\ei}{Ei}

\DeclareMathOperator{\csch}{csch}

\begin{document}

\date{\today}

\title{Thermal Anyon Interferometry in Phonon-Coupled Kitaev Spin Liquids}
\author{Kai Klocke}
\affiliation{Department of Physics, University of California, Berkeley, California 94720, USA}
\author{Joel E. Moore}
\affiliation{Department of Physics, University of California, Berkeley, California 94720, USA}
\affiliation{Materials Sciences Division, Lawrence Berkeley National Laboratory, Berkeley, California 94720, USA}
\author{Jason Alicea}
\affiliation{Department of Physics and Institute for Quantum Information and Matter,California Institute of Technology, Pasadena, California 91125, USA}
\affiliation{Walter Burke Institute for Theoretical Physics, California Institute of Technology, Pasadena, California 91125, USA}
\author{G\'abor B. Hal\'asz}
\thanks{This manuscript has been authored by UT-Battelle, LLC under Contract No. DE-AC05-00OR22725 with the U.S. Department of Energy. The United States Government retains and the publisher, by accepting the article for publication, acknowledges that the United States Government retains a non-exclusive, paid-up, irrevocable, world-wide license to publish or reproduce the published form of this manuscript, or allow others to do so, for United States Government purposes. The Department of Energy will provide public access to these results of federally sponsored research in accordance with the DOE Public Access Plan (http://energy.gov/downloads/doe-public-access-plan).}
\affiliation{Materials Science and Technology Division, Oak Ridge National Laboratory, Oak Ridge, Tennessee 37831, USA}
\affiliation{Quantum Science Center, Oak Ridge, Tennessee 37831, USA}

\begin{abstract}

Recent theoretical studies inspired by experiments on the Kitaev magnet $\alpha$-RuCl$_3$ highlight the nontrivial impact of phonons on the thermal Hall conductivity of chiral topological phases.
Here we introduce mixed mesoscopic-macroscopic devices that allow refined thermal-transport probes of non-Abelian spin liquids with Ising topological order.
These devices feature a quantum-coherent mesoscopic region with negligible phonon conductance, flanked by macroscopic lobes that facilitate efficient thermalization between chiral Majorana edge modes and bulk phonons.
We show that our devices enable $(i)$ accurate determination of the quantized thermal Hall conductivity, $(ii)$ identification of non-Abelian Ising anyons via the temperature dependence of the thermal conductance, and most interestingly $(iii)$ single-anyon detection through heat-based anyon interferometry.
Analogous results apply broadly to phonon-coupled chiral topological orders.

\end{abstract}

\maketitle

\tableofcontents

\section{Introduction}

Topologically ordered phases of matter possess a wide range of exotic properties including long-range entanglement, topological ground-state degeneracy, and anyonic quasiparticle excitations.
Perhaps most notably, the non-trivial fusion and braiding rules of non-Abelian anyons offer a route to intrinsically fault-tolerant quantum computation~\cite{Kitaev2003, TQCreview}.
Spin-orbit-coupled Mott insulators~\cite{Witczak-Krempa_2014} have recently generated a great deal of interest as potential platforms to realize topologically ordered quantum spin liquids (QSLs).
This excitement was largely spurred by a series of seminal works~\cite{Jackeli2009, Chaloupka2010, Chaloupka2013} that proposed an approximate realization of Kitaev's honeycomb model~\cite{Kitaev2006}---an exactly solvable lattice model that captures both Abelian and non-Abelian QSL phases---in insulating $4d$ and $5d$ honeycomb magnets.
Following these proposals, numerous honeycomb materials have been put forward as candidate Kitaev spin liquids~\cite{Rau_2016, Trebst_2017, Hermanns_2018, Takagi_2019}.
Among these ``Kitaev materials'', $\alpha$-RuCl$_3$~\cite{Plumb_2014, Sandilands_2015, Sears_2015, Majumder_2015, Johnson_2015, Sandilands_2016, Banerjee_2016, Banerjee_2017, Do_2017} displays particularly tantalizing behavior at low temperatures: in-plane magnetic fields suppress zigzag spin order~\cite{Kubota_2015, Leahy2017, Sears2017, Wolter2017, Baek2017, Banerjee_2018, Hentrich_2018, Jansa_2018, Widmann2019, Balz2019}, possibly giving way to the \emph{non-Abelian} QSL from the Kitaev honeycomb model~\cite{Kitaev2006}.
Strikingly, recent experiments~\cite{Kasahara2018, Tokoi, bruin2021} have reported a half-integer-quantized thermal Hall conductivity at intermediate fields $\sim 10$ T, which is consistent with the emergence of a chiral Majorana edge mode~\cite{Banerjee} hosted by the non-Abelian Kitaev spin liquid (see also Refs.~\onlinecite{Yamashita_2020, Czajka_2021} for related thermal transport experiments).

In general, thermal transport is one of the most promising experimental techniques for identifying topologically ordered chiral QSLs.
Contrary to conventional probes of magnetic systems such as neutron scattering, quantized responses in thermal transport directly reflect the universal properties of a given topological order.
For example, the quantized thermal Hall conductivity~\footnote{We note that the thermal Hall conductivity is identical to the thermal Hall conductance in two dimensions.} is proportional to the edge theory's chiral central charge~\cite{KaneFisherThermal}---fractional values of which indicate non-Abelian topological orders.
Quantized thermal transport correspondingly plays an analogous role to quantized electrical transport, which is a fundamental signature of electronic topological orders such as fractional quantum Hall states.

There is, however, a key conceptual difference between thermal and electronic transport: the chiral edge modes of QSLs are not the only heat carriers in the system.
In fact, for most realistic materials, the quantized thermal Hall conductivity resulting from the chiral edge modes, $\kappa_{xy}$, is expected to be much smaller than the longitudinal thermal conductivity due to bulk acoustic phonons, $\kappa_{xx}$.
For $\alpha$-RuCl$_3$, the ratio of the two conductivities has been experimentally found~\cite{Kasahara2018, Tokoi, bruin2021} to be $\kappa_{xy} / \kappa_{xx} \sim 10^{-3}$ at the relevant temperatures between $3$ K and $6$ K.
Given that bulk phonons not only produce large $\kappa_{xx}$ but also couple to the chiral edge modes, one may na\"ively expect that chiral edge transport is sufficiently disturbed that quantized $\kappa_{xy}$ is no longer observable.
Surprisingly, however, Refs.~\onlinecite{Ye_2018} and \onlinecite{Vinkler2018} showed that the experimentally measured thermal Hall conductivity remains (approximately) quantized in the conventional rectangular geometry~\cite{Kasahara2018, Tokoi, bruin2021}, provided the sample is sufficiently large that the edge can effectively thermalize with the bulk.
In fact, according to these works, the observation of quantized thermal transport \emph{relies on} a sufficiently large edge-bulk coupling, as the thermal leads and sensors are expected to couple predominantly to lattice vibrations (i.e., bulk phonons) rather than the chiral edge modes.

\begin{figure}
	\centering
	\includegraphics[width=\columnwidth]{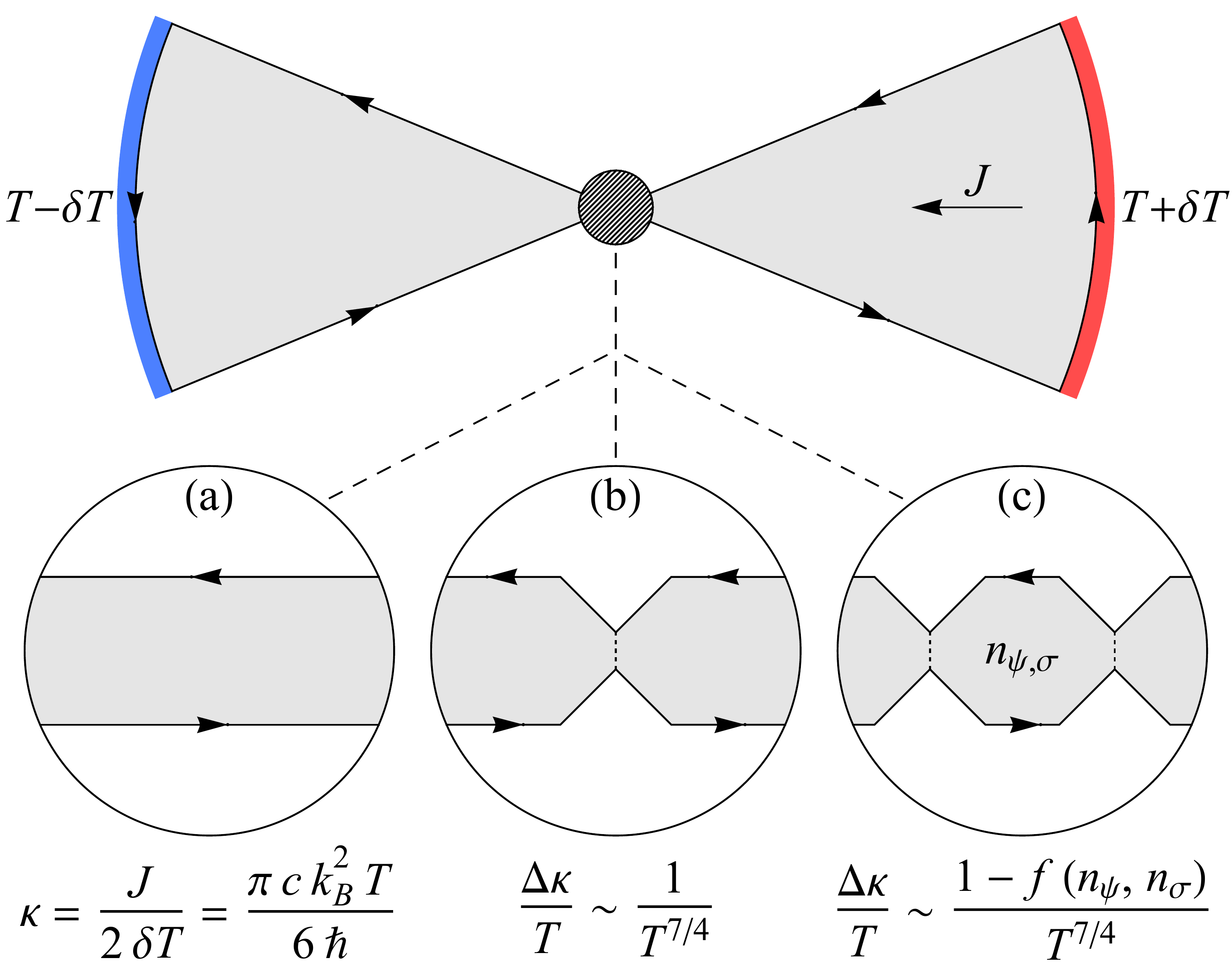}
	\caption{
	Device geometry for extracting universal characteristics of the non-Abelian Kitaev spin liquid through conventional thermal transport in the presence of phonons.
	Two macroscopic lobes are bridged by a central mesoscopic region; heat current $J$ flows from the right lobe (heated at the outer edge to temperature $T+\delta T$) to the left lobe (cooled at the outer edge to $T-\delta T$).
	The two lobes facilitate edge-phonon thermalization, while the central region acts as a bottleneck for bulk phonon transport and thus emphasizes chiral edge transport.
	The central region contains (a) zero, (b) one, or (c) two pinch points at which the edge separation becomes comparable to the bulk correlation length---allowing edge anyons to tunnel between the two chiral edge modes.
	These tunneling processes can be detected by measuring the heat current between the two lobes, or temperatures in either lobe close to the central region.
	If the phonon thermal conductance of the central region is sufficiently small, measurements of the heat current reveal the following universal spin liquid characteristics.
	(a) In the absence of any pinch points, the \emph{longitudinal} thermal conductance $\kappa$ of the device, \emph{directly} obtained as the ratio of the heat current $J$ and the temperature difference $2 \, \delta T$ between the hot and cold leads, is proportional to the chiral central charge, $c = \frac{1}{2}$, which immediately confirms the presence of non-Abelian anyons.
	(b) For a single pinch point, the dominant low-temperature correction to the thermal conductance, $\Delta \kappa$, follows a nontrivial power law with universal exponent $7/4$ as a function of the temperature $T$, which reflects the tunneling of non-Abelian Ising anyons.
	(c) For two pinch points, the correction $\Delta \kappa$ acquires an interference term, $f(n_\psi,n_\sigma)$, that is sensitive to the number of fermions ($n_{\psi}$) and Ising anyons ($n_{\sigma}$) between the two pinch points and thereby enables detection of individual bulk anyons.
	}
	\label{fig:device}
\end{figure}

Another, more recent, thread of research is concerned with employing anyonic edge interferometry to probe topological orders in chiral QSLs.
By utilizing point contacts at which edge anyons may tunnel between counterpropagating edges, this approach allows for a direct observation of anyonic statistics as well as detection of individual bulk anyons.
For fractional quantum Hall states, anyonic edge interferometry has an extensive theoretical literature~\cite{Chamon_1997, Nayak2005, Stern2006, Bonderson2006, Bonderson2006a, Kim_2006, Rosenow_2007, Halperin_2011, Rosenow_2012} and has been demonstrated in recent electrical transport experiments~\cite{Nakamura2019, Willett2019, manfra_2020}.
Adapting electrical anyon interferometry techniques to chiral QSLs, however, is nontrivial because the edge and bulk anyons of such insulating systems do not carry electric charge.
During the last year, alternative schemes have been proposed to perform anyon interferometry in QSLs by exploiting conversion between charged and neutral edge modes at a superconducting interface~\cite{Aasen_2020}, or by means of time-domain measurements using ancillary spins~\cite{Klocke_2020} (see Refs.~\onlinecite{Feldmeier_2020, Pereira_2020, Udagawa_2021} for other routes to detecting individual bulk anyons or chiral edge modes in QSLs).
Though the possibility of heat-based anyon interferometry has also been noted~\cite{Bonderson_2013}, practical implementation of the idea raises an important conceptual challenge: one must reconcile anyon braiding by edge tunneling---a process relying on phase coherence---with conventional thermometry that requires large-scale thermalization and couples principally to bulk phonons.

In this paper, we show that thermal transport is indeed a feasible route to realizing anyonic edge interferometry in a chiral QSL, as one can exploit edge-bulk thermalization while \emph{also} harvesting phase-coherent edge transport in a single device.
To this end, we consider the unconventional device geometry shown in Fig.~\ref{fig:device} which consists of a mesoscopic central region flanked by two macroscopic lobes.
Crucially, phase-coherent edge tunneling processes inside the mesoscopic region, where coupling to phonons has a negligible effect by construction, directly influence phonon thermodynamics in the macroscopic lobes.
One can thereby accurately extract universal characteristics of the underlying QSL within the framework of currently available thermal transport experiments.
In the simplest setup, wherein the upper and lower edges of the mesoscopic region decouple [Fig.~\ref{fig:device}(a)], our device enables refined measurement of the chiral central charge that is manifest in the dominant \emph{longitudinal} thermal conductance rather than a subdominant thermal Hall conductivity.
Adding a single constriction in the mesoscopic region [Fig.~\ref{fig:device}(b)] allows energy to backscatter between the upper and lower edges via anyon tunneling; remarkably, the temperature dependence of the resulting backscattered heat current reveals fingerprints of the anyonic quasiparticles hosted by the QSL.
Finally, and most interestingly, adding a second constriction [Fig.~\ref{fig:device}(c)] defines an experimentally viable thermal anyon interferometer that solves the conceptual challenge noted above: individual anyons localized in the central region can be sensitively detected, along with their braiding statistics, by measuring phonon temperatures within either lobe, or the total heat current between the two lobes.
For concreteness, we focus on the non-Abelian spin liquid relevant for Kitaev materials~\cite{Kitaev2006}; the universal characteristics accessible by measuring the total heat current and, hence, the thermal conductance of the device, are then shown in Fig.~\ref{fig:device}.
We also emphasize, however, that much of our results apply broadly to phonon-coupled chiral topological orders.

\begin{figure}
	\centering
	\includegraphics[width=\columnwidth]{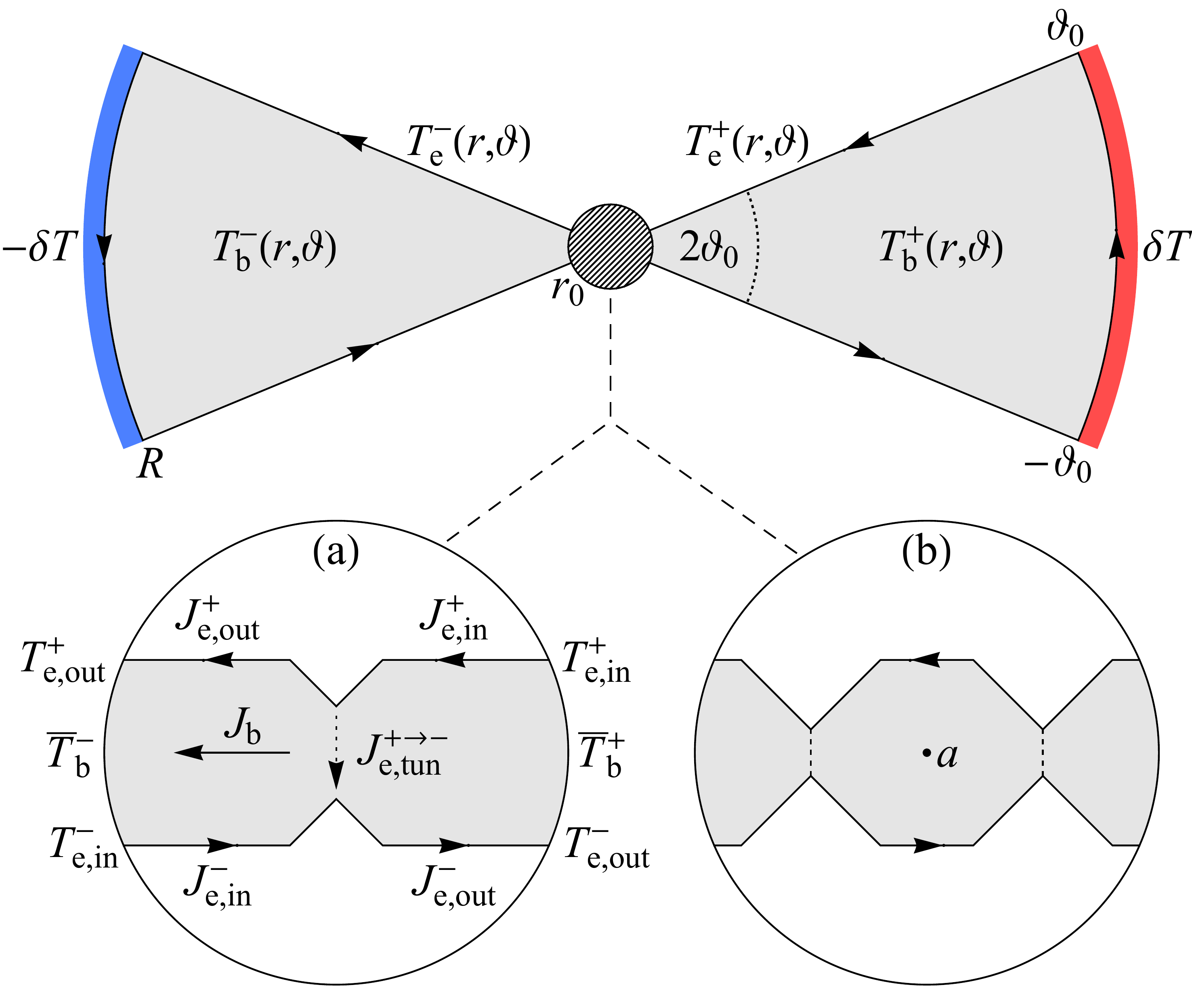}
	\caption{
	Thermal transport within a phonon-coupled chiral spin liquid in the device geometry of Fig.~\ref{fig:device}.
	All temperatures shown here, including those of the hot and cold leads ($\pm \delta T$), are small relative temperatures with respect to an average overall temperature $T_0$ that corresponds to $T$ in Fig.~\ref{fig:device}.
	There are distinct relative temperatures for the edge mode ($T_{\mathrm{e}}$) and the bulk phonons ($T_{\mathrm{b}}$).
	Each lobe is an annular section with opening angle $2 \vartheta_0$, outer radius $R$, and inner radius $r_0$.
	The central channel carries a bulk heat current $J_{\mathrm{b}}$ driven by the different mean bulk temperatures $\bar{T}_{\mathrm{b}}^{\pm}$ at its two ends, as well as chiral edge heat currents $J_{\mathrm{e}}^{\pm}$ determined by the local edge temperatures $T_{\mathrm{e}}^{\pm}$.
	For each edge, the tunneling heat current $J_{\mathrm{e}, \mathrm{tun}}^{+ \to -}$ associated with the pinch point(s) changes the edge heat current from $J_{\mathrm{e}, \mathrm{in}}^{\pm}$ to $J_{\mathrm{e}, \mathrm{out}}^{\pm} = J_{\mathrm{e}, \mathrm{in}}^{\pm} \mp J_{\mathrm{e}, \mathrm{tun}}^{+ \to -}$.
	While the notations are shown for the single-pinch case (a), they readily generalize to the double-pinch case (b), where the tunneling heat current $J_{\mathrm{e}, \mathrm{tun}}^{+ \to -}$ has contributions from both pinch points and is sensitive to bulk anyons $a$ between the two pinch points.
	}
	\label{fig:two_lobes}
\end{figure}

The rest of this paper is organized as follows.
In Sec.~\ref{sec:phonon_coupling}, we study large-scale thermal transport in the device geometry of Fig.~\ref{fig:device} and demonstrate how macroscopic temperatures and heat currents within the two lobes reflect mesoscopic edge processes in the central region.
In the framework of this calculation, the edge processes are accounted for by a single dimensionless parameter $\beta$.
In Sec.~\ref{sec:tunneling}, we concentrate on the edge processes themselves in the presence of one [Fig.~\ref{fig:device}(b)] or two [Fig.~\ref{fig:device}(c)] point contacts.
While the introduction of a single pinch point is sufficient to reveal Ising-anyon tunneling via a universal temperature dependence of $\beta$, the double-pinch geometry allows for interference effects to imprint signatures of anyon braiding on the large-scale thermal transport.
In Sec.~\ref{sec:discussion}, we discuss the experimental conditions necessary to observe the signatures we identify and the challenges one might face in implementing the proposed thermal interferometry scheme.
Finally, we close the paper with a brief summary and outlook in Sec.~\ref{sec:summary}.

\section{Thermal Transport}\label{sec:phonon_coupling}

We consider the low-temperature thermal transport of a phonon-coupled chiral spin liquid in the device geometry of Fig.~\ref{fig:device}.
In this geometry, two large lobes of radius $R$ are connected by a narrow channel in a small central region of radius $r_0 \ll R$ (see Fig.~\ref{fig:two_lobes}).
The narrow channel between the two lobes features a pinch region with one [Fig.~\ref{fig:two_lobes}(a)] or two [Fig.~\ref{fig:two_lobes}(b)] pinch points at which energy may tunnel between the counterpropagating top and bottom edges of the channel.
We assume that each lobe is an annular section with opening angle $2 \vartheta_0$, outer radius $R$, and inner radius $r_0$.
While this assumption does not affect the main results of our work, it considerably simplifies the solution of our heat-transport equations.

Importantly, heat in our system is carried by two disparate degrees of freedom: the chiral edge mode and bulk acoustic phonons.
Given that the thermal coupling between these two degrees of freedom is expected to be very small at low temperatures~\cite{Ye_2018}, we cannot assume that they are in thermal equilibrium at any given position along the edge of the system.
Therefore, we consider two separate temperatures, corresponding to the edge mode ($T_{\mathrm{e}}$) and the bulk phonons ($T_{\mathrm{b}}$).

Thermal transport can be set up by attaching the outer edges of the right and left lobes to hot and cold leads, respectively, and probed by measuring $(i)$ temperatures close to the central region (i.e., far from the leads) or $(ii)$ the total heat current between the two leads.
Crucially, while the quantum effects of interest (see Fig.~\ref{fig:device}) rely on thermal tunneling between otherwise isolated chiral edge modes, the thermal leads and the temperature probes are expected to couple to the bulk phonons instead.
Therefore, an optimal detection of the quantum effects through thermal transport requires significant edge-bulk thermalization in the two lobes but negligible edge-bulk thermalization in the central region.
Since the thermalization between the edge mode and the bulk phonons is known~\cite{Ye_2018} to be significant at length scales exceeding a characteristic edge-bulk thermalization length $\ell$, this scenario is naturally realized if the central region is mesoscopic, $r_0 \ll \ell$, while the two lobes are macroscopic, $R \gg \ell$.

\subsection{Heat-Transport Equations}

We now describe a general set of hydrodynamic equations governing the heat transport and the temperature distribution of our phonon-coupled chiral spin liquid.
For a start, we assume that the system has an average overall temperature $T_0$ and that the temperatures $T^{\pm}$ shown in Fig.~\ref{fig:two_lobes} are small variations with respect to this overall temperature.
In this regime, we can readily linearize all of our equations in the temperature variations $T^{\pm}$.
Also, we can use the inversion symmetry around the center of the system to establish the general relation $T^{-} = -T^{+}$, where $T_{\mathrm{b}, \mathrm{e}}^{+}(r,\vartheta)$ [$T_{\mathrm{b}, \mathrm{e}}^{-}(r,\vartheta)$] correspond to the right (left) lobe, while $T_{\mathrm{e}, \mathrm{in}}^{+}$ ($T_{\mathrm{e}, \mathrm{in}}^{-}$) and $T_{\mathrm{e}, \mathrm{out}}^{+}$ ($T_{\mathrm{e}, \mathrm{out}}^{-}$) correspond to the top (bottom) edges in the central region.

We first describe the macroscopic thermal transport in the two lobes by employing the hydrodynamic equations introduced in Ref.~\onlinecite{Ye_2018}.
Inversion symmetry permits us to focus on the right lobe alone.
The bulk and edge temperatures $T_{\mathrm{b}, \mathrm{e}}(r,\vartheta) \equiv T_{\mathrm{b}, \mathrm{e}}^+(r,\vartheta)$ are specified in polar coordinates, $(r,\vartheta)$, with $r_0 < r < R$ and  $|\vartheta| < \vartheta_0$.
Importantly, the bulk temperature $T_{\mathrm{b}}(r,\vartheta)$ is defined within the entire lobe, while the edge temperature $T_{\mathrm{e}}(r,\vartheta)$ is only defined along the edge.
We note that similar hydrodynamic equations were also considered in Ref.~\onlinecite{Vinkler2018} but without making a distinction between bulk and edge temperatures.

The thermal transport within the lobe comprises three different kinds of thermal currents~\cite{Ye_2018}.
First, if we assume that the acoustic phonons are diffusive, there is a bulk heat-current density (heat current per unit length) proportional to the negative gradient of the bulk temperature, $\mathbf{j}_{\mathrm{b}} = -\kappa_{\mathrm{b}} \nabla T_{\mathrm{b}}$, where the coefficient $\kappa_{\mathrm{b}} \equiv \kappa_{xx}$ is the longitudinal thermal conductivity due to the phonons \footnote{We note that $\kappa_{\mathrm{b}}$ is the \emph{two-dimensional} thermal conductivity which is the product of the usual three-dimensional thermal conductivity and the layer thickness.}.
Second, the chiral edge mode carries a counterclockwise heat current $J_{\mathrm{e}} = (\pi c / 12) (T_0 + T_{\mathrm{e}})^2$ \footnote{Here we have set $\hbar = k_B = 1$.} along the edge, where $c$ is the chiral central charge of the edge conformal field theory (CFT).
At linear order in the relative edge temperature $T_{\mathrm{e}}$, this edge heat current is given by $J_{\mathrm{e}}= J_0 + \kappa_{\mathrm{e}} T_{\mathrm{e}}$, where $J_0 = \pi c T_0^2 / 12$ is a constant edge current, while $\kappa_{\mathrm{e}} = \pi c T_0 / 6$ is the quantized thermal Hall conductivity~\cite{Kasahara2018, Tokoi, bruin2021} associated with the chiral edge mode \footnote{We emphasize that this thermal Hall conductivity corresponds to a \emph{thermally isolated} edge mode and could only be \emph{directly} measured by using thermal leads and temperature sensors that couple to the edge mode rather than the bulk phonons.}.
Third, there is a heat-current density (heat current per unit length) from the bulk phonons to the edge mode, which is driven by the temperature difference between these two degrees of freedom.
At linear order, this exchange heat-current density must take the general form $j_{\mathrm{b} \to \mathrm{e}} = \lambda (T_{\mathrm{b}} - T_{\mathrm{e}})$, where $\lambda$ is the linearized thermal coupling between the edge and the bulk.
Importantly, $\lambda$ is strongly suppressed at low temperatures and was argued in Ref.~\onlinecite{Ye_2018} to vanish as $\lambda \propto T^6$.

The equations governing the temperatures are then obtained by assuming a stationary state and imposing the conservation of energy.
First of all, the divergence of the bulk heat-current density must vanish, $\nabla \cdot \mathbf{j}_{\mathrm{b}} = 0$, which translates into a Laplace's equation for the bulk temperature, $\nabla^2 T_{\mathrm{b}} = 0$.
Using polar coordinates, this Laplace's equation then takes the separable form
\be
	 \frac{\partial^2 T_{\mathrm{b}}(r,\vartheta)} {(\partial \ln r)^2} + \frac{\partial^2 T_{\mathrm{b}}(r,\vartheta)} {\partial \vartheta^2} = 0. \label{eq:laplace}
\ee
Next, along the upper and lower edges ($\vartheta = \pm \vartheta_0$), the normal component of the bulk heat-current density must match the exchange heat-current density, $\hat{\mathbf{n}}_{\perp} \cdot \mathbf{j}_{\mathrm{b}} = j_{\mathrm{b} \to \mathrm{e}}$, where $\hat{\mathbf{n}}_{\perp}$ is the unit vector pointing in the ``outward'' direction perpendicular to the edge.
In polar coordinates, the resulting equations can be written as
\be
    -\frac{\kappa_{\mathrm{b}}}{r}\frac{\partial T_{\mathrm{b}}(r,\vartheta = \pm \vartheta_0)}{\partial\vartheta} = \pm \lambda\left[T_{\mathrm{b}}(r,\pm\vartheta_0) - T_{\mathrm{e}}(r,\pm\vartheta_0)\right]. \label{eq:exchange}
\ee
In contrast, along the outer edge ($r = R$), the presence of the lead invalidates the above constraint and imposes a fixed bulk temperature instead:
\be
	T_{\mathrm{b}}(R,\vartheta) =\delta T. \label{eq:lead}
\ee
Finally, along all three edges, the spatial variation of the edge heat current must match the exchange heat-current density, $\nabla_{\parallel} J_{\mathrm{e}} = j_{\mathrm{b} \to \mathrm{e}}$, where $\nabla_{\parallel}$ is the derivative along the edge in the counterclockwise direction.
In polar coordinates, the resulting equations read
\be
\begin{aligned}
	\frac{\kappa_{\mathrm{e}}}{R} \frac{\partial T_{\mathrm{e}}(R,\vartheta)}{\partial\vartheta} &= \lambda\left[T_{\mathrm{b}}(R,\vartheta) - T_{\mathrm{e}}(R,\vartheta)\right],\\
	\kappa_{\mathrm{e}} \frac{\partial T_{\mathrm{e}}(r,\pm \vartheta_0)}{\partial r} &= \mp \lambda\left[T_{\mathrm{b}}(r,\pm\vartheta_0) - T_{\mathrm{e}}(r,\pm\vartheta_0)\right].
\end{aligned} \label{eq:edge_current_variation}
\ee
Thus, the characteristic edge-bulk thermalization length scale is found to be $\ell = \kappa_{\mathrm{e}} / \lambda$.
We emphasize again that Eqs.~\eqref{eq:laplace}-\eqref{eq:edge_current_variation} are directly taken from Ref.~\onlinecite{Ye_2018} and adapted to our unconventional device geometry.

We next describe the mesoscopic thermal transport between the two lobes through the central region.
Since the edge-bulk thermalization is negligible in the narrow channel, it is not necessary to describe the full spatial dependence of the bulk temperature.
Instead, we use a more coarse-grained picture (see Fig.~\ref{fig:two_lobes}), and assume a general phenomenological relation, $J_{\mathrm{b}} = \kappa_{\mathrm{c}} (\bar{T}_{\mathrm{b}}^{+} - \bar{T}_{\mathrm{b}}^{-})$, where $J_{\mathrm{b}}$ is the bulk heat current, carried by the phonons, through the narrow channel, $\bar{T}_{\mathrm{b}}^{+}$ ($\bar{T}_{\mathrm{b}}^{-}$) is the mean bulk temperature at the right (left) end of the channel, and $\kappa_{\mathrm{c}}$ is the thermal conductance~\cite{footnote} of the channel.
Importantly, this relation does not assume that the phonons are diffusive in the central region.
We note, however, that, in the specific case of diffusive phonons, the channel conductance is expected to be $\kappa_{\mathrm{c}} \sim \kappa_{\mathrm{b}} W/L$, where $W$ and $L$ are the width and the length of the channel, respectively.
If the channel is narrow enough ($W \ll L$), the conductance $\kappa_{\mathrm{c}}$ is then much smaller than the conductivity $\kappa_{\mathrm{b}}$.

Matching the bulk heat currents and the bulk temperatures between the two lobes and the narrow channel, we then immediately find $J_{\mathrm{b}} = \mp r_0 \int_{-\vartheta_0}^{+\vartheta_0} \dif\vartheta \, \hat{\mathbf{r}} \cdot \mathbf{j}_{\mathrm{b}}^{\pm} (r_0,\vartheta)$ and $\bar{T}_{\mathrm{b}}^{\pm} = (2\vartheta_0)^{-1} \int_{-\vartheta_0}^{+\vartheta_0} \dif\vartheta \, T_{\mathrm{b}}^{\pm} (r_0,\vartheta)$, where $\hat{\mathbf{r}}$ is the radial unit vector.
Thus, the relation between the heat current and the mean temperature difference becomes
\be
\begin{aligned}
    &\kappa_{\mathrm{b}} r_0 \int_{-\vartheta_0}^{+\vartheta_0} \dif\vartheta \, \frac{\partial T_{\mathrm{b}}^{\pm} (r=r_0,\vartheta)}{\partial r} \\
    &\qquad = \pm \frac{\kappa_{\mathrm{c}}} {2\vartheta_0} \int_{-\vartheta_0}^{+\vartheta_0} \dif\vartheta \left[ T_{\mathrm{b}}^{+} (r_0,\vartheta) - T_{\mathrm{b}}^{-} (r_0,\vartheta) \right]. \nonumber
\end{aligned}
\ee
However, this relation is too coarse grained and does not enable a unique solution for Eqs.~\eqref{eq:laplace}-\eqref{eq:edge_current_variation}.
Therefore, we generalize this relation by imposing it on the integrands themselves rather than the integrals.
Using the inversion symmetry of the system, and remembering the notation $T_{\mathrm{b}}(r,\vartheta) \equiv T_{\mathrm{b}}^+(r,\vartheta) = -T_{\mathrm{b}}^-(r,\vartheta)$, the equation governing the bulk temperature along the interface ($r=r_0$) of each lobe and the central region is then
\be
    \frac{\partial T_{\mathrm{b}} (r=r_0,\vartheta)} {\partial \ln r} = \alpha T_{\mathrm{b}} (r_0,\vartheta), \label{eq:interface-bulk}
\ee
where $\alpha = \kappa_{\mathrm{c}} / (\kappa_{\mathrm{b}} \vartheta_0)$ is a dimensionless ratio of the channel conductance to the lobe conductance.
We emphasize that this generalized relation is equivalent to the original coarse-grained one if the temperature variations are sufficiently small along the interface (which is expected for small-enough $r_0$ and/or $\vartheta_0$).

Due to the negligible edge-bulk thermalization in the central region, the edge temperatures along the narrow channel can only change as a result of thermal tunneling between the counterpropagating top and bottom edges at the pinch points.
Thus, if we consider the entire pinch region with one or two pinch points (see Fig.~\ref{fig:two_lobes}) as a single unit, the only relevant edge temperatures and heat currents are the incoming ($T_{\mathrm{e}, \mathrm{in}}^{\pm}$ and $J_{\mathrm{e}, \mathrm{in}}^{\pm}$) and the outgoing ($T_{\mathrm{e}, \mathrm{out}}^{\pm}$ and $J_{\mathrm{e}, \mathrm{out}}^{\pm}$) ones~\footnote{We implicitly assume that the outgoing edge thermalizes with itself and reaches thermal equilibrium by the time it leaves the central region}.
Given that there is no quantum coherence between the two incoming edges, we assume that, for each edge, the tunneling heat current to the other edge, $J_{\mathrm{e}, \mathrm{tun}} (T)$, is fully determined by the absolute incoming temperature, $T = T_0 + T_{\mathrm{e}, \mathrm{in}}$, of the given edge.
At linear order in the relative temperatures $T_{\mathrm{e}, \mathrm{in}}^{\pm}$, the net tunneling current from the top edge to the bottom edge (see Fig.~\ref{fig:two_lobes}) is then $J_{\mathrm{e}, \mathrm{tun}}^{+ \to -} = \kappa_{\mathrm{e}, \mathrm{tun}} [T_{\mathrm{e}, \mathrm{in}}^+ - T_{\mathrm{e}, \mathrm{in}}^-]$, where $\kappa_{\mathrm{e}, \mathrm{tun}} = (dJ_{\mathrm{e}, \mathrm{tun}} / dT)|_{T = T_0}$ is a tunneling thermal conductance~\cite{footnote}.
Since the incoming and outgoing heat currents are related to each other as $J_{\mathrm{e}, \mathrm{out}}^{\pm} = J_{\mathrm{e}, \mathrm{in}}^{\pm} \mp J_{\mathrm{e}, \mathrm{tun}}^{+ \to -}$ and to the incoming and outgoing temperatures according to $J_{\mathrm{e}, \mathrm{in}}^{\pm} = J_0 + \kappa_{\mathrm{e}} T_{\mathrm{e}, \mathrm{in}}^{\pm}$ and $J_{\mathrm{e}, \mathrm{out}}^{\pm} = J_0 + \kappa_{\mathrm{e}} T_{\mathrm{e}, \mathrm{out}}^{\pm}$, the incoming and outgoing temperatures are related by
\be
    T_{\mathrm{e}, \mathrm{out}}^{\pm} = T_{\mathrm{e}, \mathrm{in}}^{\pm} \mp \beta [T_{\mathrm{e}, \mathrm{in}}^+ - T_{\mathrm{e}, \mathrm{in}}^-], \nonumber
\ee
where $\beta = \kappa_{\mathrm{e}, \mathrm{tun}} / \kappa_{\mathrm{e}}$ is a dimensionless parameter satisfying $0 <\beta < 1$.
Physically, $\beta$ is the fraction of excess energy in the hotter edge that tunnels to the colder edge in the entire pinch region.
As we explore in Sec.~\ref{sec:tunneling}, this quantity depends on the number of pinch points in the pinch region and, if there are two pinch points, on the total anyon content in between them (see Fig.~\ref{fig:two_lobes}).
Finally, with the identifications $T_{\mathrm{e}, \mathrm{in}}^{\pm} = T_{\mathrm{e}}^{\pm}(r_0,\vartheta_0) \equiv \pm T_{\mathrm{e}} (r_0,\vartheta_0)$ and $T_{\mathrm{e}, \mathrm{out}}^{\pm} = T_{\mathrm{e}}^{\mp}(r_0,-\vartheta_0) \equiv \mp T_{\mathrm{e}}(r_0,-\vartheta_0)$, the equation governing the edge temperatures at the interface ($r=r_0$) of each lobe and the central region reads
\be
	T_{\mathrm{e}}(r_0,-\vartheta_0) = (2\beta-1) \, T_{\mathrm{e}}(r_0,\vartheta_0). \label{eq:interface-edge}
\ee
Importantly, Eqs.~\eqref{eq:interface-bulk} and \eqref{eq:interface-edge} are exclusively for the bulk and edge temperatures of a single lobe, and they are only affected by the rest of the system through the parameters $\alpha$ and $\beta$.
Together with Eqs.~\eqref{eq:laplace}-\eqref{eq:edge_current_variation}, they uniquely determine the temperature profile of the given lobe.

\subsection{Perturbative Solution}

Now we solve Eqs.~\eqref{eq:laplace}-\eqref{eq:interface-edge} for the temperature profile of each lobe by employing the perturbative approach introduced in Ref.~\onlinecite{Ye_2018}.
As we will later find, this perturbative approach is convergent in our device geometry whenever $\kappa_{\mathrm{e}} \ll \kappa_{\mathrm{b}} \vartheta_0$.
To start with, we write the bulk and edge temperatures in series expansions as $T_{\mathrm{b}, \mathrm{e}} = \sum_{n=0}^{\infty} T_{\mathrm{b}, \mathrm{e}}^{(n)}$, where $T_{\mathrm{b}, \mathrm{e}}^{(0)}$ are the unperturbed solutions in the absence of edge-bulk coupling ($\lambda=0$), while $T_{\mathrm{b}, \mathrm{e}}^{(n)}$ with $n>0$ are perturbative corrections due to finite $\lambda$.
The unperturbed solutions are given by
\be
\begin{aligned}
	T_{\mathrm{b}}^{(0)}(r,\vartheta) &= \frac{\delta T[1 + \alpha\ln(r/r_0)]} {1 + \alpha\ln(R/r_0)}, \\
	T_{\mathrm{e}}^{(0)}(r,\vartheta) &= 0,
\end{aligned} \label{eq:T-0}
\ee
where the edge temperature vanishes because the hot and cold leads are assumed to couple exclusively to the bulk phonons.
The perturbative corrections are then found by means of an iterative procedure.
For each iteration step $n>0$, we first obtain the edge temperature by solving the ordinary differential equations [see Eq.~\eqref{eq:edge_current_variation}]
\be
\begin{aligned}
	\frac{\partial T_{\mathrm{e}}^{(n)} (R,\vartheta)}{\partial\vartheta} &= \frac{R}{\ell} \left[T_{\mathrm{b}}^{(n-1)} (R,\vartheta) - T_{\mathrm{e}}^{(n)} (R,\vartheta)\right],\\
	\frac{\partial T_{\mathrm{e}}^{(n)} (r,\pm \vartheta_0)}{\partial r} &= \mp \frac{1}{\ell} \left[T_{\mathrm{b}}^{(n-1)} (r,\pm\vartheta_0) - T_{\mathrm{e}}^{(n)} (r,\pm\vartheta_0)\right],
\end{aligned} \label{eq:edge_current_variation-n}
\ee
together with the boundary condition [see Eq.~\eqref{eq:interface-edge}]
\be
	T_{\mathrm{e}}^{(n)} (r_0,-\vartheta_0) = (2\beta-1) \, T_{\mathrm{e}}^{(n)} (r_0,\vartheta_0). \label{eq:interface-edge-n}
\ee
Next, we find the bulk temperature by solving Laplace's equation within the lobe [see Eq.~\eqref{eq:laplace}],
\be
	 \frac{\partial^2 T_{\mathrm{b}}^{(n)} (r,\vartheta)} {(\partial \ln r)^2} + \frac{\partial^2 T_{\mathrm{b}}^{(n)} (r,\vartheta)} {\partial \vartheta^2} = 0, \label{eq:laplace-n}
\ee
together with Neumann boundary conditions along the upper and lower edges at $\vartheta = \pm \vartheta_0$ [see Eq.~\eqref{eq:exchange}],
\be
\begin{aligned}
    &\frac{\partial T_{\mathrm{b}}^{(n)} (r,\vartheta = \pm \vartheta_0)}{\partial\vartheta} \\
    &\qquad = \mp \frac{\lambda r} {\kappa_{\mathrm{b}}} \left[T_{\mathrm{b}}^{(n-1)} (r,\pm\vartheta_0) - T_{\mathrm{e}}^{(n)} (r,\pm\vartheta_0)\right], \label{eq:exchange-n}
\end{aligned}
\ee
Dirichlet boundary conditions along the outer edge at $r = R$ [see Eq.~\eqref{eq:lead}],
\be
	T_{\mathrm{b}}^{(n)} (R,\vartheta) = 0, \label{eq:lead-n}
\ee
and homogeneous boundary conditions along the interface with the central region at $r = r_0$ [see Eq.~\eqref{eq:interface-bulk}],
\be
    \frac{\partial T_{\mathrm{b}}^{(n)} (r=r_0,\vartheta)} {\partial \ln r} = \alpha T_{\mathrm{b}}^{(n)} (r_0,\vartheta). \label{eq:interface-bulk-n}
\ee
The iterative procedure is convergent if the perturbative corrections $T_{\mathrm{b}, \mathrm{e}}^{(n)}$ are progressively smaller.

To understand how the thermal transport may be sensitive to edge tunneling processes in the central region, we seek the $\beta$-dependent components of the leading-order corrections $T_{\mathrm{b}, \mathrm{e}}^{(1)}$ to the bulk and edge temperatures.
For simplicity, we assume that the characteristic edge-bulk thermalization length, $\ell = \kappa_{\mathrm{e}} / \lambda$, is much larger than the central region, $\ell \gg r_0$, but much smaller than the lobe width, $\ell \ll R \vartheta_0 \lesssim R$.
In this regime, the edge temperature fully thermalizes to $\delta T$ along the outer edge of the lobe but retains a dependence on $\beta$ close to the central region.
By solving Eqs.~\eqref{eq:edge_current_variation-n} and \eqref{eq:interface-edge-n} with $n=1$ (see Appendix~\ref{app:phonon_appendix}), the edge temperatures along the upper and lower edges take the forms $T_{\mathrm{e}}^{(1)} (r,\vartheta_0) = \tilde{T}_{\mathrm{e}}^{(1)} (r,\vartheta_0)$ and $T_{\mathrm{e}}^{(1)} (r,-\vartheta_0) = \tilde{T}_{\mathrm{e}}^{(1)} (r,-\vartheta_0) + \beta \hat{T}_{\mathrm{e}}^{(1)} (r,-\vartheta_0)$, where $\tilde{T}_{\mathrm{e}}^{(1)} (r,\pm \vartheta_0)$ are independent of $\beta$, while
\be
    \hat{T}_{\mathrm{e}}^{(1)} (r,-\vartheta_0) = \frac{2 \, \delta T \left\{ 1 + \alpha \left[ \ln(\ell/r_0) - \gamma \right] \right\} e^{-r/\ell}} {1 + \alpha\ln(R/r_0)} \label{eq:T-e}
\ee
with $\gamma\approx 0.577$ the Euler-Mascheroni constant.
The solution of Eqs.~\eqref{eq:laplace-n}-\eqref{eq:interface-bulk-n} for the bulk temperature can then be written as $T_{\mathrm{b}}^{(1)} (r,\vartheta) = \tilde{T}_{\mathrm{b}}^{(1)} (r,\vartheta) + \beta \hat{T}_{\mathrm{b}}^{(1)} (r,\vartheta)$, where $\tilde{T}_{\mathrm{b}}^{(1)} (r,\vartheta)$ and $\hat{T}_{\mathrm{b}}^{(1)} (r,\vartheta)$ are independent of $\beta$.
Assuming that the temperature probe couples to the bulk phonons, the sensitivity of a local temperature measurement to tunneling processes in the central region is characterized by $\hat{T}_{\mathrm{b}}^{(1)} (r,\vartheta) = \partial T_{\mathrm{b}}^{(1)} (r,\vartheta) / \partial \beta$.
For simplicity, we focus on the angular average of this temperature sensitivity, $\langle \hat{T}_{\mathrm{b}}^{(1)} (r) \rangle \equiv (2\vartheta_0)^{-1} \int_{-\vartheta_0}^{+\vartheta_0} \dif \vartheta \, \hat{T}_{\mathrm{b}}^{(1)} (r,\vartheta)$, which provides a lower bound on the maximal temperature sensitivity at a given radius $r$ within the lobe.
In Appendix~\ref{app:phonon_appendix}, we solve Eqs.~\eqref{eq:laplace-n}-\eqref{eq:interface-bulk-n} with $n=1$ to derive approximate expressions for this average temperature sensitivity in the limits of large and small $\alpha = \kappa_{\mathrm{c}} / (\kappa_{\mathrm{b}} \vartheta_0)$:
\begin{widetext}
\be
	\langle \hat{T}_{\mathrm{b}}^{(1)} (r) \rangle = \frac{\partial \langle T_{\mathrm{b}}^{(1)} (r) \rangle} {\partial \beta} \approx
	\begin{cases}
		\frac{2}{\pi^2}\left[\ln(\ell / r_0) - \gamma\right]\frac{\kappa_{\mathrm{e}} \delta T}{\kappa_{\mathrm{b}} \vartheta_0} \sin\left[\frac{\pi\ln(R/\ell)}{\ln(R/r_0)}\right]\sin\left[\frac{\pi\ln(R/r)}{\ln(R/r_0)}\right] & (\alpha \gg 1), \\
		\frac{8\ln(R/r_0)}{\pi^2}\frac{\kappa_{\mathrm{e}} \delta T}{\kappa_{\mathrm{b}} \vartheta_0} \sin\left[\frac{\pi\ln(R/\ell)}{2\ln(R/r_0)}\right]\sin\left[\frac{\pi\ln(R/r)}{2\ln(R/r_0)}\right] & (\alpha \ll 1).
	\end{cases}\label{eq:phonon_correction}
\ee
\end{widetext}
From the dependence of $\langle \hat{T}_{\mathrm{b}}^{(1)} (r) \rangle$ on the radius $r$, we conclude that the temperature sensitivity is maximized at $r \sim \sqrt{Rr_0}$ for $\alpha \gg 1$ and at $r\sim r_0$ for $\alpha \ll 1$.
In both limits, the maximal temperature sensitivity is on the order of $\kappa_{\mathrm{e}} \delta T / (\kappa_{\mathrm{b}} \vartheta_0)$.
This result shows that the iterative procedure is convergent for $\kappa_{\mathrm{e}} \ll \kappa_{\mathrm{b}} \vartheta_0$ and that, given fixed $\kappa_{\mathrm{e}}$ and $\kappa_{\mathrm{b}}$, the temperature sensitivity is significantly enhanced for $\vartheta_0 \ll 1$.
In this limit, the angular average $\langle \hat{T}_{\mathrm{b}}^{(1)} (r) \rangle$ also provides an accurate lower bound on the maximal temperature sensitivity (corresponding to $\vartheta = -\vartheta_0$) at the given radius $r$.
Thus, we find that, by appropriately tailoring the geometry of the sample, one may significantly enhance the visibility of corrections to the thermal transport which are, in turn, sensitive to edge tunneling processes in the central region.

In addition to local temperature measurements, these edge tunneling processes can also be detected by measuring the total heat current between the hot and the cold leads.
Evaluating this heat current at the interface of the right lobe and the central region, it is given by $J = \kappa_{\mathrm{e}} [T_{\mathrm{e}, \mathrm{in}}^+ - T_{\mathrm{e}, \mathrm{out}}^-] + J_{\mathrm{b}}$, where the two terms correspond to the edges and the bulk of the narrow channel (see Fig.~\ref{fig:two_lobes}), respectively.
In terms of the bulk and edge temperatures $T_{\mathrm{b}, \mathrm{e}}$ of the lobe at the interface, the total heat current then becomes
\be
    J = \kappa_{\mathrm{e}} \left[ T_{\mathrm{e}} (r_0, \vartheta_0) - T_{\mathrm{e}} (r_0, -\vartheta_0) \right] + 2\kappa_{\mathrm{c}} \langle T_{\mathrm{b}} (r_0) \rangle.
\ee
Finally, up to the leading-order perturbative corrections, $T_{\mathrm{b}, \mathrm{e}}^{(1)}$, the sensitivity of the heat current to tunneling processes in the central region is
\be
    \frac{\partial J} {\partial \beta} = -\kappa_{\mathrm{e}} \hat{T}_{\mathrm{e}}^{(1)} (r_0, -\vartheta_0) + 2\kappa_{\mathrm{c}} \langle \hat{T}_{\mathrm{b}}^{(1)} (r_0) \rangle.
\ee
For $\alpha \ll 1$, the first term, from Eq.~\eqref{eq:T-e}, is approximately $-2\kappa_{\mathrm{e}} \delta T$, while the second term, according to Eq.~\eqref{eq:phonon_correction}, is positive and on the order of $\alpha \kappa_{\mathrm{e}} \delta T \ll \kappa_{\mathrm{e}} \delta T$.
Therefore, we obtain that the heat-current sensitivity in this limit is $\partial J / \partial \beta \approx -2\kappa_{\mathrm{e}} \delta T$.
Approaching $\alpha \sim 1$, the negative first term slightly decreases in magnitude, while the positive second term increases to the same order of magnitude, $\kappa_{\mathrm{e}} \delta T$, as the first term.
Given that $\partial J / \partial \beta$ must be negative on physical grounds (since stronger tunneling means smaller energy transfer between the two lobes), we then deduce that the heat-current sensitivity must be suppressed for $\alpha \gg 1$, even though the approximate expression for $\langle \hat{T}_{\mathrm{b}}^{(1)} (r_0) \rangle$ in Eq.~\eqref{eq:phonon_correction} is not accurate enough (simply zero) in this limit.
We emphasize that this result is further confirmed by the simple heat-resistor picture introduced in the next subsection.

\begin{figure*}
	\centering
	\includegraphics[width=2\columnwidth]{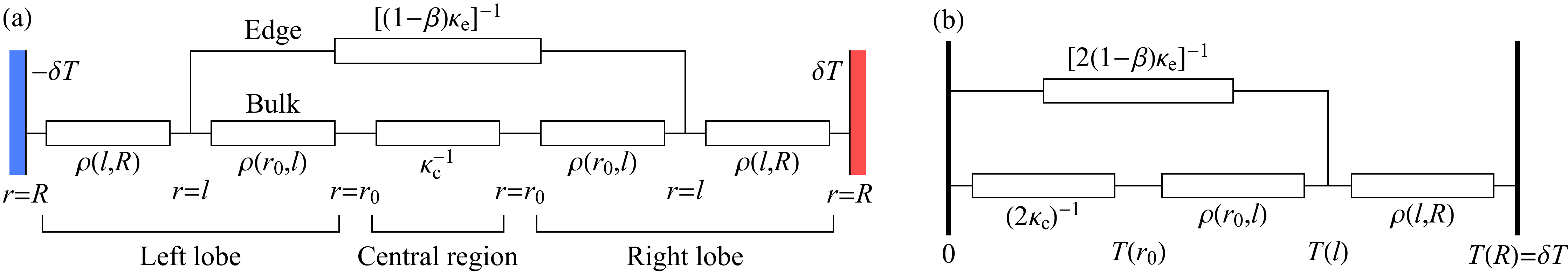}
	\caption{
	(a) Effective heat-resistor network of the phonon-coupled chiral spin liquid.
	Each resistor represents a heat-carrying component of the system, with the resistors in the upper (lower) row accounting for the edge mode (bulk phonons).
	The right and left leads are fixed at temperatures $\delta T$ and $-\delta T$, respectively.
	(b) Reduced heat-resistor network representing the right half of the system.
	The right lead is fixed at temperature $\delta T$, while the center of the system is fixed at temperature $0$.
	}
	\label{fig:resistor}
\end{figure*}

\subsection{Heat-Resistor Picture}

To understand the thermal transport in a less rigorous but more intuitive manner, we now consider the effective heat-resistor network shown in Fig.~\ref{fig:resistor}(a).
Each heat resistor represents a heat-carrying component of our phonon-coupled chiral spin liquid and is analogous to an electrical resistor.
In this analogy, temperature (heat current) then corresponds to voltage (electrical current).
We note that the simple heat-resistor picture is complementary to the perturbative approach in the previous subsection and, as a result of its one-dimensional nature, is expected to be most reliable for $\vartheta_0 \ll 1$.

The heat-resistor network in Fig.~\ref{fig:resistor}(a) accounts for the entire system between the hot and cold leads.
In terms of bulk phonon transport, the heat resistance of the central region is $\kappa_{\mathrm{c}}^{-1}$, while each section of either lobe between radii $r_1$ and $r_2 > r_1$ corresponds to heat resistance
\be
    \rho (r_1, r_2) = \int_{r_1}^{r_2} \frac{\dif r} {2 \kappa_{\mathrm{b}} r \vartheta_0} = \frac{\ln (r_2 / r_1)} {2 \kappa_{\mathrm{b}} \vartheta_0}.
\ee
For radii $r > \ell$ in either lobe, the edge mode thermalizes well with the bulk phonons and, assuming $\kappa_{\mathrm{e}} \lesssim \kappa_{\mathrm{b}} \vartheta_0$, its effect on the thermal transport can be neglected.
In contrast, within the central region and for radii $r < \ell$ in either lobe, the edge modes are thermally decoupled from the bulk phonons.
Therefore, they facilitate direct thermal coupling between radius $r = \ell$ in the right lobe and radius $r = \ell$ in the left lobe.
If the corresponding temperatures in the two respective lobes are $T^{\pm} (\ell)$, the top and bottom edges (see Fig.~\ref{fig:two_lobes}) then carry a net heat current $J_{\mathrm{e}, \mathrm{in}}^+ - J_{\mathrm{e}, \mathrm{out}}^- = (1-\beta) \kappa_{\mathrm{e}} [T^{+} (\ell) - T^{-} (\ell)]$ from the right lobe to the left lobe.
Hence, the two edges can be represented with an effective heat resistance $[(1-\beta) \kappa_{\mathrm{e}}]^{-1}$ between radii $r = \ell$ in the two lobes.

To analyze the heat-resistor network in Fig.~\ref{fig:resistor}(a), we first exploit the inversion symmetry around the center of the system to obtain the reduced heat-resistor network in Fig.~\ref{fig:resistor}(b).
This network only contains the right half of the system between the right lead fixed at temperature $\delta T$ and the center fixed at temperature $0$.
It is then clear that the temperature most sensitive to the tunneling parameter $\beta$ is $T(\ell)$ at radius $r = \ell$ within the lobe.
From Fig.~\ref{fig:resistor}(b), this temperature is given by
\be
    T(\ell) = \frac{\delta T \left[ \rho(\ell, R) \right]^{-1}} {2 (1-\beta) \kappa_{\mathrm{e}} + \left[ \rho(\ell) \right]^{-1} + \left[ \rho(\ell, R) \right]^{-1}},
\ee
while the corresponding temperature sensitivity becomes
\be
    \frac{\partial T(\ell)} {\partial \beta} = \frac{2 \kappa_{\mathrm{e}} \delta T \left[ \rho(\ell, R) \right]^{-1}} {\left\{ 2 (1-\beta) \kappa_{\mathrm{e}} + \left[ \rho(\ell) \right]^{-1} + \left[ \rho(\ell, R) \right]^{-1} \right\}^2}, \label{eq:resistor-1}
\ee
where $\rho(\ell) = (2 \kappa_{\mathrm{c}})^{-1} + \rho(r_0, \ell)$.
Taking $\kappa_{\mathrm{e}} \ll \kappa_{\mathrm{b}} \vartheta_0$, which corresponds to the perturbative approach in the previous subsection, the temperature sensitivities for large and small $\alpha = \kappa_{\mathrm{c}} / (\kappa_{\mathrm{b}} \vartheta_0)$ are then found to be
\be
    \frac{\partial T(\ell)} {\partial \beta} \approx
	\begin{cases}
		\frac{\kappa_{\mathrm{e}} \delta T \ln (R / \ell) \left[ \ln (\ell / r_0) \right]^2} {\kappa_{\mathrm{b}} \vartheta_0 \left[ \ln (R / r_0) \right]^2} & (\alpha \gg 1), \\

		\frac{\kappa_{\mathrm{e}} \delta T \ln (R / \ell)} {\kappa_{\mathrm{b}} \vartheta_0} & (\alpha \ll 1).
	\end{cases} \label{eq:resistor-2}
\ee
Remarkably, Eqs.~\eqref{eq:phonon_correction} and \eqref{eq:resistor-2}, obtained from the perturbative approach and the heat-resistor picture, respectively, give the same order of magnitude, $\kappa_{\mathrm{e}} \delta T / (\kappa_{\mathrm{b}} \vartheta_0)$, for the temperature sensitivity and the same qualitative dependence on the edge-bulk thermalization length $\ell$.
In particular, $\partial T(\ell) / \partial \beta$ vanishes for both $\ell \to R$ and $\ell \to r_0$ in the $\alpha \gg 1$ limit, whereas it vanishes for $\ell \to R$ but is maximal for $\ell \to r_0$ in the $\alpha \ll 1$ limit.
By analyzing the general result in Eq.~\eqref{eq:resistor-1}, we can further deduce that the temperature sensitivity is maximized if $[\rho(\ell)]^{-1}$ is very small while $[\rho(\ell, R)]^{-1}$ is similar to $\kappa_{\mathrm{e}}$, which translates into $\kappa_{\mathrm{c}} \ll \kappa_{\mathrm{e}} \sim \kappa_{\mathrm{b}} \vartheta_0$.
In this case, the temperature sensitivity reaches the largest order of magnitude that is theoretically possible: $\partial T(\ell) / \partial \beta \sim \delta T$.

The heat-resistor network in Fig.~\ref{fig:resistor}(b) can also be used to compute the total heat current leaving the right lead: $J = [\rho(\ell, R)]^{-1} [T(R) - T(\ell)]$.
The sensitivity of the heat current to the tunneling parameter $\beta$ is then
\be
    \frac{\partial J} {\partial \beta} = -\left[ \rho(\ell, R) \right]^{-1} \frac{\partial T(\ell)} {\partial \beta}, \label{eq:resistor-3}
\ee
and, in the limit of $\kappa_{\mathrm{e}} \ll \kappa_{\mathrm{b}} \vartheta_0$, we readily obtain
\be
    \frac{\partial J} {\partial \beta} \approx
	\begin{cases}
		-\frac{2 \kappa_{\mathrm{e}} \delta T \left[ \ln (\ell / r_0) \right]^2} {\left[ \ln (R / r_0) \right]^2} & (\alpha \gg 1), \\

		-2 \kappa_{\mathrm{e}} \delta T & (\alpha \ll 1).
	\end{cases} \label{eq:resistor-4}
\ee
Once again, these results are consistent with those of the perturbative approach.
Also, from Eqs.~\eqref{eq:resistor-1} and \eqref{eq:resistor-3}, we find that the magnitude of the heat-current sensitivity is generally maximized when $\kappa_{\mathrm{e}}$ and $[\rho(\ell)]^{-1}$ are both much smaller than $[\rho(\ell, R)]^{-1}$, which precisely corresponds to the second case of Eq.~\eqref{eq:resistor-4}.
Finally, we point out that, for $\kappa_{\mathrm{c}} \ll \kappa_{\mathrm{e}} \ll \kappa_{\mathrm{b}} \vartheta_0$, the heat current itself takes the simple form $J \approx 2 (1-\beta) \kappa_{\mathrm{e}} \delta T$.
Therefore, in the absence of any pinch points in the central region ($\beta = 0$), measuring the heat current can be used to directly extract the edge thermal Hall conductivity $\kappa_{\mathrm{e}}$ and, hence, the chiral central charge of the corresponding edge CFT.

\section{Tunneling at Point Contacts}\label{sec:tunneling}

The temperature profiles calculated in the previous section relied upon the parameter $\beta$ describing the fraction of energy that jumps across the point contacts in the narrow central region.
This quantity may be directly calculated by a perturbative treatment of the quasiparticle tunneling processes that shuttle between the upper and lower edges.
To this end, we consider the specific case of the Ising topological order realized in a non-Abelian Kitaev spin liquid phase~\cite{Kitaev2006}.
Here the edge exhibits central charge $c = 1/2$ and hosts a single chiral Majorana fermion $\gamma(x)$ governed by the Hamiltonian $H_0 = -iv\int_x \gamma \partial_x \gamma$, with $v$ the edge velocity (which we subsequently set to $v = 1$ in this section).
The bulk supports three kinds of gapped quasiparticles: trivial bosons ($\mathbb{I}$), Majorana fermions ($\psi$), and Ising anyons ($\sigma$).
The latter two are mutual semions and satisfy the non-trivial fusion rules $\sigma \times \sigma = \mathbb{I} + \psi$, and $\sigma \times \psi = \sigma$, and $\psi \times \psi = \mathbb{I}$.

\begin{figure}
	\centering
	\includegraphics[width=.9\columnwidth]{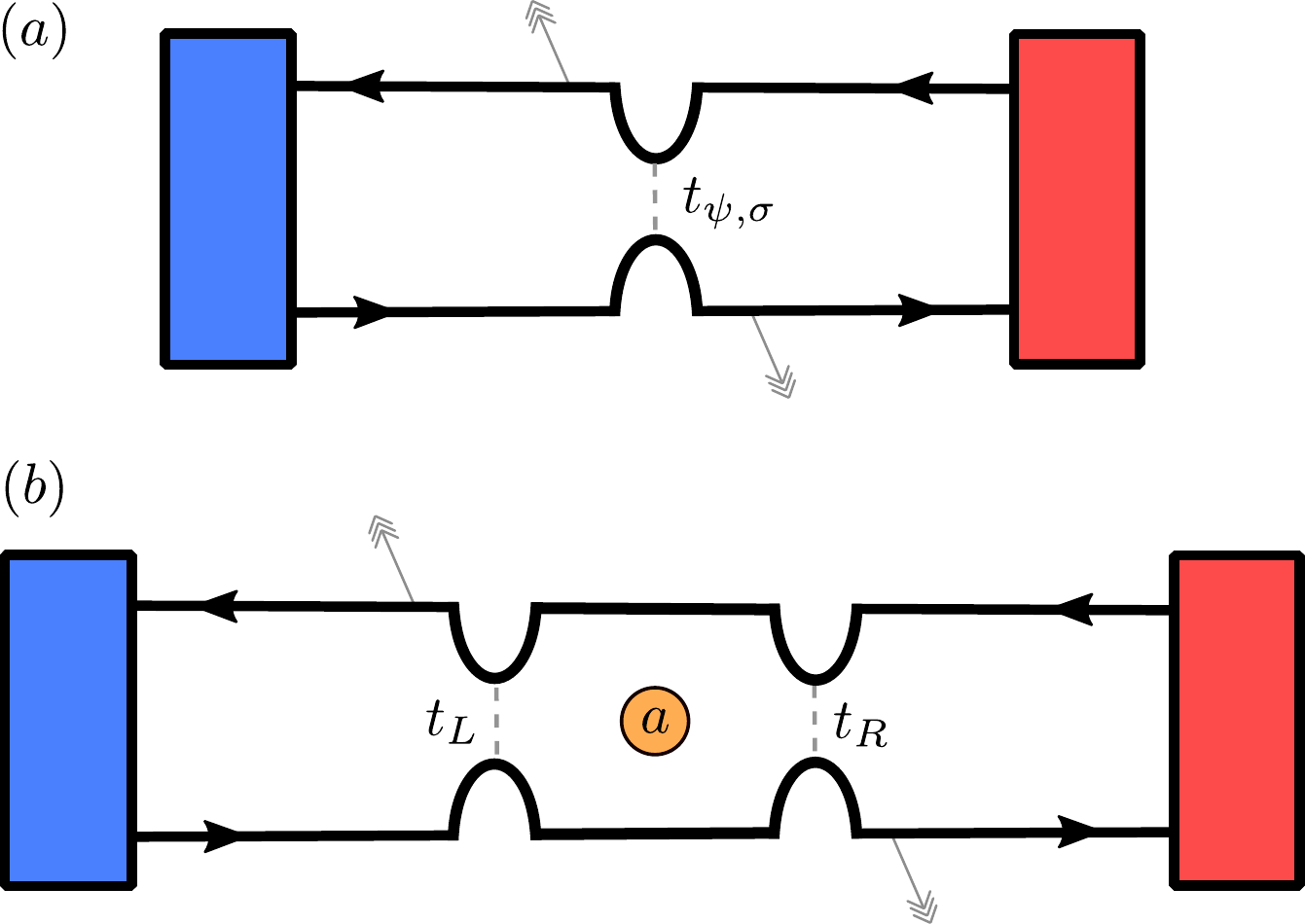}
	\caption{
	(a) Single pinch and (b) double pinch geometries for a chiral spin liquid heated from the right and cooled from the left.
	In the double pinch geometry, we allow for a bulk quasiparticle of type $a$ in the enclosed region.
	We imagine measuring the heat current immediately after the pinch point(s), at the locations marked with light gray arrows.
	}
	\label{fig:tunneling_geometries}
\end{figure}

Consider then introducing a pinch into the spin liquid that brings opposing edges close to one another.
At such a point contact, both fermions and Ising anyons may tunnel across, as captured by the perturbing Hamiltonian
\be
    H_\text{tun} = -it_\gamma \gamma(x_{\mathrm{top}})\gamma(x_{\mathrm{bot}}) + e^{-i\pi/16}t_\sigma \sigma(x_{\mathrm{top}})\sigma(x_{\mathrm{bot}}).
\ee
Here $x_{\mathrm{top}}$ ($x_{\mathrm{bot}}$) is the coordinate of the pinch on the top (bottom) edge---with $x$ values increasing along the propagation direction---and $t_{\gamma}$ ($t_{\sigma}$) is the tunneling amplitude for fermions (Ising anyons).
The phenomenology of such tunneling has been studied extensively~\cite{Fendley2009, Fendley2007, Fendley_2006}, and one may straightforwardly evaluate the resulting corrections to thermal transport.
In the interest of drawing a distinction between signatures inherent to quasiparticle tunneling versus those arising from quasiparticle braiding, we consider both single-pinch and double-pinch geometries (see Fig.~\ref{fig:tunneling_geometries}).

Mirroring the analysis from Sec.~\ref{sec:phonon_coupling}, we couple the spin liquid to heating and cooling elements so as to establish a temperature differential between the opposing ends.
We emphasize, however, that each heating or cooling element in Fig.~\ref{fig:tunneling_geometries} corresponds to an entire lobe connected to the central region (see Fig.~\ref{fig:device}) and accounts for the actual thermal lead coupling to the bulk as well as the edge-bulk thermalization.
Thus, these effective heating and cooling elements directly set the temperatures of the incoming edges.
We can further assume without loss of generality that the hot end is held at temperature $T$ while the cold end is held at zero temperature.
We may then neglect any heat current originating from the bottom edge where it couples to the cooling element.

Since we are interested in the fraction of energy that continues through the central region, we may calculate the heat current along the top edge.
Fourier transforming the free Hamiltonian gives a linear spectrum $\varepsilon(k) = k$.
The heating element then excites momentum modes whose population follows from the Fermi-Dirac distribution $n(k,T) = [e^{\varepsilon(k) / T} + 1]^{-1}$.
We assume that the heat current is measured along the top edge shortly after the point contact (see the probe marker in Fig.~\ref{fig:tunneling_geometries}).
The heat current may be explicitly written as
\be
    I(T) = \int_0^\Lambda \frac{\dif k}{2\pi} \varepsilon(k) n(k,T) \abs{A(k)}^2,
\ee
where $\Lambda$ is a momentum cutoff and $A(k)$ is the transmission amplitude along the top edge.
The linearized thermal conductance of the edge is then given by $\kappa = \partial_T I(T)$.
Since we are dealing with the heat-current component that continues along the top edge, this conductance corresponds to $\kappa = \kappa_{\mathrm{e}} (1 - \beta)$ in our earlier notation.
Crucially, at the level of our treatment, temperature appears in the expression for the current only via the Fermi-Dirac distribution.
That is, we consider only the zero-temperature correlators within the $c=\tfrac12$ CFT when evaluating $\abs{A(k)}$.
A comparable result for a finite temperature edge can be found in Ref.~\onlinecite{Nilsson_2010}.

In the absence of any tunneling, $\abs{A(k)}$ is of course simply unity and one readily recovers the quantized result $\kappa = \pi T / 12$ (i.e., just $\kappa_{\mathrm{e}}$).
For finite tunneling amplitude ($t_{\gamma / \sigma} \not= 0$), one anticipates that energy can escape to the bottom edge, and so the heat current---and similarly $\abs{A(k)}$---will be reduced relative to the unperturbed case.
As we will show, $\abs{A(k)}$ then encodes both universal signatures of the quasiparticles involved in tunneling as well as information about the anyonic content of the bulk.

\subsection{Single Pinch Geometry}
Consider first a geometry with a single pinch at which quasiparticle tunneling may occur [Fig.~\ref{fig:tunneling_geometries}(a)].
In this case, it is convenient to define the coordinates so that $x_{\mathrm{top}} = x_{\mathrm{bot}} = 0$.
This geometry involves no braiding of edge excitations about bulk quasiparticles, so there are no interferometric effects.
Nonetheless, we may still glean some signature of the flavors of quasiparticles by universal corrections to the thermal transport.

\subsubsection{Fermion Tunneling}
Fermion tunneling is a quadratic perturbation, so we may treat its effects exactly by considering the Heisenberg equations of motion for the fermion fields $\gamma_{\mathrm{top}}(x,t)$ and $\gamma_{\mathrm{bot}}(x,t)$ on the top and bottom edges, respectively:
\be
\begin{aligned}
	i\partial_t \gamma_{\mathrm{top}}(x,t) &= -i\partial_x \gamma_{\mathrm{top}}(x,t) + it_\gamma \delta(x)\gamma_{\mathrm{bot}}(0,t),\\
	i\partial_t \gamma_{\mathrm{bot}}(x,t) &= -i\partial_x \gamma_{\mathrm{bot}}(x,t) - it_\gamma \delta(x)\gamma_{\mathrm{top}}(0,t).
\end{aligned}
\ee
Integrating over an infinitesimal window enclosing $x=0$, we find
\be
    \begin{aligned}
        \gamma_{\mathrm{top}}(0^+) &= -\frac{t_\gamma^2 - 4}{t_\gamma^2+4}\gamma_{\mathrm{top}}(0^-) + \frac{4t_\gamma}{t^2_\gamma + 4}\gamma_{\mathrm{bot}}(0^-),\\
        \gamma_{\mathrm{bot}}(0^+) &= -\frac{4t_\gamma}{t_\gamma^2 + 4}\gamma_{\mathrm{top}}(0^-) - \frac{t_\gamma^2 - 4}{t_\gamma^2 + 4}\gamma_{\mathrm{bot}}(0^-),
    \end{aligned}\label{eq:fermion_single_pinch_soln}
\ee
where $x=0^-$ ($x=0^+$) denotes the position immediately  before (after) the constriction and we suppressed the time dependence for compactness.
These expressions encode the reflection and transmission coefficients at the point contact for a wavepacket with components originating on either the upper or lower edges.
Let us focus on a plane wave with momentum $k$ that is incident on the constriction from the top edge.
The amplitude for transmission along the top edge then satisfies
\be
	\abs{A(k)}^2 =  \abs{\frac{t_\gamma^2 - 4}{t_\gamma^2 + 4}}^2 \approx 1 - t_\gamma^2.
\ee
Since the amplitude is momentum-independent, the corrected thermal conductance is simply rescaled downwards,
\be
	\kappa = \kappa_{\mathrm{e}} \abs{\frac{t_\gamma^2  - 4}{t_\gamma^2 + 4}}^2 \approx \kappa_{\mathrm{e}} \left(1 - t_\gamma^2\right),
\ee
as one might expect from the fact that fermion tunneling is a marginal perturbation.
With just fermion tunneling, one can continuously tune from perfect transmission ($\abs{A(k)} = 1$) to perfect reflection ($\abs{A(k)} = 0$) at the point contact by modulating the tunneling strength $t_\gamma$.
Since this correction introduces no additional temperature dependence to the edge conductance, it may not be a readily distinguishable feature in noisy measurements when $t_\gamma$ is perturbatively small.
In the two-pinch interferometer geometry examined later, however, fermion tunneling does induce nontrivial corrections associated with braiding.

\subsubsection{Ising Anyon Tunneling}

We now consider the more relevant tunneling process that may take place at point contacts in this non-Abelian spin liquid phase---Ising anyon tunneling.
For a geometry as shown in Fig.~\ref{fig:tunneling_geometries}(a), the upper and lower edges are treated as independent chiral edges coupled only through the point contact.
This treatment allows all correlation functions to be simplified by a cluster decomposition between the two edges.
Such a decomposition is equivalent to considering the system as a single continuous edge but taking the limit where the circumference between the upper and lower edges is taken to infinity.
This is essentially the scenario one has in mind for a setup as in Fig.~\ref{fig:device} where the lobes adjacent to the central region are macroscopic (i.e., much larger than the coherence and thermalization lengths).

If we measure the temperature on the upper edge shortly after the point contact, then the relevant Green's function in the interaction picture takes the form
\be
G(t) = \langle \gamma_{\mathrm{top}}(L,t)\gamma_{\mathrm{top}}(-L,0) \rangle,
\ee
where $L > 0$ is an arbitrary coordinate so that we are considering fermions starting near the hot reservoir and ending near the cold reservoir.
The transmission amplitude $A(k)$ then is related to this Green's function in momentum space (see Appendix~\ref{app:cft_appendix}), and we thus seek the leading correction to the Green's function arising from the Ising anyon tunneling Hamiltonian.

At first order in Ising anyon tunneling, one essentially has to compute a correlator like $\langle \sigma \rangle_{\mathrm{bot}} \langle \gamma \sigma \gamma \rangle_{\mathrm{top}}$ where the subscripts indicate that the correlators are evaluated on the top or bottom edge.
Both terms must vanish since the nontrivial primary fields exhibit vanishing ground-state expectation values (e.g., $\bra{0} \sigma \ket{0} = 0$).
Put simply, tunneling a single Ising anyon cannot move a fermion from the top edge to the bottom edge, and so cannot affect the heat current along the edge.

The first non-trivial correction then comes at second order in Ising anyon tunneling,
\be
    \sim \int \dif s_1\dif s_2 \langle \gamma_{\mathrm{top}}(L,t) H_\sigma(s_1) H_\sigma(s_2) \gamma_{\mathrm{top}}(-L,0) \rangle,
\ee
where implicitly one may employ a cluster decomposition.
Here we have adopted the shorthand $H_\sigma(t) = t_\sigma \sigma_{\mathrm{top}}(0,t)\sigma_{\mathrm{bot}}(0,t)$ for the Ising anyon tunneling Hamiltonian in the interaction picture.
As we show in Appendix~\ref{app:cft_appendix}, the leading order correction to $\abs{A(k)}^2$ due to Ising anyon tunneling diverges as $t_{\sigma}^2 k^{-7/4}$ as $k\rightarrow 0$.
While the corresponding correction to the heat current $I(T)$ vanishes as $T^{1/4}$, the correction to the edge conductance, $\Delta \kappa \sim -t_\sigma^2 T^{-3/4}$, diverges at temperature $T = 0$.
This divergence comports with the observation that Ising anyon tunneling is a relevant perturbation under which the system flows to a split bar where all energy is scattered to the bottom edge at the point contact.

In the interest of treating Ising anyon tunneling perturbatively, we therefore assume that the temperature $T$ is finite and large enough that the correction to the edge conductance remains small.
The necessary temperature $T$ may still be small relative to other energy scales (e.g., the bulk gap), but this assumption ensures that we are not operating the device in the split bar regime.
For the sake of notational brevity, we define an effective renormalized Ising anyon tunneling strength $\tilde t_\sigma \propto t_\sigma T^{-7/8}$, where the precise value of the proportionality constant is given in Appendix~\ref{app:cft_appendix}.
In terms of this effective tunneling amplitude, the correction to the thermal conductance takes on the following simple form:
\be
	\Delta \kappa = -\kappa_{\mathrm{e}} \tilde t_{\sigma}^2 \propto -\kappa_{\mathrm{e}} t_\sigma^2 T^{-7/4}.
\ee
This $-7/4$ power law has been discussed by Nilsson and Akhmerov~\cite{Nilsson_2010} in the context of electrical conductance measurements for a topological insulator in proximity with a superconductor.
In both scenarios, this exponent comes from the scaling dimension of the tunneling operator that corrects the fermionic edge current.
To the best of our knowledge, this power law is not reproduced by other less exotic phenomenology, and so provides a unique signature of fractional quasiparticles tunneling at the point contact in a single-pinch geometry.

\subsection{Double Pinch Geometry}

We now turn our attention to the double pinch geometry depicted in Fig.~\ref{fig:tunneling_geometries}(b).
Here, second order tunneling processes can transport an edge excitation all the way around the enclosed (bulk) region by first tunneling to the bottom edge via the left constriction, and then back to the top edge via the right constriction.
Such processes not only change the path length traversed by an excitation---which is imprinted on the dynamical phase acquired---but also \emph{braid} the edge anyon around any quasiparticles residing in the bulk.
Interference between the different paths edge anyons may take then depends on the braiding statistics with the encircled bulk quasiparticles, in turn affecting the transmission probability $\abs{A(k)}^2$.
This dependence opens the door for using thermal transport measurements to detect individual bulk quasiparticles.

Let $x_1, x_2$ be the coordinates of the left and right pinch on the top edge.
On the bottom edge we invert the coordinate system so that $x_1$ and $x_2$ are now the coordinates of the right and left pinches, respectively.
The tunneling Hamiltonian then becomes
\be
\begin{aligned}
H_\text{tun} = & -it_{L,\gamma} \gamma_{\mathrm{top}}(x_1)\gamma_{\mathrm{bot}}(x_2) - it_{R,\gamma} \gamma_{\mathrm{top}}(x_2)\gamma_{\mathrm{bot}}(x_1) \\
&+ e^{-i\pi/16}t_{L,\sigma}\sigma_{\mathrm{top}}(x_1)\sigma_{\mathrm{bot}}(x_2) \\
&+ e^{-i\pi/16}t_{R,\sigma}\sigma_{\mathrm{top}}(x_2)\sigma_{\mathrm{bot}}(x_1).
\end{aligned}
\ee
Under the assumption that tunneling at either contact has comparable amplitude ($t_L \sim t_R$), we seek corrections to the thermal conductance at up to second order in tunneling, which is sufficient to capture interference effects.

\subsubsection{Fermion Tunneling}
At second order in fermion tunneling, we now anticipate an order $\mathcal{O}(t_{L,\gamma}t_{R,\gamma})$ correction that corresponds to a fermion encircling the entire bulk region.  Since fermions and Ising anyons are mutual semions, such processes are sensitive to the parity of the number of Ising anyons in the bulk.  That is, edge fermions that encircle $n_\sigma$ bulk Ising anyons acquire a braiding phase $(-1)^{n_\sigma}$.  Edge fermions encircling either bulk bosons or bulk fermions, by contrast, do not acquire any nontrivial braiding phases.

\begin{figure}[h]
	\centering
	\includegraphics[width=\columnwidth]{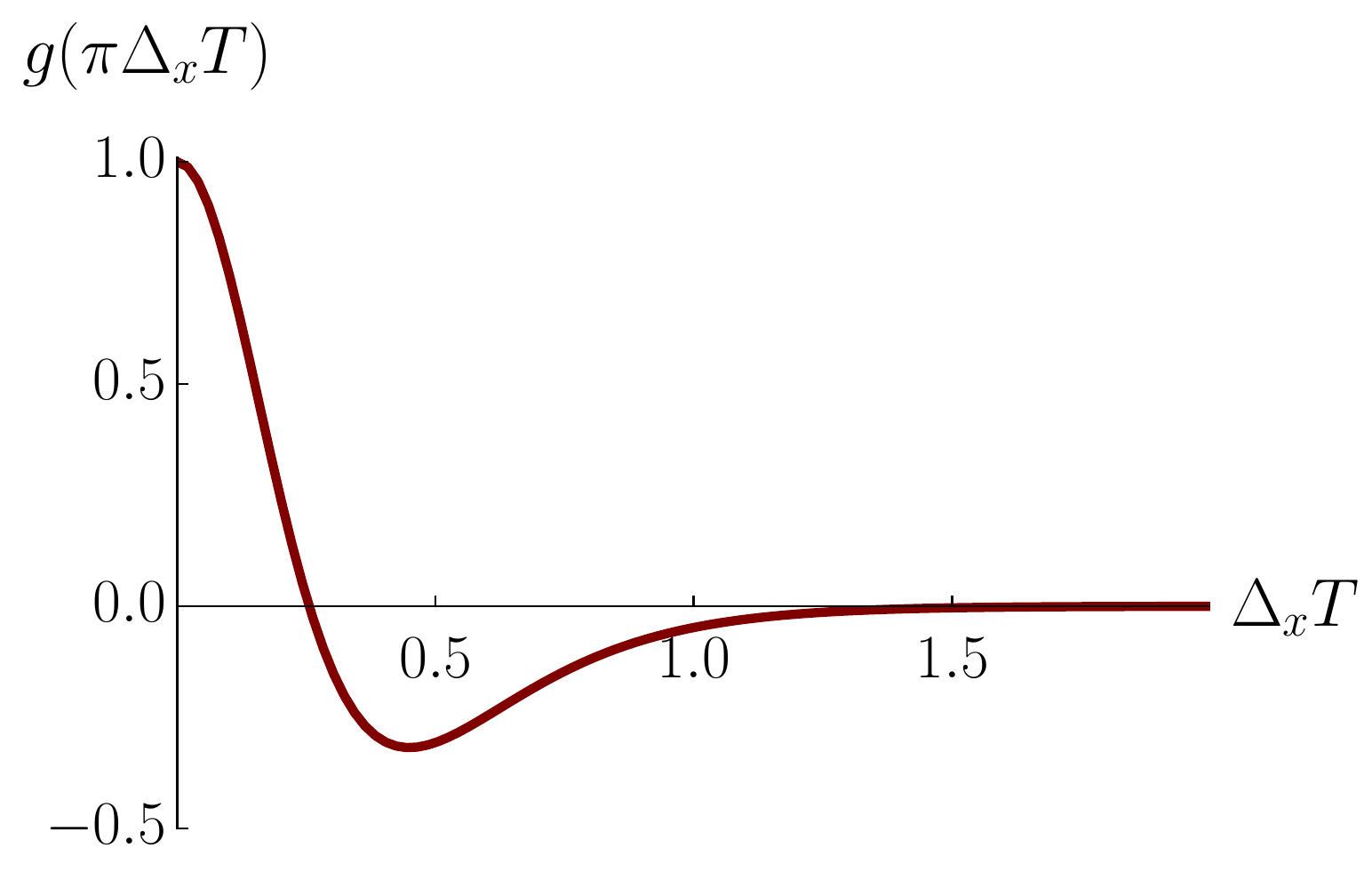}
	\caption{
	Dimensionless function $g(\pi \Delta_x T)$ for the fermion braiding correction to the thermal conductance against temperature $T$ and pinch separation $\Delta_x$ (or braiding path length difference $2\Delta_x$) in a two-pinch geometry.
	For $\Delta_x T \gg 1$, this correction is exponentially suppressed; in the opposite limit, $g(\pi \Delta_x T)$ exhibits a quadratic deviation away from unity.
	}
	\label{fig:fermion_braiding_correction}
\end{figure}

Let us then calculate the transmission amplitude in the two pinch geometry.
Once again fermion tunneling may be treated exactly by considering the equations of motion
\be
\begin{aligned}
	i\partial_t \gamma_{\mathrm{top}}(x,t) =& -i\partial_x \gamma_{\mathrm{top}}(x,t) + it_{L,\gamma}\delta(x-x_1)\gamma_{\mathrm{bot}}(x_2,t) \\ &+ it_{R,\gamma}\delta(x-x_2)\gamma_{\mathrm{bot}}(x_1,t),\\
	i\partial_t \gamma_{\mathrm{bot}}(x,t) =& -i\partial_x \gamma_{\mathrm{bot}}(x,t) - it_{L,\gamma}\delta(x-x_2)\gamma_{\mathrm{top}}(x_1,t)\\ &- it_{R,\gamma}\delta(x-x_1)\gamma_{\mathrm{top}}(x_2,t).
\end{aligned}
\ee
As before, we consider plane waves propagating on the edge.
Between the constrictions, the fermions propagate freely and so we may write
$\gamma_{\mathrm{top / bot}}(x_2^-,t) = e^{ik\Delta_x}\gamma_{\mathrm{top / bot}}(x_1^+,t)$ for momentum $k$ and with $\Delta_x = x_2 - x_1$ the separation between the pinches.
Dropping the explicit time dependence for simplicity, one finds
\be
    \begin{aligned}
        e^{-ik\Delta_x} \gamma_{\mathrm{top}}(x_2^+) \approx \gamma_{\mathrm{bot}}(x_1^-)\left(e^{ik\Delta_x}t_{L,\gamma} + e^{-ik\Delta_x}t_{R,\gamma}\right) \\
        + \gamma_{\mathrm{top}}(x_1^-)\left[1 - \frac{t_{L,\gamma}^2 + t_{R,\gamma}^2}{2} - (-1)^{n_\sigma} e^{i2k\Delta_x}t_{R,\gamma}t_{L,\gamma}\right],
    \end{aligned}
\ee
where we have explicitly included the statistical braiding phase anticipated for the $\mathcal{O}(t_{L,\gamma}t_{R,\gamma})$ term.
For a wavepacket incident along the upper edge, the transmission amplitude follows as
\be
    \abs{A(k)}^2 \approx 1 - \abs{t_{L,\gamma} + (-1)^{n_\sigma} e^{i2k\Delta_x} t_{R,\gamma}}^2.
\ee
The crucial difference from the single pinch geometry is that the $\mathcal{O}(t_{L,\gamma}t_{R,\gamma})$ braiding term carries a phase dependence on both the bulk anyon content ($n_\sigma$) as well as the momentum $k$.
The latter distinction imparts  nontrivial temperature dependence in the thermal transport.
In particular, the corrected thermal conductance of the edge is now given by
\be
    \kappa = \kappa_{\mathrm{e}}\left[1 - t_{L,\gamma}^2 - t_{R,\gamma}^2 - 2 (-1)^{n_\sigma} t_{L,\gamma}t_{R,\gamma} g(\Delta_x \pi T)\right].
    \label{eq:fermion_twoPinch_trivial}
\ee
The dimensionless function $g(x)$ describing the dependence of the fermion braiding term on temperature and pinch separation is
\be
    \begin{aligned}
        g(\Delta_x \pi T) &= \frac{1}{\kappa_{\mathrm{e}}} \partial_T \int_0^\infty \frac{\dif k}{2\pi} \varepsilon(k) n(k,T) \cos(2k\Delta_x) \\
        &= 3\csch^3(2\Delta_x \pi T)\biggl[\sinh(4\Delta_x \pi T) \\ & {\,\,\,\,\,\,} - \Delta_x \pi T\left(3 + \cosh(4\Delta_x \pi T)\right)\biggr].
    \end{aligned}
\ee
At zero temperature $g(0) = 1$, so for equal tunneling amplitudes $t_{L,\gamma} = t_{R,\gamma}$ and an odd number of Ising anyons in the bulk, this braiding term may negate the trivial $\mathcal{O}(t_{R/L, \gamma}^2)$ terms in Eq.~\eqref{eq:fermion_twoPinch_trivial}.
For small but finite temperature we find that $g(x) \approx 1 - \tfrac{14}{5}x^2$, giving an additional quadratic temperature dependence to the conductance correction that may be useful in distinguishing fermion braiding effects.
Going to large temperature or pinch separation, $g(x\gg1)$ is exponentially suppressed and fermion braiding becomes unobservable.
Considering the full $T$-dependence, one finds that the braiding correction can be tuned through zero by appropriately varying the temperature; see Fig.~\ref{fig:fermion_braiding_correction}.

In this analysis we assumed that the separation of the pinches $x_2 - x_1$ was the same on the top and bottom edges.
More generally, $2\Delta_x$ corresponds to the circumference of the enclosed region.
This dependence on the circumference provides an additional tuning parameter by which one might optimize the geometry to maximize the visibility of corrections to the thermal conductance.

\subsubsection{Ising Anyon Tunneling}

Next we include Ising anyon tunneling in the double pinch geometry.
To get some effect beyond that found in the single pinch geometry, we should consider processes wherein an Ising anyon braids around the bulk, analogous to the braiding processes for fermions examined above.
In this scenario, we may envision that a Majorana fermion incident on the constrictions splinters into two Ising anyons, one of which tunnels to the bottom edge.
This tunneled Ising anyon later returns to the top edge by tunneling across at the other constriction.
Such processes take a \emph{single} Ising anyon around  quasiparticles residing in the bulk.
Recall that taking an Ising anyon around a Majorana fermion yields a statistical phase factor of $-1$.
Then for $n_\psi$ bulk Majorana fermions contained between the constrictions, this braiding process yields a phase $(-1)^{n_\psi}$.
If the number of enclosed bulk Ising anyons $n_\sigma$ is odd, then the braid non-trivially rotates the many-body wavefunction---killing the interference term.
Putting these effects together, we must attach a factor of $(-1)^{n_\psi}\left[1 + (-1)^{n_\sigma}\right]/2$ to corrections arising from $\mathcal{O}(t_{L,\sigma}t_{R,\sigma})$ Ising anyon braiding processes.  Notice that when $n_\sigma$ is odd the braiding phase becomes independent of $n_\psi$; this result is indeed required by the fusion rule $\sigma \times \psi = \sigma$.

Based on our earlier analysis, we employ a semi-classical treatment of the Ising anyon tunneling in a two pinch geometry.
Again assuming that the effective tunneling amplitudes are $\tilde t_{L/R, \sigma} \propto t_{L/R, \sigma}T^{-7/8}$, we obtain the following correction to the thermal conductance:
\be
    \Delta\kappa = -\kappa_{\mathrm{e}}\left\{\tilde t_{L,\sigma}^2 + \tilde t_{R,\sigma}^2 +  (-1)^{n_\psi} \left[1 + (-1)^{n_\sigma}\right]\tilde t_{L,\sigma} \tilde t_{R,\sigma}\right\}.
\ee
Unlike the fermion braiding correction, the $\mathcal{O}(t_{L,\sigma}t_{R,\sigma})$ term above is sensitive both to fermions \emph{and} Ising anyons in the bulk, allowing for detection of individual bulk quasiparticles of either type.

\subsection{General Result}

Recall that the conductance $\kappa$ we derive here is equal to $\kappa_{\mathrm{e}}(1-\beta)$, where $\beta$ is the parameter describing the fraction of energy which tunnels across the constrictions.
Collecting our two-pinch-geometry results, we obtain
\be
\begin{aligned}
  	    &\beta(\Delta_x, T) \, \approx \, t_{L,\gamma}^2 + t_{R,\gamma}^2 + 2(-1)^{n_\sigma}t_{L,\gamma}t_{R,\gamma}g(\Delta_x \pi T) \\
      	 &+T^{-7/4}\biggl\{t_{L,\sigma}^2 + t_{R,\sigma}^2 + (-1)^{n_\psi}\left[1 + (-1)^{n_\sigma}\right]t_{L,\sigma}t_{R,\sigma} \biggr\}.
  	\end{aligned}\label{eq:tunneling_results_combined}
\ee
In a single pinch geometry, this result simplifies further to $\beta(T) = t_{\gamma}^2 + t_{\sigma}^2 T^{-7/4}$.
Note that we have opted to use $t_{\sigma}$ rather than $\tilde t_\sigma$ in order to emphasize the unique temperature dependence here.
In doing so, we have absorbed some proportionality constant into the bare tunneling amplitudes, the value of which is derived in Appendix~\ref{app:cft_appendix}.
To connect these results to our earlier analysis of heat exchange between the bulk phonons and the edge mode, $\beta$ should be evaluated with respect to the overall system temperature $T_0$, assuming that the deviations $\delta T$ are small.

It is worth noting that the Ising anyon tunneling terms should have an additional functional dependence when the temperature or pinch separation are taken to be very large.
Much as the fermion braiding correction $g(x)$ was exponentially suppressed for $x \gg 1$, we anticipate the Ising anyon tunneling corrections are also exponentially suppressed at high temperature \cite{Nilsson_2010}.
Within the temperature range where interference is feasible, however, Eq.~\eqref{eq:tunneling_results_combined} captures the essential corrections to the thermal transport due to quasiparticle tunneling in a two pinch geometry.

While the particular expression for $\beta$ derived here was for the case of Ising topological order ($c=\tfrac12$) in a Kitaev spin liquid, a similar analysis could be carried out for other chiral topological orders.
In general, the expression for $\beta$ in a single pinch geometry will always feature a temperature dependence $\beta \propto T^{-\eta}$ where $\eta = 2 - 4h$ is determined by the conformal dimension $h$ of the most relevant edge operator in the theory.
Similarly, the details of the interference corrections arising in a multi-pinch geometry will then of course depend on the braiding statistics of the quasiparticles in the particular phase.

\section{Discussion}\label{sec:discussion}

In the previous two sections, we described how thermal transport in the unconventional device geometry of Fig.~\ref{fig:device} can be used to probe the anyonic excitations of the non-Abelian Kitaev spin liquid.
Our main focus was anyonic edge interferometry, using two point contacts, which can detect individual anyons and directly observe their nontrivial braiding statistics.
In addition, we considered simpler measurements, requiring at most one point contact, for unambiguously identifying the underlying spin liquid via both the quantized thermal Hall conductivity~\cite{Kasahara2018, Tokoi, bruin2021} extracted in a more direct way and ``smoking-gun'' signatures of Ising-anyon tunneling.
Indeed, it was described in Sec.~\ref{sec:tunneling} how the tunneling parameter $\beta$ is determined by the edge tunneling processes in the central region, while it was explained in Sec.~\ref{sec:phonon_coupling} how $\beta$ (or a change in $\beta$) can be detected in the bulk thermal transport by measuring appropriate temperatures or heat currents.
In this section, we discuss potential challenges in the experimental observation of these signatures and provide simple guidelines for implementing our proposal in realistic candidate materials, such as $\alpha$-RuCl$_3$.

\subsection{Thermalization Length}

For effectively probing the quantum-coherent edge processes inside the central region via bulk thermal transport within the two lobes (see Fig.~\ref{fig:two_lobes}), the thermalization length $\ell$ between the edge mode and the bulk phonons must satisfy $r_0 \ll \ell \ll R$.
On the one hand, if the central region is larger than the thermalization length, $r_0 \gg \ell$, thermalization with bulk phonons may disrupt the quantum coherence of the edge processes.
On the other hand, if the lobes are smaller than the thermalization length, $R \ll \ell$, the edge mode is essentially decoupled from the bulk thermal leads and sensors.

The thermalization length is given by $\ell = \kappa_{\mathrm{e}} / \lambda$, where $\kappa_{\mathrm{e}} \propto T$ is the thermal Hall conductance of the edge mode, and $\lambda$ is the linearized thermal coupling between the edge mode and the bulk phonons.
Since $\lambda$ was argued~\cite{Ye_2018} to be proportional to $T^6$, the thermalization length is then expected to diverge as $\ell \propto T^{-5}$ at low temperatures.
Due to this strong temperature dependence of the thermalization length, our proposal is only applicable within a reasonable temperature range if the two characteristic sizes $r_0$ and $R$ differ by orders of magnitude.
Nevertheless, for $r_0 \sim$ 1 $\mu$m and $R \sim 1$ cm~\cite{Kasahara2018}, the range of applicability spans almost a decade in temperature, which is sufficient for observing all of our proposed signatures, including a reliable extraction of the universal power law $\beta \propto T^{-7/4}$ that corresponds to Ising-anyon tunneling.

Importantly, this range of applicability must also correspond to temperatures that are experimentally accessible but sufficiently small to be consistent with the other constraints described in the following subsections.
However, we remark that, according to Ref.~\onlinecite{Ye_2018}, the dominant low-temperature mechanism for the edge-bulk coupling $\lambda \propto T^6$ relies on the presence of disorder.
Therefore, the range of applicability may be shifted to lower temperatures if an appropriate amount of disorder is artificially introduced in the material to reduce the thermalization length without destroying the underlying spin liquid.

\subsection{Edge Coherence Length}

For observing anyonic edge interferometry in the presence of thermal fluctuations at the edge, the separation between the two pinch points, $\Delta_x$, must not exceed the thermal edge coherence length, $\xi \sim \hbar v / (k_B T)$, where $v$ is the edge-mode velocity~\cite{Chamon_1997}.
This constraint is directly manifest in our result for the fermion tunneling where the interference term in $\beta$ [see Eq.~\eqref{eq:tunneling_results_combined}] is proportional to the function $g[\pi k_B T \Delta_x / (\hbar v)]$~\footnote{We have restored appropriate factors of $\hbar$, $k_B$, and $v$.} that is exponentially suppressed for large arguments (see Fig.~\ref{fig:fermion_braiding_correction}).
While our calculation of the Ising-anyon tunneling is based on zero-temperature correlators and only applies at low temperatures, $\Delta_x \ll \xi$~\footnote{At the same time, we also assume that the temperature is high enough that the tunneling is still perturbative.}, we nevertheless expect a similar exponential suppression for the Ising-anyon tunneling in the high-temperature $\Delta_x \gg \xi$ limit~\cite{Nilsson_2010}.

Considering $\alpha$-RuCl$_3$, if we assume that the edge velocity is controlled by the Kitaev interaction, $J \sim 100$ K, we can estimate the edge velocity as $v \sim J a / \hbar$, where $a \sim 1$ nm is the lattice constant.
The edge coherence length is then given by $\xi \sim J a / (k_B T)$ and, for a pinch separation $\Delta_x$, the temperature must satisfy $k_B T \lesssim J a / \Delta_x$.
Thus, taking $\Delta_x \sim 100$ nm, we find that anyon interferometry must be performed at temperatures below $1$ K.
We note that this constraint may be more stringent if the edge velocity is instead controlled by the bulk energy gap which is suggested by Ref.~\onlinecite{Kasahara2018} to be $\Delta \approx 5$ K.
Therefore, to minimize disruptions due to thermal fluctuations at the edge, it is ideal to keep the pinch separation $\Delta_x$ as small as possible.
At the same time, however, the pinch separation must also be much larger than the bulk correlation length to avoid finite-size effects.
Based on the exactly solvable Kitaev model, this correlation length is comparable to the lattice constant if the bulk gap is a sizeable fraction of the Kitaev interaction.
Finally, we emphasize that the edge coherence length is only relevant for anyon interferometry and does not affect the simpler measurements using at most one pinch point.

\subsection{Fluctuating Bulk Anyons}

To detect individual bulk anyons through anyonic edge interferometry, we assume that the bulk anyons are localized at appropriate pinning sites in a stable way.
For example, it is known that spin vacancies in the exactly solvable Kitaev model bind gauge fluxes~\cite{Willans_2010, Willans_2011, Kao_2021}, which correspond to Ising anyons in the non-Abelian spin liquid.
Ideally, one would be able to control the total anyon content between the two pinch points \emph{in situ} and observe the resulting changes in $\beta$ by measuring temperatures or heat currents.
However, in the presence of thermally excited bulk anyons, uncontrolled anyon motion may lead to repeated switches in the total anyon content during the (presumably long) time scale of such a measurement, thereby washing out any interference effect resulting from the anyons localized at pinning sites.

We first present a sufficient (but not necessary) condition for observing anyon interferometry in the presence of thermally excited bulk anyons.
In general, for a bulk energy gap $\Delta$, the density of such thermal anyons is expected to be $\rho \sim \exp [-\Delta / (k_B T)]$.
The average number of thermal anyons in the region between the two pinch points is then $\nu \sim N \rho$, where $N = A / a^2$ is the size of the region in terms of its area $A$ and the lattice constant $a$.
In turn, if $\nu \ll 1$, thermal anyons are sufficiently rare that they cannot be disruptive to anyon interferometry.
Thus, our sufficient condition for observing anyon interferometry becomes $k_B T \ll \Delta / \ln N$, which is only slightly more stringent than the standard condition $k_B T \ll \Delta$ for probing universal low-energy properties.
Indeed, if we assume $\Delta \approx 5$ K for $\alpha$-RuCl$_3$~\cite{Kasahara2018} and take $A \sim (100$ nm$)^2$ as well as $a \sim 1$ nm, we find that the sufficient condition is satisfied for temperatures below $\approx 0.5$ K.

However, even if this sufficient condition is not satisfied and there are many thermal anyons in the region between the two pinch points, anyon interferometry may still be observable if the total anyon content fluctuates sufficiently slowly with respect to the time scale of the measurement.
While it is challenging to estimate the relevant time scales, the fluctuation rate is limited by the fact that the total anyon content can only change if an anyon is created or annihilated at the edge or if it goes through one of the pinch points.
In turn, for a sufficiently small fluctuation rate, we anticipate that the measured temperature or heat current exhibits telegraph noise~\cite{Aasen_2020}, repeatedly switching between a finite number of allowed values, which correspond to different total anyon contents [see Eq.~\eqref{eq:tunneling_results_combined}], as a function of time.

Finally, we point out that distinct fingerprints of anyon interference may be statistically observable even if the measurement is obtained by accumulating signal over a time scale that is long compared to the time scales of fermion ($n_{\psi}$) and Ising-anyon ($n_{\sigma}$) fluctuations.
To start with a simple problem, we assume that the only fluctations are in the fermion parity, $n_{\psi} = \{0,1\}$, so that the measured quantity $O$ (whether that be temperature or heat current) can take only two different instantaneous values, $O_{n_{\psi} = 0}$ and $O_{n_{\psi} = 1}$.
A standard stochastic process, the Goldstein-Kac process~\cite{Goldstein_1951, Kac_1974}, gives a solvable model for this problem if the time between two flips is assumed to be much longer than the time taken in the flipping process itself such that repeated \emph{instantaneous} measurements result in telegraph noise.
Taking a flip rate $\mu$, we define the integrated measurement over a time interval $\tau$
to be $\int_0^\tau O(t) \dif t = \tau\bar O + x$, where $\bar O \equiv (O_0 + O_1)/2$ is the mean value.
Then, the probability for the integrated measurement to deviate from the mean value by $x$ is
\begin{eqnarray}
    p(x,\tau) &=& e^{-\mu \tau} \left[ \delta(x - \chi \tau) + \delta(x + \chi \tau) \right] \cr && + {e^{-\mu \tau} \over 2 \chi} \Big[ \mu I_0(\mu \sqrt{\chi^2 \tau^2 -x^2} / \chi) \cr && + {\chi I_1 (\mu \sqrt{\chi^2 \tau^2 -x^2} / \chi) \over \sqrt{\chi^2 \tau^2 -x^2}}\Big] \theta(\chi \tau - |x|), \qquad
    \label{eq:kacdist}
\end{eqnarray}
where $\chi = |O_0 - O_1|/2$ is the standard deviation, $\theta$ is the Heaviside step function, and $I_{0,1}$ are modified Bessel functions of the first kind.
This probability distribution is normalized for each integration time $\tau$ and consists of two parts: the $\delta$-function pieces come from the probability that the system has not flipped from its initial state, and the remainder is a smooth function that dominates and approaches a Gaussian for $\mu \tau \gg 1$.
Repeating the given measurement many times and for various integration times $\tau$, the results can then be fit to Eq.~\eqref{eq:kacdist} to determine the underlying value of $\chi$ which, in turn, can be compared to the predictions in Eq.~\eqref{eq:tunneling_results_combined}.

If there are fluctuations in both $n_\sigma$ and $n_\psi$, the model becomes a four-state Markov process with two distinct rates $\mu_\sigma$ and $\mu_\psi$ for the two kinds of flips between four different states labeled by $(n_\sigma, n_\psi)$.
Recognizing that the measured quantity $O$ takes identical instantaneous values $O_{(n_\sigma, n_\psi)}$ for the topologically equivalent states $(1,0)$ and $(1,1)$, the variance of an integrated measurement is then
\begin{eqnarray}
    \langle x^2(\tau) \rangle &=& \tau^2 \bigg\{ \left[ {O_{(0,0)} - O_{(0,1)} \over 4} \right]^2 \phi (\mu_\psi \tau) \cr
    &&+ \left[ {O_{(0,0)} - O_{(0,1)} \over 4} \right]^2 \phi (\{ \mu_\psi + \mu_\sigma \} \tau) \cr
    &&+ \left[ {O_{(0,0)} + O_{(0,1)} -2O_{(1,0)} \over 4} \right]^2 \phi (\mu_\sigma \tau) \bigg\}, \qquad \label{eq:markov_four_state}
\end{eqnarray}
where we define the dimensionless function
\be
    \phi(x) = \frac{1}{x} - \frac{1-e^{-2x}}{2x^2}.
\ee
As in the two-state model, repeated measurements for a range of integration times $\tau$ allow extraction of the instantaneous values $O_{(n_\sigma, n_\psi)}$ which are directly comparable with the predictions of Eq.~\eqref{eq:tunneling_results_combined}.
We further note that three distinct rates, $\mu_\psi$, $\mu_\sigma$, and $\mu_\psi + \mu_\sigma$, appear in Eq.~\eqref{eq:markov_four_state}.
The need for three distinct rates to describe the $\tau$-dependence of the variance $\langle x^2(\tau) \rangle$ can differentiate the noise of combined fermion and Ising-anyon flips from other possible sources of a single telegraph noise, for which the variance would simply be
\begin{eqnarray}
    \langle x^2(\tau) \rangle = \tau^2 \left({O_0 - O_1 \over 2}\right)^2 \phi (\mu \tau). \label{eq:markov_two_state}
\end{eqnarray}
Note that Eq.~\eqref{eq:markov_two_state} is also consistent with integration over the full probability distribution in Eq.~\eqref{eq:kacdist}.

\subsection{Multi-Layer Samples}

In this subsection, we briefly discuss how our results for a single two-dimensional layer generalize to more realistic multi-layer systems.
For sufficiently weak inter-layer interactions, each layer remains an independent spin liquid with its own chiral edge mode and bulk anyons.
Therefore, the edge tunneling processes in the central region are controlled by the individual layers, and we can define a distinct tunneling parameter $\beta_i$ for each of the $n$ layers labeled by $i = 1, \ldots, n$.
In contrast, the phonons can move freely between the different layers and result in a bulk inter-layer thermalization within the two lobes.
Thus, using the heat-resistor picture in Sec.~\ref{sec:phonon_coupling}, the temperatures and heat currents measured in an $n$-layer system are expected to be $T_n (\{\beta_i\}) = T_1 (\bar{\beta})$ and $J_n (\{\beta_i\}) = n J_1 (\bar{\beta})$, respectively, where $\bar{\beta} = \frac{1}{n} \sum_{i=1}^n \beta_i$ is the mean tunneling coefficient, while $T_1 (\beta)$ and $J_1 (\beta)$ are the corresponding single-layer results for tunneling coefficient $\beta$.

For the simpler measurements using at most one point contact, the tunneling parameters are identical in all $n$ layers.
The measured temperatures are then not affected at all by the presence of $n > 1$ layers, while the measured heat currents are simply multiplied by $n$.
Hence, for extracting the quantized thermal Hall conductivity $\kappa_{\mathrm{e}}$ or the universal power law $\beta \propto T^{-7/4}$ reflecting Ising-anyon tunneling, it is advantageous to detect the edge tunneling processes via the total heat current between the two leads, as multi-layer systems can then actually result in a more accurate measurement.
To strengthen this point, we also recall that in the limit of $\kappa_{\mathrm{c}} \ll \kappa_{\mathrm{e}} \ll \kappa_{\mathrm{b}} \vartheta_0$, the total heat current of a single layer takes the simple form $J_1 (\beta) = 2 (1-\beta) \kappa_{\mathrm{e}} \delta T$, which enables direct extraction of $\kappa_{\mathrm{e}}$ in the absence of any point contacts and $\beta \propto T^{-7/4}$ in the presence of a single point contact.

For anyonic edge interferometry, however, the distinct layers are expected to have different tunneling parameters $\beta_i$, unless one can simultaneously stabilize the same kind of anyon in all layers.
Thus, we expect that anyon interferometry is generally more observable in few-layer systems where different combinations of $\{\beta_i\}$ still result in only a small number of distinct $\bar{\beta}$.
In particular, a few-layer system with a small number of discrete values could still be modeled with an expanded version of the Markov-chain model in the preceding subsection.
However, such few-layer systems present a number of notable experimental challenges.
First, the system must be thermally isolated, as significant thermal coupling to, for example, a substrate may weaken the sensitivity to $\bar{\beta}$.
Second, it may not be straightforward to measure the temperature of such a few-layer system or to detect small changes in the heat current on the order of $\kappa_{\mathrm{e}} \delta T$.
Third, a few-layer system is expected to be more susceptible to crystallographic defects, such as stacking faults, that may have a detrimental effect on the underlying spin liquid.

\section{Summary and Outlook}\label{sec:summary}

In this paper, we have proposed that anyonic edge interferometry can be realized in the non-Abelian Kitaev spin liquid by performing conventional thermal transport measurements in the unconventional device geometry depicted in Fig.~\ref{fig:device}.
Individual anyons inside the mesoscopic central region can be detected by interference patterns of edge tunneling processes which, in turn, affect the bulk thermal transport within the two macroscopic lobes due to edge-phonon coupling.

To corroborate our proposal, we have synthesized two complementary calculations, each capturing some universal properties of the system under consideration.
On the one hand, we have considered a set of linearized hydrodynamic equations~\cite{Ye_2018, Vinkler2018} to describe the large-scale thermal transport mediated by bulk acoustic phonons coupled to a chiral edge mode.
By obtaining a perturbative solution to these equations and interpreting the results in terms of an effective heat-resistor network, we have understood how to optimize the geometry and the probe placement in order to maximize the sensitivity of the total heat current or a locally measured temperature to coherent edge tunneling processes.
On the other hand, we have utilized the CFT approach~\cite{Aasen_2020, Klocke_2020} to investigate the edge tunneling processes themselves and understand how they can be used to detect individual anyons or extract universal properties of the underlying spin liquid.
This treatment highlights unique signatures arising from bare tunneling at a single point contact and from braiding by tunneling in the presence of multiple point contacts.

While the main focus of this work has been the detection of individual anyons through edge interferometry, we emphasize that the device geometry in Fig.~\ref{fig:device} can also be used to demonstrate the existence of anyons via simpler measurements that do not rely on anyon braiding.
If the narrow channel in the central region has a sufficiently small bulk thermal conductance, the heat exchange between the two lobes is mediated almost exclusively by the edge mode.
In the absence of any point contacts, the total heat current $J$ is then directly proportional to the chiral central charge of the edge theory.
Hence, the chiral central charge, $c = \frac{1}{2}$, which immediately reveals the presence of non-Abelian anyons, can be extracted from the dominant \emph{longitudinal} thermal conductance rather than a subdominant thermal Hall conductivity~\cite{Kasahara2018, Tokoi, bruin2021}.
Furthermore, in the presence of a single point contact, the leading-order correction to the total heat current follows a nontrivial power law, $\delta J / T \propto T^{-7/4}$, as a function of the temperature.
The universal exponent $7/4$ is innately tied to the conformal dimension of the non-Abelian Ising anyons and would be difficult to reproduce from less exotic physics.
Most importantly, these non-interferometry measurements are within reach of the currently available thermal transport experiments~\cite{Kasahara2018, Tokoi, bruin2021} in existing candidate materials like $\alpha$-RuCl$_3$.

Finally, we point out that our proposed measurements straightforwardly generalize to all kinds of chiral topological orders, including Abelian and non-Abelian chiral spin liquids, as well as their electronic counterparts such as fractional quantum Hall states.
Even though thermal transport is experimentally more challenging than electrical transport, it provides complementary information on such electronic topological orders, for example, by giving direct access to the chiral central charge~\cite{Banerjee}.

\begin{acknowledgments}
We thank Chengyun Hua and Alan Tennant for helpful discussions.
This material is based upon work supported by the U.S. Department of Energy, Office of Science, National Quantum Information Science Research Centers, Quantum Science Center.
J.A.~additionally acknowledges support from the Army Research Office under Grant
Award W911NF17- 1-0323; the National Science Foundation through grant DMR-1723367; the Caltech Institute
for Quantum Information and Matter, an NSF Physics
Frontiers Center with support of the Gordon and Betty
Moore Foundation through Grant GBMF1250; and the
Walter Burke Institute for Theoretical Physics at Caltech.  J.E.M.~acknowledges additional support from a Simons Investigatorship.

\emph{Note added.---}At the time of our arXiv submission, we became aware of a related manuscript~\cite{Wei2021} that focuses on complementary aspects of the same problem and was completed independently from ours.
\end{acknowledgments}

\bibliography{thermal_interferometry}

\begin{thebibliography}{75}%
\makeatletter
\providecommand \@ifxundefined [1]{%
 \@ifx{#1\undefined}
}%
\providecommand \@ifnum [1]{%
 \ifnum #1\expandafter \@firstoftwo
 \else \expandafter \@secondoftwo
 \fi
}%
\providecommand \@ifx [1]{%
 \ifx #1\expandafter \@firstoftwo
 \else \expandafter \@secondoftwo
 \fi
}%
\providecommand \natexlab [1]{#1}%
\providecommand \enquote  [1]{``#1''}%
\providecommand \bibnamefont  [1]{#1}%
\providecommand \bibfnamefont [1]{#1}%
\providecommand \citenamefont [1]{#1}%
\providecommand \href@noop [0]{\@secondoftwo}%
\providecommand \href [0]{\begingroup \@sanitize@url \@href}%
\providecommand \@href[1]{\@@startlink{#1}\@@href}%
\providecommand \@@href[1]{\endgroup#1\@@endlink}%
\providecommand \@sanitize@url [0]{\catcode `\\12\catcode `\$12\catcode
  `\&12\catcode `\#12\catcode `\^12\catcode `\_12\catcode `\%12\relax}%
\providecommand \@@startlink[1]{}%
\providecommand \@@endlink[0]{}%
\providecommand \url  [0]{\begingroup\@sanitize@url \@url }%
\providecommand \@url [1]{\endgroup\@href {#1}{\urlprefix }}%
\providecommand \urlprefix  [0]{URL }%
\providecommand \Eprint [0]{\href }%
\providecommand \doibase [0]{https://doi.org/}%
\providecommand \selectlanguage [0]{\@gobble}%
\providecommand \bibinfo  [0]{\@secondoftwo}%
\providecommand \bibfield  [0]{\@secondoftwo}%
\providecommand \translation [1]{[#1]}%
\providecommand \BibitemOpen [0]{}%
\providecommand \bibitemStop [0]{}%
\providecommand \bibitemNoStop [0]{.\EOS\space}%
\providecommand \EOS [0]{\spacefactor3000\relax}%
\providecommand \BibitemShut  [1]{\csname bibitem#1\endcsname}%
\let\auto@bib@innerbib\@empty
\bibitem [{\citenamefont {Kitaev}(2003)}]{Kitaev2003}%
  \BibitemOpen
  \bibfield  {author} {\bibinfo {author} {\bibfnamefont {A.~Y.}\ \bibnamefont
  {Kitaev}},\ }\bibfield  {title} {\bibinfo {title} {{Fault-tolerant quantum
  computation by anyons}},\ }\href
  {https://doi.org/10.1016/S0003-4916(02)00018-0} {\bibfield  {journal}
  {\bibinfo  {journal} {Ann.\ Phys.}\ }\textbf {\bibinfo {volume} {303}},\
  \bibinfo {pages} {2} (\bibinfo {year} {2003})}\BibitemShut {NoStop}%
\bibitem [{\citenamefont {Nayak}\ \emph {et~al.}(2008)\citenamefont {Nayak},
  \citenamefont {Simon}, \citenamefont {Stern}, \citenamefont {Freedman},\ and\
  \citenamefont {Das~Sarma}}]{TQCreview}%
  \BibitemOpen
  \bibfield  {author} {\bibinfo {author} {\bibfnamefont {C.}~\bibnamefont
  {Nayak}}, \bibinfo {author} {\bibfnamefont {S.~H.}\ \bibnamefont {Simon}},
  \bibinfo {author} {\bibfnamefont {A.}~\bibnamefont {Stern}}, \bibinfo
  {author} {\bibfnamefont {M.}~\bibnamefont {Freedman}},\ and\ \bibinfo
  {author} {\bibfnamefont {S.}~\bibnamefont {Das~Sarma}},\ }\bibfield  {title}
  {\bibinfo {title} {Non-{Abelian} anyons and topological quantum
  computation},\ }\href {https://doi.org/10.1103/RevModPhys.80.1083} {\bibfield
   {journal} {\bibinfo  {journal} {Rev. Mod. Phys.}\ }\textbf {\bibinfo
  {volume} {80}},\ \bibinfo {pages} {1083} (\bibinfo {year}
  {2008})}\BibitemShut {NoStop}%
\bibitem [{\citenamefont {Witczak-Krempa}\ \emph {et~al.}(2014)\citenamefont
  {Witczak-Krempa}, \citenamefont {Chen}, \citenamefont {Kim},\ and\
  \citenamefont {Balents}}]{Witczak-Krempa_2014}%
  \BibitemOpen
  \bibfield  {author} {\bibinfo {author} {\bibfnamefont {W.}~\bibnamefont
  {Witczak-Krempa}}, \bibinfo {author} {\bibfnamefont {G.}~\bibnamefont
  {Chen}}, \bibinfo {author} {\bibfnamefont {Y.~B.}\ \bibnamefont {Kim}},\ and\
  \bibinfo {author} {\bibfnamefont {L.}~\bibnamefont {Balents}},\ }\bibfield
  {title} {\bibinfo {title} {Correlated quantum phenomena in the strong
  spin-orbit regime},\ }\href
  {https://doi.org/10.1146/annurev-conmatphys-020911-125138} {\bibfield
  {journal} {\bibinfo  {journal} {Annual Review of Condensed Matter Physics}\
  }\textbf {\bibinfo {volume} {5}},\ \bibinfo {pages} {57} (\bibinfo {year}
  {2014})}\BibitemShut {NoStop}%
\bibitem [{\citenamefont {Jackeli}\ and\ \citenamefont
  {Khaliullin}(2009)}]{Jackeli2009}%
  \BibitemOpen
  \bibfield  {author} {\bibinfo {author} {\bibfnamefont {G.}~\bibnamefont
  {Jackeli}}\ and\ \bibinfo {author} {\bibfnamefont {G.}~\bibnamefont
  {Khaliullin}},\ }\bibfield  {title} {\bibinfo {title} {Mott insulators in the
  strong spin-orbit coupling limit: From {H}eisenberg to a quantum compass and
  {K}itaev models},\ }\href {https://doi.org/10.1103/PhysRevLett.102.017205}
  {\bibfield  {journal} {\bibinfo  {journal} {Phys. Rev. Lett.}\ }\textbf
  {\bibinfo {volume} {102}},\ \bibinfo {pages} {017205} (\bibinfo {year}
  {2009})}\BibitemShut {NoStop}%
\bibitem [{\citenamefont {Chaloupka}\ \emph {et~al.}(2010)\citenamefont
  {Chaloupka}, \citenamefont {Jackeli},\ and\ \citenamefont
  {Khaliullin}}]{Chaloupka2010}%
  \BibitemOpen
  \bibfield  {author} {\bibinfo {author} {\bibfnamefont {J.}~\bibnamefont
  {Chaloupka}}, \bibinfo {author} {\bibfnamefont {G.}~\bibnamefont {Jackeli}},\
  and\ \bibinfo {author} {\bibfnamefont {G.}~\bibnamefont {Khaliullin}},\
  }\bibfield  {title} {\bibinfo {title} {Kitaev-heisenberg model on a honeycomb
  lattice: Possible exotic phases in {Iridium Oxides
  ${A}_{2}{\mathrm{IrO}}_{3}$}},\ }\href
  {https://doi.org/10.1103/PhysRevLett.105.027204} {\bibfield  {journal}
  {\bibinfo  {journal} {Phys. Rev. Lett.}\ }\textbf {\bibinfo {volume} {105}},\
  \bibinfo {pages} {027204} (\bibinfo {year} {2010})}\BibitemShut {NoStop}%
\bibitem [{\citenamefont {Chaloupka}\ \emph {et~al.}(2013)\citenamefont
  {Chaloupka}, \citenamefont {Jackeli},\ and\ \citenamefont
  {Khaliullin}}]{Chaloupka2013}%
  \BibitemOpen
  \bibfield  {author} {\bibinfo {author} {\bibfnamefont {J.}~\bibnamefont
  {Chaloupka}}, \bibinfo {author} {\bibfnamefont {G.}~\bibnamefont {Jackeli}},\
  and\ \bibinfo {author} {\bibfnamefont {G.}~\bibnamefont {Khaliullin}},\
  }\bibfield  {title} {\bibinfo {title} {Zigzag magnetic order in the {Iridium
  Oxide ${\mathrm{Na}}_{2}{\mathrm{IrO}}_{3}$}},\ }\href
  {https://doi.org/10.1103/PhysRevLett.110.097204} {\bibfield  {journal}
  {\bibinfo  {journal} {Phys. Rev. Lett.}\ }\textbf {\bibinfo {volume} {110}},\
  \bibinfo {pages} {097204} (\bibinfo {year} {2013})}\BibitemShut {NoStop}%
\bibitem [{\citenamefont {Kitaev}(2006)}]{Kitaev2006}%
  \BibitemOpen
  \bibfield  {author} {\bibinfo {author} {\bibfnamefont {A.}~\bibnamefont
  {Kitaev}},\ }\bibfield  {title} {\bibinfo {title} {Anyons in an exactly
  solved model and beyond},\ }\href {https://doi.org/10.1016/j.aop.2005.10.005}
  {\bibfield  {journal} {\bibinfo  {journal} {Annals of Physics}\ }\textbf
  {\bibinfo {volume} {321}},\ \bibinfo {pages} {2 } (\bibinfo {year}
  {2006})}\BibitemShut {NoStop}%
\bibitem [{\citenamefont {Rau}\ \emph {et~al.}(2016)\citenamefont {Rau},
  \citenamefont {Lee},\ and\ \citenamefont {Kee}}]{Rau_2016}%
  \BibitemOpen
  \bibfield  {author} {\bibinfo {author} {\bibfnamefont {J.~G.}\ \bibnamefont
  {Rau}}, \bibinfo {author} {\bibfnamefont {E.~K.-H.}\ \bibnamefont {Lee}},\
  and\ \bibinfo {author} {\bibfnamefont {H.-Y.}\ \bibnamefont {Kee}},\
  }\bibfield  {title} {\bibinfo {title} {Spin-orbit physics giving rise to
  novel phases in correlated systems: {I}ridates and related materials},\
  }\href {https://doi.org/10.1146/annurev-conmatphys-031115-011319} {\bibfield
  {journal} {\bibinfo  {journal} {Annual Review of Condensed Matter Physics}\
  }\textbf {\bibinfo {volume} {7}},\ \bibinfo {pages} {195} (\bibinfo {year}
  {2016})}\BibitemShut {NoStop}%
\bibitem [{\citenamefont {{Trebst}}(2017)}]{Trebst_2017}%
  \BibitemOpen
  \bibfield  {author} {\bibinfo {author} {\bibfnamefont {S.}~\bibnamefont
  {{Trebst}}},\ }\bibfield  {title} {\bibinfo {title} {{Kitaev Materials}},\
  }\href@noop {} {\bibfield  {journal} {\bibinfo  {journal} {ArXiv e-prints}\ }
  (\bibinfo {year} {2017})},\ \Eprint {https://arxiv.org/abs/1701.07056}
  {arXiv:1701.07056 [cond-mat.str-el]} \BibitemShut {NoStop}%
\bibitem [{\citenamefont {Hermanns}\ \emph {et~al.}(2018)\citenamefont
  {Hermanns}, \citenamefont {Kimchi},\ and\ \citenamefont
  {Knolle}}]{Hermanns_2018}%
  \BibitemOpen
  \bibfield  {author} {\bibinfo {author} {\bibfnamefont {M.}~\bibnamefont
  {Hermanns}}, \bibinfo {author} {\bibfnamefont {I.}~\bibnamefont {Kimchi}},\
  and\ \bibinfo {author} {\bibfnamefont {J.}~\bibnamefont {Knolle}},\
  }\bibfield  {title} {\bibinfo {title} {Physics of the {K}itaev model:
  Fractionalization, dynamic correlations, and material connections},\ }\href
  {https://doi.org/10.1146/annurev-conmatphys-033117-053934} {\bibfield
  {journal} {\bibinfo  {journal} {Annual Review of Condensed Matter Physics}\
  }\textbf {\bibinfo {volume} {9}},\ \bibinfo {pages} {17} (\bibinfo {year}
  {2018})}\BibitemShut {NoStop}%
\bibitem [{\citenamefont {Takagi}\ \emph {et~al.}(2019)\citenamefont {Takagi},
  \citenamefont {Takayama}, \citenamefont {Jackeli}, \citenamefont
  {Khaliullin},\ and\ \citenamefont {Nagler}}]{Takagi_2019}%
  \BibitemOpen
  \bibfield  {author} {\bibinfo {author} {\bibfnamefont {H.}~\bibnamefont
  {Takagi}}, \bibinfo {author} {\bibfnamefont {T.}~\bibnamefont {Takayama}},
  \bibinfo {author} {\bibfnamefont {G.}~\bibnamefont {Jackeli}}, \bibinfo
  {author} {\bibfnamefont {G.}~\bibnamefont {Khaliullin}},\ and\ \bibinfo
  {author} {\bibfnamefont {S.~E.}\ \bibnamefont {Nagler}},\ }\bibfield  {title}
  {\bibinfo {title} {Concept and realization of {K}itaev quantum spin
  liquids},\ }\href {https://doi.org/10.1038/s42254-019-0038-2} {\bibfield
  {journal} {\bibinfo  {journal} {Nature Reviews Physics}\ }\textbf {\bibinfo
  {volume} {1}},\ \bibinfo {pages} {264} (\bibinfo {year} {2019})}\BibitemShut
  {NoStop}%
\bibitem [{\citenamefont {Plumb}\ \emph {et~al.}(2014)\citenamefont {Plumb},
  \citenamefont {Clancy}, \citenamefont {Sandilands}, \citenamefont {Shankar},
  \citenamefont {Hu}, \citenamefont {Burch}, \citenamefont {Kee},\ and\
  \citenamefont {Kim}}]{Plumb_2014}%
  \BibitemOpen
  \bibfield  {author} {\bibinfo {author} {\bibfnamefont {K.~W.}\ \bibnamefont
  {Plumb}}, \bibinfo {author} {\bibfnamefont {J.~P.}\ \bibnamefont {Clancy}},
  \bibinfo {author} {\bibfnamefont {L.~J.}\ \bibnamefont {Sandilands}},
  \bibinfo {author} {\bibfnamefont {V.~V.}\ \bibnamefont {Shankar}}, \bibinfo
  {author} {\bibfnamefont {Y.~F.}\ \bibnamefont {Hu}}, \bibinfo {author}
  {\bibfnamefont {K.~S.}\ \bibnamefont {Burch}}, \bibinfo {author}
  {\bibfnamefont {H.-Y.}\ \bibnamefont {Kee}},\ and\ \bibinfo {author}
  {\bibfnamefont {Y.-J.}\ \bibnamefont {Kim}},\ }\bibfield  {title} {\bibinfo
  {title} {{$\ensuremath{\alpha}{\mathrm{-RuCl}}_{3}$}: A spin-orbit assisted
  {M}ott insulator on a honeycomb lattice},\ }\href
  {https://doi.org/10.1103/PhysRevB.90.041112} {\bibfield  {journal} {\bibinfo
  {journal} {Phys. Rev. B}\ }\textbf {\bibinfo {volume} {90}},\ \bibinfo
  {pages} {041112} (\bibinfo {year} {2014})}\BibitemShut {NoStop}%
\bibitem [{\citenamefont {Sandilands}\ \emph {et~al.}(2015)\citenamefont
  {Sandilands}, \citenamefont {Tian}, \citenamefont {Plumb}, \citenamefont
  {Kim},\ and\ \citenamefont {Burch}}]{Sandilands_2015}%
  \BibitemOpen
  \bibfield  {author} {\bibinfo {author} {\bibfnamefont {L.~J.}\ \bibnamefont
  {Sandilands}}, \bibinfo {author} {\bibfnamefont {Y.}~\bibnamefont {Tian}},
  \bibinfo {author} {\bibfnamefont {K.~W.}\ \bibnamefont {Plumb}}, \bibinfo
  {author} {\bibfnamefont {Y.-J.}\ \bibnamefont {Kim}},\ and\ \bibinfo {author}
  {\bibfnamefont {K.~S.}\ \bibnamefont {Burch}},\ }\bibfield  {title} {\bibinfo
  {title} {Scattering continuum and possible fractionalized excitations in
  {$\ensuremath{\alpha}\text{\ensuremath{-}}{\mathrm{RuCl}}_{3}$}},\ }\href
  {https://doi.org/10.1103/PhysRevLett.114.147201} {\bibfield  {journal}
  {\bibinfo  {journal} {Phys. Rev. Lett.}\ }\textbf {\bibinfo {volume} {114}},\
  \bibinfo {pages} {147201} (\bibinfo {year} {2015})}\BibitemShut {NoStop}%
\bibitem [{\citenamefont {Sears}\ \emph {et~al.}(2015)\citenamefont {Sears},
  \citenamefont {Songvilay}, \citenamefont {Plumb}, \citenamefont {Clancy},
  \citenamefont {Qiu}, \citenamefont {Zhao}, \citenamefont {Parshall},\ and\
  \citenamefont {Kim}}]{Sears_2015}%
  \BibitemOpen
  \bibfield  {author} {\bibinfo {author} {\bibfnamefont {J.~A.}\ \bibnamefont
  {Sears}}, \bibinfo {author} {\bibfnamefont {M.}~\bibnamefont {Songvilay}},
  \bibinfo {author} {\bibfnamefont {K.~W.}\ \bibnamefont {Plumb}}, \bibinfo
  {author} {\bibfnamefont {J.~P.}\ \bibnamefont {Clancy}}, \bibinfo {author}
  {\bibfnamefont {Y.}~\bibnamefont {Qiu}}, \bibinfo {author} {\bibfnamefont
  {Y.}~\bibnamefont {Zhao}}, \bibinfo {author} {\bibfnamefont {D.}~\bibnamefont
  {Parshall}},\ and\ \bibinfo {author} {\bibfnamefont {Y.-J.}\ \bibnamefont
  {Kim}},\ }\bibfield  {title} {\bibinfo {title} {Magnetic order in
  {$\ensuremath{\alpha}\ensuremath{-}{\text{RuCl}}_{3}$}: A honeycomb-lattice
  quantum magnet with strong spin-orbit coupling},\ }\href
  {https://doi.org/10.1103/PhysRevB.91.144420} {\bibfield  {journal} {\bibinfo
  {journal} {Phys. Rev. B}\ }\textbf {\bibinfo {volume} {91}},\ \bibinfo
  {pages} {144420} (\bibinfo {year} {2015})}\BibitemShut {NoStop}%
\bibitem [{\citenamefont {Majumder}\ \emph {et~al.}(2015)\citenamefont
  {Majumder}, \citenamefont {Schmidt}, \citenamefont {Rosner}, \citenamefont
  {Tsirlin}, \citenamefont {Yasuoka},\ and\ \citenamefont
  {Baenitz}}]{Majumder_2015}%
  \BibitemOpen
  \bibfield  {author} {\bibinfo {author} {\bibfnamefont {M.}~\bibnamefont
  {Majumder}}, \bibinfo {author} {\bibfnamefont {M.}~\bibnamefont {Schmidt}},
  \bibinfo {author} {\bibfnamefont {H.}~\bibnamefont {Rosner}}, \bibinfo
  {author} {\bibfnamefont {A.~A.}\ \bibnamefont {Tsirlin}}, \bibinfo {author}
  {\bibfnamefont {H.}~\bibnamefont {Yasuoka}},\ and\ \bibinfo {author}
  {\bibfnamefont {M.}~\bibnamefont {Baenitz}},\ }\bibfield  {title} {\bibinfo
  {title} {Anisotropic {${\mathrm{Ru}}^{3+} 4{d}^{5}$} magnetism in the
  {$\ensuremath{\alpha}\ensuremath{-}{\mathrm{RuCl}}_{3}$} honeycomb system:
  Susceptibility, specific heat, and zero-field {NMR}},\ }\href
  {https://doi.org/10.1103/PhysRevB.91.180401} {\bibfield  {journal} {\bibinfo
  {journal} {Phys. Rev. B}\ }\textbf {\bibinfo {volume} {91}},\ \bibinfo
  {pages} {180401} (\bibinfo {year} {2015})}\BibitemShut {NoStop}%
\bibitem [{\citenamefont {Johnson}\ \emph {et~al.}(2015)\citenamefont
  {Johnson}, \citenamefont {Williams}, \citenamefont {Haghighirad},
  \citenamefont {Singleton}, \citenamefont {Zapf}, \citenamefont {Manuel},
  \citenamefont {Mazin}, \citenamefont {Li}, \citenamefont {Jeschke},
  \citenamefont {Valent\'{\i}},\ and\ \citenamefont {Coldea}}]{Johnson_2015}%
  \BibitemOpen
  \bibfield  {author} {\bibinfo {author} {\bibfnamefont {R.~D.}\ \bibnamefont
  {Johnson}}, \bibinfo {author} {\bibfnamefont {S.~C.}\ \bibnamefont
  {Williams}}, \bibinfo {author} {\bibfnamefont {A.~A.}\ \bibnamefont
  {Haghighirad}}, \bibinfo {author} {\bibfnamefont {J.}~\bibnamefont
  {Singleton}}, \bibinfo {author} {\bibfnamefont {V.}~\bibnamefont {Zapf}},
  \bibinfo {author} {\bibfnamefont {P.}~\bibnamefont {Manuel}}, \bibinfo
  {author} {\bibfnamefont {I.~I.}\ \bibnamefont {Mazin}}, \bibinfo {author}
  {\bibfnamefont {Y.}~\bibnamefont {Li}}, \bibinfo {author} {\bibfnamefont
  {H.~O.}\ \bibnamefont {Jeschke}}, \bibinfo {author} {\bibfnamefont
  {R.}~\bibnamefont {Valent\'{\i}}},\ and\ \bibinfo {author} {\bibfnamefont
  {R.}~\bibnamefont {Coldea}},\ }\bibfield  {title} {\bibinfo {title}
  {Monoclinic crystal structure of
  {$\ensuremath{\alpha}\ensuremath{-}{\mathrm{RuCl}}_{3}$} and the zigzag
  antiferromagnetic ground state},\ }\href
  {https://doi.org/10.1103/PhysRevB.92.235119} {\bibfield  {journal} {\bibinfo
  {journal} {Phys. Rev. B}\ }\textbf {\bibinfo {volume} {92}},\ \bibinfo
  {pages} {235119} (\bibinfo {year} {2015})}\BibitemShut {NoStop}%
\bibitem [{\citenamefont {Sandilands}\ \emph {et~al.}(2016)\citenamefont
  {Sandilands}, \citenamefont {Tian}, \citenamefont {Reijnders}, \citenamefont
  {Kim}, \citenamefont {Plumb}, \citenamefont {Kim}, \citenamefont {Kee},\ and\
  \citenamefont {Burch}}]{Sandilands_2016}%
  \BibitemOpen
  \bibfield  {author} {\bibinfo {author} {\bibfnamefont {L.~J.}\ \bibnamefont
  {Sandilands}}, \bibinfo {author} {\bibfnamefont {Y.}~\bibnamefont {Tian}},
  \bibinfo {author} {\bibfnamefont {A.~A.}\ \bibnamefont {Reijnders}}, \bibinfo
  {author} {\bibfnamefont {H.-S.}\ \bibnamefont {Kim}}, \bibinfo {author}
  {\bibfnamefont {K.~W.}\ \bibnamefont {Plumb}}, \bibinfo {author}
  {\bibfnamefont {Y.-J.}\ \bibnamefont {Kim}}, \bibinfo {author} {\bibfnamefont
  {H.-Y.}\ \bibnamefont {Kee}},\ and\ \bibinfo {author} {\bibfnamefont {K.~S.}\
  \bibnamefont {Burch}},\ }\bibfield  {title} {\bibinfo {title} {Spin-orbit
  excitations and electronic structure of the putative {K}itaev magnet
  {$\ensuremath{\alpha}\ensuremath{-}{\mathrm{RuCl}}_{3}$}},\ }\href
  {https://doi.org/10.1103/PhysRevB.93.075144} {\bibfield  {journal} {\bibinfo
  {journal} {Phys. Rev. B}\ }\textbf {\bibinfo {volume} {93}},\ \bibinfo
  {pages} {075144} (\bibinfo {year} {2016})}\BibitemShut {NoStop}%
\bibitem [{\citenamefont {Banerjee}\ \emph {et~al.}(2016)\citenamefont
  {Banerjee}, \citenamefont {Bridges}, \citenamefont {Yan}, \citenamefont
  {Aczel}, \citenamefont {Li}, \citenamefont {Stone}, \citenamefont {Granroth},
  \citenamefont {Lumsden}, \citenamefont {Yiu}, \citenamefont {Knolle},
  \citenamefont {Bhattacharjee}, \citenamefont {Kovrizhin}, \citenamefont
  {Moessner}, \citenamefont {Tennant}, \citenamefont {G.},\ and\ \citenamefont
  {Nagler}}]{Banerjee_2016}%
  \BibitemOpen
  \bibfield  {author} {\bibinfo {author} {\bibfnamefont {A.}~\bibnamefont
  {Banerjee}}, \bibinfo {author} {\bibfnamefont {C.~A.}\ \bibnamefont
  {Bridges}}, \bibinfo {author} {\bibfnamefont {J.-Q.}\ \bibnamefont {Yan}},
  \bibinfo {author} {\bibfnamefont {A.~A.}\ \bibnamefont {Aczel}}, \bibinfo
  {author} {\bibfnamefont {L.}~\bibnamefont {Li}}, \bibinfo {author}
  {\bibfnamefont {M.~B.}\ \bibnamefont {Stone}}, \bibinfo {author}
  {\bibfnamefont {G.~E.}\ \bibnamefont {Granroth}}, \bibinfo {author}
  {\bibfnamefont {M.~D.}\ \bibnamefont {Lumsden}}, \bibinfo {author}
  {\bibfnamefont {Y.}~\bibnamefont {Yiu}}, \bibinfo {author} {\bibfnamefont
  {J.}~\bibnamefont {Knolle}}, \bibinfo {author} {\bibfnamefont
  {S.}~\bibnamefont {Bhattacharjee}}, \bibinfo {author} {\bibfnamefont {D.~L.}\
  \bibnamefont {Kovrizhin}}, \bibinfo {author} {\bibfnamefont {R.}~\bibnamefont
  {Moessner}}, \bibinfo {author} {\bibfnamefont {D.~A.}\ \bibnamefont
  {Tennant}}, \bibinfo {author} {\bibfnamefont {M.~D.}\ \bibnamefont {G.}},\
  and\ \bibinfo {author} {\bibfnamefont {S.~E.}\ \bibnamefont {Nagler}},\
  }\bibfield  {title} {\bibinfo {title} {{Proximate Kitaev quantum spin liquid
  behaviour in a honeycomb magnet}},\ }\href {https://doi.org/10.1038/nmat4604}
  {\bibfield  {journal} {\bibinfo  {journal} {Nature materials}\ ,\ \bibinfo
  {pages} {733–740}} (\bibinfo {year} {2016})}\BibitemShut {NoStop}%
\bibitem [{\citenamefont {Banerjee}\ \emph {et~al.}(2017)\citenamefont
  {Banerjee}, \citenamefont {Yan}, \citenamefont {Knolle}, \citenamefont
  {Bridges}, \citenamefont {Stone}, \citenamefont {Lumsden}, \citenamefont
  {Mandrus}, \citenamefont {Tennant}, \citenamefont {Moessner},\ and\
  \citenamefont {Nagler}}]{Banerjee_2017}%
  \BibitemOpen
  \bibfield  {author} {\bibinfo {author} {\bibfnamefont {A.}~\bibnamefont
  {Banerjee}}, \bibinfo {author} {\bibfnamefont {J.}~\bibnamefont {Yan}},
  \bibinfo {author} {\bibfnamefont {J.}~\bibnamefont {Knolle}}, \bibinfo
  {author} {\bibfnamefont {C.~A.}\ \bibnamefont {Bridges}}, \bibinfo {author}
  {\bibfnamefont {M.~B.}\ \bibnamefont {Stone}}, \bibinfo {author}
  {\bibfnamefont {M.~D.}\ \bibnamefont {Lumsden}}, \bibinfo {author}
  {\bibfnamefont {D.~G.}\ \bibnamefont {Mandrus}}, \bibinfo {author}
  {\bibfnamefont {D.~A.}\ \bibnamefont {Tennant}}, \bibinfo {author}
  {\bibfnamefont {R.}~\bibnamefont {Moessner}},\ and\ \bibinfo {author}
  {\bibfnamefont {S.~E.}\ \bibnamefont {Nagler}},\ }\bibfield  {title}
  {\bibinfo {title} {Neutron scattering in the proximate quantum spin liquid
  {$\alpha\mathrm{-RuCl}_3$}},\ }\href
  {https://doi.org/10.1126/science.aah6015} {\bibfield  {journal} {\bibinfo
  {journal} {Science}\ }\textbf {\bibinfo {volume} {356}},\ \bibinfo {pages}
  {1055} (\bibinfo {year} {2017})}\BibitemShut {NoStop}%
\bibitem [{\citenamefont {Do}\ \emph {et~al.}(2017)\citenamefont {Do},
  \citenamefont {Park}, \citenamefont {Yoshitake}, \citenamefont {Nasu},
  \citenamefont {Motome}, \citenamefont {Kwon}, \citenamefont {Adroja},
  \citenamefont {Voneshen}, \citenamefont {Kim}, \citenamefont {Jang},
  \citenamefont {Park}, \citenamefont {Choi},\ and\ \citenamefont
  {Ji}}]{Do_2017}%
  \BibitemOpen
  \bibfield  {author} {\bibinfo {author} {\bibfnamefont {S.-H.}\ \bibnamefont
  {Do}}, \bibinfo {author} {\bibfnamefont {S.-Y.}\ \bibnamefont {Park}},
  \bibinfo {author} {\bibfnamefont {J.}~\bibnamefont {Yoshitake}}, \bibinfo
  {author} {\bibfnamefont {J.}~\bibnamefont {Nasu}}, \bibinfo {author}
  {\bibfnamefont {Y.}~\bibnamefont {Motome}}, \bibinfo {author} {\bibfnamefont
  {Y.~S.}\ \bibnamefont {Kwon}}, \bibinfo {author} {\bibfnamefont {D.~T.}\
  \bibnamefont {Adroja}}, \bibinfo {author} {\bibfnamefont {D.~J.}\
  \bibnamefont {Voneshen}}, \bibinfo {author} {\bibfnamefont {K.}~\bibnamefont
  {Kim}}, \bibinfo {author} {\bibfnamefont {T.-H.}\ \bibnamefont {Jang}},
  \bibinfo {author} {\bibfnamefont {J.-H.}\ \bibnamefont {Park}}, \bibinfo
  {author} {\bibfnamefont {K.-Y.}\ \bibnamefont {Choi}},\ and\ \bibinfo
  {author} {\bibfnamefont {S.}~\bibnamefont {Ji}},\ }\bibfield  {title}
  {\bibinfo {title} {Majorana fermions in the {K}itaev quantum spin system
  {$\alpha\mathrm{-RuCl}_3$}},\ }\href {https://doi.org/10.1038/nphys4264}
  {\bibfield  {journal} {\bibinfo  {journal} {Nature Physics}\ }\textbf
  {\bibinfo {volume} {13}},\ \bibinfo {pages} {1079} (\bibinfo {year}
  {2017})}\BibitemShut {NoStop}%
\bibitem [{\citenamefont {Kubota}\ \emph {et~al.}(2015)\citenamefont {Kubota},
  \citenamefont {Tanaka}, \citenamefont {Ono}, \citenamefont {Narumi},\ and\
  \citenamefont {Kindo}}]{Kubota_2015}%
  \BibitemOpen
  \bibfield  {author} {\bibinfo {author} {\bibfnamefont {Y.}~\bibnamefont
  {Kubota}}, \bibinfo {author} {\bibfnamefont {H.}~\bibnamefont {Tanaka}},
  \bibinfo {author} {\bibfnamefont {T.}~\bibnamefont {Ono}}, \bibinfo {author}
  {\bibfnamefont {Y.}~\bibnamefont {Narumi}},\ and\ \bibinfo {author}
  {\bibfnamefont {K.}~\bibnamefont {Kindo}},\ }\bibfield  {title} {\bibinfo
  {title} {Successive magnetic phase transitions in
  {$\ensuremath{\alpha}\ensuremath{-}{\mathrm{RuCl}}_{3}$}: X{Y}-like
  frustrated magnet on the honeycomb lattice},\ }\href
  {https://doi.org/10.1103/PhysRevB.91.094422} {\bibfield  {journal} {\bibinfo
  {journal} {Phys. Rev. B}\ }\textbf {\bibinfo {volume} {91}},\ \bibinfo
  {pages} {094422} (\bibinfo {year} {2015})}\BibitemShut {NoStop}%
\bibitem [{\citenamefont {Leahy}\ \emph {et~al.}(2017)\citenamefont {Leahy},
  \citenamefont {Pocs}, \citenamefont {Siegfried}, \citenamefont {Graf},
  \citenamefont {Do}, \citenamefont {Choi}, \citenamefont {Normand},\ and\
  \citenamefont {Lee}}]{Leahy2017}%
  \BibitemOpen
  \bibfield  {author} {\bibinfo {author} {\bibfnamefont {I.~A.}\ \bibnamefont
  {Leahy}}, \bibinfo {author} {\bibfnamefont {C.~A.}\ \bibnamefont {Pocs}},
  \bibinfo {author} {\bibfnamefont {P.~E.}\ \bibnamefont {Siegfried}}, \bibinfo
  {author} {\bibfnamefont {D.}~\bibnamefont {Graf}}, \bibinfo {author}
  {\bibfnamefont {S.-H.}\ \bibnamefont {Do}}, \bibinfo {author} {\bibfnamefont
  {K.-Y.}\ \bibnamefont {Choi}}, \bibinfo {author} {\bibfnamefont
  {B.}~\bibnamefont {Normand}},\ and\ \bibinfo {author} {\bibfnamefont
  {M.}~\bibnamefont {Lee}},\ }\bibfield  {title} {\bibinfo {title} {Anomalous
  thermal conductivity and magnetic torque response in the honeycomb magnet
  {$\alpha$-$\mathrm{RuCl}_3$}},\ }\href
  {https://doi.org/10.1103/PhysRevLett.118.187203} {\bibfield  {journal}
  {\bibinfo  {journal} {Phys. Rev. Lett.}\ }\textbf {\bibinfo {volume} {118}},\
  \bibinfo {pages} {187203} (\bibinfo {year} {2017})}\BibitemShut {NoStop}%
\bibitem [{\citenamefont {Sears}\ \emph {et~al.}(2017)\citenamefont {Sears},
  \citenamefont {Zhao}, \citenamefont {Xu}, \citenamefont {Lynn},\ and\
  \citenamefont {Kim}}]{Sears2017}%
  \BibitemOpen
  \bibfield  {author} {\bibinfo {author} {\bibfnamefont {J.~A.}\ \bibnamefont
  {Sears}}, \bibinfo {author} {\bibfnamefont {Y.}~\bibnamefont {Zhao}},
  \bibinfo {author} {\bibfnamefont {Z.}~\bibnamefont {Xu}}, \bibinfo {author}
  {\bibfnamefont {J.~W.}\ \bibnamefont {Lynn}},\ and\ \bibinfo {author}
  {\bibfnamefont {Y.-J.}\ \bibnamefont {Kim}},\ }\bibfield  {title} {\bibinfo
  {title} {Phase diagram of {$\alpha$-$\mathrm{RuCl}_3$} in an in-plane
  magnetic field},\ }\href {https://doi.org/10.1103/PhysRevB.95.180411}
  {\bibfield  {journal} {\bibinfo  {journal} {Phys. Rev. B}\ }\textbf {\bibinfo
  {volume} {95}},\ \bibinfo {pages} {180411} (\bibinfo {year}
  {2017})}\BibitemShut {NoStop}%
\bibitem [{\citenamefont {Wolter}\ \emph {et~al.}(2017)\citenamefont {Wolter},
  \citenamefont {Corredor}, \citenamefont {Janssen}, \citenamefont {Nenkov},
  \citenamefont {Sch\"onecker}, \citenamefont {Do}, \citenamefont {Choi},
  \citenamefont {Albrecht}, \citenamefont {Hunger}, \citenamefont {Doert},
  \citenamefont {Vojta},\ and\ \citenamefont {B\"uchner}}]{Wolter2017}%
  \BibitemOpen
  \bibfield  {author} {\bibinfo {author} {\bibfnamefont {A.~U.~B.}\
  \bibnamefont {Wolter}}, \bibinfo {author} {\bibfnamefont {L.~T.}\
  \bibnamefont {Corredor}}, \bibinfo {author} {\bibfnamefont {L.}~\bibnamefont
  {Janssen}}, \bibinfo {author} {\bibfnamefont {K.}~\bibnamefont {Nenkov}},
  \bibinfo {author} {\bibfnamefont {S.}~\bibnamefont {Sch\"onecker}}, \bibinfo
  {author} {\bibfnamefont {S.-H.}\ \bibnamefont {Do}}, \bibinfo {author}
  {\bibfnamefont {K.-Y.}\ \bibnamefont {Choi}}, \bibinfo {author}
  {\bibfnamefont {R.}~\bibnamefont {Albrecht}}, \bibinfo {author}
  {\bibfnamefont {J.}~\bibnamefont {Hunger}}, \bibinfo {author} {\bibfnamefont
  {T.}~\bibnamefont {Doert}}, \bibinfo {author} {\bibfnamefont
  {M.}~\bibnamefont {Vojta}},\ and\ \bibinfo {author} {\bibfnamefont
  {B.}~\bibnamefont {B\"uchner}},\ }\bibfield  {title} {\bibinfo {title}
  {Field-induced quantum criticality in the {Kitaev} system
  {$\alpha$-$\mathrm{RuCl}_3$}},\ }\href
  {https://doi.org/10.1103/PhysRevB.96.041405} {\bibfield  {journal} {\bibinfo
  {journal} {Phys. Rev. B}\ }\textbf {\bibinfo {volume} {96}},\ \bibinfo
  {pages} {041405} (\bibinfo {year} {2017})}\BibitemShut {NoStop}%
\bibitem [{\citenamefont {Baek}\ \emph {et~al.}(2017)\citenamefont {Baek},
  \citenamefont {Do}, \citenamefont {Choi}, \citenamefont {Kwon}, \citenamefont
  {Wolter}, \citenamefont {Nishimoto}, \citenamefont {van~den Brink},\ and\
  \citenamefont {B\"uchner}}]{Baek2017}%
  \BibitemOpen
  \bibfield  {author} {\bibinfo {author} {\bibfnamefont {S.-H.}\ \bibnamefont
  {Baek}}, \bibinfo {author} {\bibfnamefont {S.-H.}\ \bibnamefont {Do}},
  \bibinfo {author} {\bibfnamefont {K.-Y.}\ \bibnamefont {Choi}}, \bibinfo
  {author} {\bibfnamefont {Y.~S.}\ \bibnamefont {Kwon}}, \bibinfo {author}
  {\bibfnamefont {A.~U.~B.}\ \bibnamefont {Wolter}}, \bibinfo {author}
  {\bibfnamefont {S.}~\bibnamefont {Nishimoto}}, \bibinfo {author}
  {\bibfnamefont {J.}~\bibnamefont {van~den Brink}},\ and\ \bibinfo {author}
  {\bibfnamefont {B.}~\bibnamefont {B\"uchner}},\ }\bibfield  {title} {\bibinfo
  {title} {Evidence for a field-induced quantum spin liquid in
  {$\alpha$-$\mathrm{RuCl}_3$}},\ }\href
  {https://doi.org/10.1103/PhysRevLett.119.037201} {\bibfield  {journal}
  {\bibinfo  {journal} {Phys. Rev. Lett.}\ }\textbf {\bibinfo {volume} {119}},\
  \bibinfo {pages} {037201} (\bibinfo {year} {2017})}\BibitemShut {NoStop}%
\bibitem [{\citenamefont {Banerjee}\ \emph
  {et~al.}(2018{\natexlab{a}})\citenamefont {Banerjee}, \citenamefont
  {Lampen-Kelley}, \citenamefont {Knolle}, \citenamefont {Balz}, \citenamefont
  {Aczel}, \citenamefont {Winn}, \citenamefont {Liu}, \citenamefont
  {Pajerowski}, \citenamefont {Yan}, \citenamefont {Bridges}, \citenamefont
  {Savici}, \citenamefont {Chakoumakos}, \citenamefont {Lumsden}, \citenamefont
  {Tennant}, \citenamefont {Moessner}, \citenamefont {Mandrus},\ and\
  \citenamefont {Nagler}}]{Banerjee_2018}%
  \BibitemOpen
  \bibfield  {author} {\bibinfo {author} {\bibfnamefont {A.}~\bibnamefont
  {Banerjee}}, \bibinfo {author} {\bibfnamefont {P.}~\bibnamefont
  {Lampen-Kelley}}, \bibinfo {author} {\bibfnamefont {J.}~\bibnamefont
  {Knolle}}, \bibinfo {author} {\bibfnamefont {C.}~\bibnamefont {Balz}},
  \bibinfo {author} {\bibfnamefont {A.~A.}\ \bibnamefont {Aczel}}, \bibinfo
  {author} {\bibfnamefont {B.}~\bibnamefont {Winn}}, \bibinfo {author}
  {\bibfnamefont {Y.}~\bibnamefont {Liu}}, \bibinfo {author} {\bibfnamefont
  {D.}~\bibnamefont {Pajerowski}}, \bibinfo {author} {\bibfnamefont
  {J.}~\bibnamefont {Yan}}, \bibinfo {author} {\bibfnamefont {C.~A.}\
  \bibnamefont {Bridges}}, \bibinfo {author} {\bibfnamefont {A.~T.}\
  \bibnamefont {Savici}}, \bibinfo {author} {\bibfnamefont {B.~C.}\
  \bibnamefont {Chakoumakos}}, \bibinfo {author} {\bibfnamefont {M.~D.}\
  \bibnamefont {Lumsden}}, \bibinfo {author} {\bibfnamefont {D.~A.}\
  \bibnamefont {Tennant}}, \bibinfo {author} {\bibfnamefont {R.}~\bibnamefont
  {Moessner}}, \bibinfo {author} {\bibfnamefont {D.~G.}\ \bibnamefont
  {Mandrus}},\ and\ \bibinfo {author} {\bibfnamefont {S.~E.}\ \bibnamefont
  {Nagler}},\ }\bibfield  {title} {\bibinfo {title} {Excitations in the
  field-induced quantum spin liquid state of {$\alpha\mathrm{-RuCl}_3$}},\
  }\href {https://doi.org/10.1038/s41535-018-0079-2} {\bibfield  {journal}
  {\bibinfo  {journal} {npj Quantum Materials}\ }\textbf {\bibinfo {volume}
  {3}},\ \bibinfo {pages} {8} (\bibinfo {year}
  {2018}{\natexlab{a}})}\BibitemShut {NoStop}%
\bibitem [{\citenamefont {Hentrich}\ \emph {et~al.}(2018)\citenamefont
  {Hentrich}, \citenamefont {Wolter}, \citenamefont {Zotos}, \citenamefont
  {Brenig}, \citenamefont {Nowak}, \citenamefont {Isaeva}, \citenamefont
  {Doert}, \citenamefont {Banerjee}, \citenamefont {Lampen-Kelley},
  \citenamefont {Mandrus}, \citenamefont {Nagler}, \citenamefont {Sears},
  \citenamefont {Kim}, \citenamefont {B\"uchner},\ and\ \citenamefont
  {Hess}}]{Hentrich_2018}%
  \BibitemOpen
  \bibfield  {author} {\bibinfo {author} {\bibfnamefont {R.}~\bibnamefont
  {Hentrich}}, \bibinfo {author} {\bibfnamefont {A.~U.~B.}\ \bibnamefont
  {Wolter}}, \bibinfo {author} {\bibfnamefont {X.}~\bibnamefont {Zotos}},
  \bibinfo {author} {\bibfnamefont {W.}~\bibnamefont {Brenig}}, \bibinfo
  {author} {\bibfnamefont {D.}~\bibnamefont {Nowak}}, \bibinfo {author}
  {\bibfnamefont {A.}~\bibnamefont {Isaeva}}, \bibinfo {author} {\bibfnamefont
  {T.}~\bibnamefont {Doert}}, \bibinfo {author} {\bibfnamefont
  {A.}~\bibnamefont {Banerjee}}, \bibinfo {author} {\bibfnamefont
  {P.}~\bibnamefont {Lampen-Kelley}}, \bibinfo {author} {\bibfnamefont {D.~G.}\
  \bibnamefont {Mandrus}}, \bibinfo {author} {\bibfnamefont {S.~E.}\
  \bibnamefont {Nagler}}, \bibinfo {author} {\bibfnamefont {J.}~\bibnamefont
  {Sears}}, \bibinfo {author} {\bibfnamefont {Y.-J.}\ \bibnamefont {Kim}},
  \bibinfo {author} {\bibfnamefont {B.}~\bibnamefont {B\"uchner}},\ and\
  \bibinfo {author} {\bibfnamefont {C.}~\bibnamefont {Hess}},\ }\bibfield
  {title} {\bibinfo {title} {Unusual phonon heat transport in
  {$\ensuremath{\alpha}\text{\ensuremath{-}}{\mathrm{RuCl}}_{3}$}: Strong
  spin-phonon scattering and field-induced spin gap},\ }\href
  {https://doi.org/10.1103/PhysRevLett.120.117204} {\bibfield  {journal}
  {\bibinfo  {journal} {Phys. Rev. Lett.}\ }\textbf {\bibinfo {volume} {120}},\
  \bibinfo {pages} {117204} (\bibinfo {year} {2018})}\BibitemShut {NoStop}%
\bibitem [{\citenamefont {Jan{\v s}a}\ \emph {et~al.}(2018)\citenamefont
  {Jan{\v s}a}, \citenamefont {Zorko}, \citenamefont {Gomil{\v s}ek},
  \citenamefont {Pregelj}, \citenamefont {Kr{\"a}mer}, \citenamefont {Biner},
  \citenamefont {Biffin}, \citenamefont {R{\"u}egg},\ and\ \citenamefont
  {Klanj{\v s}ek}}]{Jansa_2018}%
  \BibitemOpen
  \bibfield  {author} {\bibinfo {author} {\bibfnamefont {N.}~\bibnamefont
  {Jan{\v s}a}}, \bibinfo {author} {\bibfnamefont {A.}~\bibnamefont {Zorko}},
  \bibinfo {author} {\bibfnamefont {M.}~\bibnamefont {Gomil{\v s}ek}}, \bibinfo
  {author} {\bibfnamefont {M.}~\bibnamefont {Pregelj}}, \bibinfo {author}
  {\bibfnamefont {K.~W.}\ \bibnamefont {Kr{\"a}mer}}, \bibinfo {author}
  {\bibfnamefont {D.}~\bibnamefont {Biner}}, \bibinfo {author} {\bibfnamefont
  {A.}~\bibnamefont {Biffin}}, \bibinfo {author} {\bibfnamefont
  {C.}~\bibnamefont {R{\"u}egg}},\ and\ \bibinfo {author} {\bibfnamefont
  {M.}~\bibnamefont {Klanj{\v s}ek}},\ }\bibfield  {title} {\bibinfo {title}
  {Observation of two types of fractional excitation in the {K}itaev honeycomb
  magnet},\ }\href {https://doi.org/10.1038/s41567-018-0129-5} {\bibfield
  {journal} {\bibinfo  {journal} {Nature Physics}\ }\textbf {\bibinfo {volume}
  {14}},\ \bibinfo {pages} {786} (\bibinfo {year} {2018})}\BibitemShut
  {NoStop}%
\bibitem [{\citenamefont {Widmann}\ \emph {et~al.}(2019)\citenamefont
  {Widmann}, \citenamefont {Tsurkan}, \citenamefont {Prishchenko},
  \citenamefont {Mazurenko}, \citenamefont {Tsirlin},\ and\ \citenamefont
  {Loidl}}]{Widmann2019}%
  \BibitemOpen
  \bibfield  {author} {\bibinfo {author} {\bibfnamefont {S.}~\bibnamefont
  {Widmann}}, \bibinfo {author} {\bibfnamefont {V.}~\bibnamefont {Tsurkan}},
  \bibinfo {author} {\bibfnamefont {D.~A.}\ \bibnamefont {Prishchenko}},
  \bibinfo {author} {\bibfnamefont {V.~G.}\ \bibnamefont {Mazurenko}}, \bibinfo
  {author} {\bibfnamefont {A.~A.}\ \bibnamefont {Tsirlin}},\ and\ \bibinfo
  {author} {\bibfnamefont {A.}~\bibnamefont {Loidl}},\ }\bibfield  {title}
  {\bibinfo {title} {Thermodynamic evidence of fractionalized excitations in
  {$\alpha$-$\mathrm{RuCl}_3$}},\ }\href
  {https://doi.org/10.1103/PhysRevB.99.094415} {\bibfield  {journal} {\bibinfo
  {journal} {Phys. Rev. B}\ }\textbf {\bibinfo {volume} {99}},\ \bibinfo
  {pages} {094415} (\bibinfo {year} {2019})}\BibitemShut {NoStop}%
\bibitem [{\citenamefont {Balz}\ \emph {et~al.}(2019)\citenamefont {Balz},
  \citenamefont {Lampen-Kelley}, \citenamefont {Banerjee}, \citenamefont {Yan},
  \citenamefont {Lu}, \citenamefont {Hu}, \citenamefont {Yadav}, \citenamefont
  {Takano}, \citenamefont {Liu}, \citenamefont {Tennant}, \citenamefont
  {Lumsden}, \citenamefont {Mandrus},\ and\ \citenamefont {Nagler}}]{Balz2019}%
  \BibitemOpen
  \bibfield  {author} {\bibinfo {author} {\bibfnamefont {C.}~\bibnamefont
  {Balz}}, \bibinfo {author} {\bibfnamefont {P.}~\bibnamefont {Lampen-Kelley}},
  \bibinfo {author} {\bibfnamefont {A.}~\bibnamefont {Banerjee}}, \bibinfo
  {author} {\bibfnamefont {J.}~\bibnamefont {Yan}}, \bibinfo {author}
  {\bibfnamefont {Z.}~\bibnamefont {Lu}}, \bibinfo {author} {\bibfnamefont
  {X.}~\bibnamefont {Hu}}, \bibinfo {author} {\bibfnamefont {S.~M.}\
  \bibnamefont {Yadav}}, \bibinfo {author} {\bibfnamefont {Y.}~\bibnamefont
  {Takano}}, \bibinfo {author} {\bibfnamefont {Y.}~\bibnamefont {Liu}},
  \bibinfo {author} {\bibfnamefont {D.~A.}\ \bibnamefont {Tennant}}, \bibinfo
  {author} {\bibfnamefont {M.~D.}\ \bibnamefont {Lumsden}}, \bibinfo {author}
  {\bibfnamefont {D.}~\bibnamefont {Mandrus}},\ and\ \bibinfo {author}
  {\bibfnamefont {S.~E.}\ \bibnamefont {Nagler}},\ }\bibfield  {title}
  {\bibinfo {title} {Finite field regime for a quantum spin liquid in
  {$\alpha$-$\mathrm{RuCl}_3$}},\ }\href
  {https://doi.org/10.1103/PhysRevB.100.060405} {\bibfield  {journal} {\bibinfo
   {journal} {Phys. Rev. B}\ }\textbf {\bibinfo {volume} {100}},\ \bibinfo
  {pages} {060405} (\bibinfo {year} {2019})}\BibitemShut {NoStop}%
\bibitem [{\citenamefont {Kasahara}\ \emph {et~al.}(2018)\citenamefont
  {Kasahara}, \citenamefont {Ohnishi}, \citenamefont {Mizukami}, \citenamefont
  {Tanaka}, \citenamefont {Ma}, \citenamefont {Sugii}, \citenamefont {Kurita},
  \citenamefont {Tanaka}, \citenamefont {Nasu}, \citenamefont {Motome},
  \citenamefont {Shibauchi},\ and\ \citenamefont {Matsuda}}]{Kasahara2018}%
  \BibitemOpen
  \bibfield  {author} {\bibinfo {author} {\bibfnamefont {Y.}~\bibnamefont
  {Kasahara}}, \bibinfo {author} {\bibfnamefont {T.}~\bibnamefont {Ohnishi}},
  \bibinfo {author} {\bibfnamefont {Y.}~\bibnamefont {Mizukami}}, \bibinfo
  {author} {\bibfnamefont {O.}~\bibnamefont {Tanaka}}, \bibinfo {author}
  {\bibfnamefont {S.}~\bibnamefont {Ma}}, \bibinfo {author} {\bibfnamefont
  {K.}~\bibnamefont {Sugii}}, \bibinfo {author} {\bibfnamefont
  {N.}~\bibnamefont {Kurita}}, \bibinfo {author} {\bibfnamefont
  {H.}~\bibnamefont {Tanaka}}, \bibinfo {author} {\bibfnamefont
  {J.}~\bibnamefont {Nasu}}, \bibinfo {author} {\bibfnamefont {Y.}~\bibnamefont
  {Motome}}, \bibinfo {author} {\bibfnamefont {T.}~\bibnamefont {Shibauchi}},\
  and\ \bibinfo {author} {\bibfnamefont {Y.}~\bibnamefont {Matsuda}},\
  }\bibfield  {title} {\bibinfo {title} {{M}ajorana quantization and
  half-integer thermal quantum {H}all effect in a {K}itaev spin liquid},\
  }\href {https://doi.org/10.1038/s41586-018-0274-0} {\bibfield  {journal}
  {\bibinfo  {journal} {Nature}\ }\textbf {\bibinfo {volume} {559}},\ \bibinfo
  {pages} {227} (\bibinfo {year} {2018})}\BibitemShut {NoStop}%
\bibitem [{\citenamefont {Yokoi}\ \emph {et~al.}(2020)\citenamefont {Yokoi},
  \citenamefont {Ma}, \citenamefont {Kasahara}, \citenamefont {Kasahara},
  \citenamefont {Shibauchi}, \citenamefont {Kurita}, \citenamefont {Tanaka},
  \citenamefont {Nasu}, \citenamefont {Motome}, \citenamefont {Hickey},
  \citenamefont {Trebst},\ and\ \citenamefont {Matsuda}}]{Tokoi}%
  \BibitemOpen
  \bibfield  {author} {\bibinfo {author} {\bibfnamefont {T.}~\bibnamefont
  {Yokoi}}, \bibinfo {author} {\bibfnamefont {S.}~\bibnamefont {Ma}}, \bibinfo
  {author} {\bibfnamefont {Y.}~\bibnamefont {Kasahara}}, \bibinfo {author}
  {\bibfnamefont {S.}~\bibnamefont {Kasahara}}, \bibinfo {author}
  {\bibfnamefont {T.}~\bibnamefont {Shibauchi}}, \bibinfo {author}
  {\bibfnamefont {N.}~\bibnamefont {Kurita}}, \bibinfo {author} {\bibfnamefont
  {H.}~\bibnamefont {Tanaka}}, \bibinfo {author} {\bibfnamefont
  {J.}~\bibnamefont {Nasu}}, \bibinfo {author} {\bibfnamefont {Y.}~\bibnamefont
  {Motome}}, \bibinfo {author} {\bibfnamefont {C.}~\bibnamefont {Hickey}},
  \bibinfo {author} {\bibfnamefont {S.}~\bibnamefont {Trebst}},\ and\ \bibinfo
  {author} {\bibfnamefont {Y.}~\bibnamefont {Matsuda}},\ }\bibfield  {title}
  {\bibinfo {title} {Half-integer quantized anomalous thermal {Hall} effect in
  the {Kitaev} material $\alpha$-$\mathrm{RuCl}_3$},\ }\Eprint
  {https://arxiv.org/abs/2001.01899} {arXiv:2001.01899}  (\bibinfo {year}
  {2020}),\ \bibinfo {note} {unpublished}\BibitemShut {NoStop}%
\bibitem [{\citenamefont {Bruin}\ \emph {et~al.}(2021)\citenamefont {Bruin},
  \citenamefont {Claus}, \citenamefont {Matsumoto}, \citenamefont {Kurita},
  \citenamefont {Tanaka},\ and\ \citenamefont {Takagi}}]{bruin2021}%
  \BibitemOpen
  \bibfield  {author} {\bibinfo {author} {\bibfnamefont {J.~A.~N.}\
  \bibnamefont {Bruin}}, \bibinfo {author} {\bibfnamefont {R.~R.}\ \bibnamefont
  {Claus}}, \bibinfo {author} {\bibfnamefont {Y.}~\bibnamefont {Matsumoto}},
  \bibinfo {author} {\bibfnamefont {N.}~\bibnamefont {Kurita}}, \bibinfo
  {author} {\bibfnamefont {H.}~\bibnamefont {Tanaka}},\ and\ \bibinfo {author}
  {\bibfnamefont {H.}~\bibnamefont {Takagi}},\ }\href@noop {} {\bibinfo {title}
  {Robustness of the thermal {H}all effect close to half-quantization in a
  field-induced spin liquid state}} (\bibinfo {year} {2021}),\ \Eprint
  {https://arxiv.org/abs/2104.12184} {arXiv:2104.12184 [cond-mat.str-el]}
  \BibitemShut {NoStop}%
\bibitem [{\citenamefont {Banerjee}\ \emph
  {et~al.}(2018{\natexlab{b}})\citenamefont {Banerjee}, \citenamefont
  {Heiblum}, \citenamefont {Umansky}, \citenamefont {Feldman}, \citenamefont
  {Oreg},\ and\ \citenamefont {Stern}}]{Banerjee}%
  \BibitemOpen
  \bibfield  {author} {\bibinfo {author} {\bibfnamefont {M.}~\bibnamefont
  {Banerjee}}, \bibinfo {author} {\bibfnamefont {M.}~\bibnamefont {Heiblum}},
  \bibinfo {author} {\bibfnamefont {V.}~\bibnamefont {Umansky}}, \bibinfo
  {author} {\bibfnamefont {D.~E.}\ \bibnamefont {Feldman}}, \bibinfo {author}
  {\bibfnamefont {Y.}~\bibnamefont {Oreg}},\ and\ \bibinfo {author}
  {\bibfnamefont {A.}~\bibnamefont {Stern}},\ }\bibfield  {title} {\bibinfo
  {title} {Observation of half-integer thermal {Hall} conductance},\ }\href
  {https://doi.org/10.1038/s41586-018-0184-1} {\bibfield  {journal} {\bibinfo
  {journal} {Nature}\ }\textbf {\bibinfo {volume} {559}},\ \bibinfo {pages}
  {205} (\bibinfo {year} {2018}{\natexlab{b}})}\BibitemShut {NoStop}%
\bibitem [{\citenamefont {Yamashita}\ \emph {et~al.}(2020)\citenamefont
  {Yamashita}, \citenamefont {Gouchi}, \citenamefont {Uwatoko}, \citenamefont
  {Kurita},\ and\ \citenamefont {Tanaka}}]{Yamashita_2020}%
  \BibitemOpen
  \bibfield  {author} {\bibinfo {author} {\bibfnamefont {M.}~\bibnamefont
  {Yamashita}}, \bibinfo {author} {\bibfnamefont {J.}~\bibnamefont {Gouchi}},
  \bibinfo {author} {\bibfnamefont {Y.}~\bibnamefont {Uwatoko}}, \bibinfo
  {author} {\bibfnamefont {N.}~\bibnamefont {Kurita}},\ and\ \bibinfo {author}
  {\bibfnamefont {H.}~\bibnamefont {Tanaka}},\ }\bibfield  {title} {\bibinfo
  {title} {Sample dependence of half-integer quantized thermal {H}all effect in
  the {K}itaev spin-liquid candidate {$\alpha\mathrm{-RuCl}_3$}},\ }\bibfield
  {journal} {\bibinfo  {journal} {Physical Review B}\ }\textbf {\bibinfo
  {volume} {102}},\ \href {https://doi.org/10.1103/physrevb.102.220404}
  {10.1103/physrevb.102.220404} (\bibinfo {year} {2020})\BibitemShut {NoStop}%
\bibitem [{\citenamefont {Czajka}\ \emph {et~al.}(2021)\citenamefont {Czajka},
  \citenamefont {Gao}, \citenamefont {Hirschberger}, \citenamefont
  {Lampen-Kelley}, \citenamefont {Banerjee}, \citenamefont {Yan}, \citenamefont
  {Mandrus}, \citenamefont {Nagler},\ and\ \citenamefont {Ong}}]{Czajka_2021}%
  \BibitemOpen
  \bibfield  {author} {\bibinfo {author} {\bibfnamefont {P.}~\bibnamefont
  {Czajka}}, \bibinfo {author} {\bibfnamefont {T.}~\bibnamefont {Gao}},
  \bibinfo {author} {\bibfnamefont {M.}~\bibnamefont {Hirschberger}}, \bibinfo
  {author} {\bibfnamefont {P.}~\bibnamefont {Lampen-Kelley}}, \bibinfo {author}
  {\bibfnamefont {A.}~\bibnamefont {Banerjee}}, \bibinfo {author}
  {\bibfnamefont {J.}~\bibnamefont {Yan}}, \bibinfo {author} {\bibfnamefont
  {D.~G.}\ \bibnamefont {Mandrus}}, \bibinfo {author} {\bibfnamefont {S.~E.}\
  \bibnamefont {Nagler}},\ and\ \bibinfo {author} {\bibfnamefont {N.~P.}\
  \bibnamefont {Ong}},\ }\href@noop {} {\bibinfo {title} {Oscillations of the
  thermal conductivity observed in the spin-liquid state of
  {$\alpha$-RuCl$_3$}}} (\bibinfo {year} {2021}),\ \Eprint
  {https://arxiv.org/abs/2102.11410} {arXiv:2102.11410 [cond-mat.str-el]}
  \BibitemShut {NoStop}%
\bibitem [{Note1()}]{Note1}%
  \BibitemOpen
  \bibinfo {note} {We note that the thermal Hall conductivity is identical to
  the thermal Hall conductance in two dimensions.}\BibitemShut {Stop}%
\bibitem [{\citenamefont {Kane}\ and\ \citenamefont
  {Fisher}(1997)}]{KaneFisherThermal}%
  \BibitemOpen
  \bibfield  {author} {\bibinfo {author} {\bibfnamefont {C.~L.}\ \bibnamefont
  {Kane}}\ and\ \bibinfo {author} {\bibfnamefont {M.~P.~A.}\ \bibnamefont
  {Fisher}},\ }\bibfield  {title} {\bibinfo {title} {Quantized thermal
  transport in the fractional quantum {Hall} effect},\ }\href
  {https://doi.org/10.1103/PhysRevB.55.15832} {\bibfield  {journal} {\bibinfo
  {journal} {Phys. Rev. B}\ }\textbf {\bibinfo {volume} {55}},\ \bibinfo
  {pages} {15832} (\bibinfo {year} {1997})}\BibitemShut {NoStop}%
\bibitem [{\citenamefont {Ye}\ \emph {et~al.}(2018)\citenamefont {Ye},
  \citenamefont {Hal\'asz}, \citenamefont {Savary},\ and\ \citenamefont
  {Balents}}]{Ye_2018}%
  \BibitemOpen
  \bibfield  {author} {\bibinfo {author} {\bibfnamefont {M.}~\bibnamefont
  {Ye}}, \bibinfo {author} {\bibfnamefont {G.~B.}\ \bibnamefont {Hal\'asz}},
  \bibinfo {author} {\bibfnamefont {L.}~\bibnamefont {Savary}},\ and\ \bibinfo
  {author} {\bibfnamefont {L.}~\bibnamefont {Balents}},\ }\bibfield  {title}
  {\bibinfo {title} {Quantization of the thermal {H}all conductivity at small
  {H}all angles},\ }\href {https://doi.org/10.1103/PhysRevLett.121.147201}
  {\bibfield  {journal} {\bibinfo  {journal} {Phys. Rev. Lett.}\ }\textbf
  {\bibinfo {volume} {121}},\ \bibinfo {pages} {147201} (\bibinfo {year}
  {2018})}\BibitemShut {NoStop}%
\bibitem [{\citenamefont {Vinkler-Aviv}\ and\ \citenamefont
  {Rosch}(2018)}]{Vinkler2018}%
  \BibitemOpen
  \bibfield  {author} {\bibinfo {author} {\bibfnamefont {Y.}~\bibnamefont
  {Vinkler-Aviv}}\ and\ \bibinfo {author} {\bibfnamefont {A.}~\bibnamefont
  {Rosch}},\ }\bibfield  {title} {\bibinfo {title} {Approximately quantized
  thermal {H}all effect of chiral liquids coupled to phonons},\ }\href
  {https://doi.org/10.1103/PhysRevX.8.031032} {\bibfield  {journal} {\bibinfo
  {journal} {Phys. Rev. X}\ }\textbf {\bibinfo {volume} {8}},\ \bibinfo {pages}
  {031032} (\bibinfo {year} {2018})}\BibitemShut {NoStop}%
\bibitem [{\citenamefont {de~C.~Chamon}\ \emph {et~al.}(1997)\citenamefont
  {de~C.~Chamon}, \citenamefont {Freed}, \citenamefont {Kivelson},
  \citenamefont {Sondhi},\ and\ \citenamefont {Wen}}]{Chamon_1997}%
  \BibitemOpen
  \bibfield  {author} {\bibinfo {author} {\bibfnamefont {C.}~\bibnamefont
  {de~C.~Chamon}}, \bibinfo {author} {\bibfnamefont {D.~E.}\ \bibnamefont
  {Freed}}, \bibinfo {author} {\bibfnamefont {S.~A.}\ \bibnamefont {Kivelson}},
  \bibinfo {author} {\bibfnamefont {S.~L.}\ \bibnamefont {Sondhi}},\ and\
  \bibinfo {author} {\bibfnamefont {X.~G.}\ \bibnamefont {Wen}},\ }\bibfield
  {title} {\bibinfo {title} {Two point-contact interferometer for quantum
  {H}all systems},\ }\href {https://doi.org/10.1103/PhysRevB.55.2331}
  {\bibfield  {journal} {\bibinfo  {journal} {Phys. Rev. B}\ }\textbf {\bibinfo
  {volume} {55}},\ \bibinfo {pages} {2331} (\bibinfo {year}
  {1997})}\BibitemShut {NoStop}%
\bibitem [{\citenamefont {Das~Sarma}\ \emph {et~al.}(2005)\citenamefont
  {Das~Sarma}, \citenamefont {Freedman},\ and\ \citenamefont
  {Nayak}}]{Nayak2005}%
  \BibitemOpen
  \bibfield  {author} {\bibinfo {author} {\bibfnamefont {S.}~\bibnamefont
  {Das~Sarma}}, \bibinfo {author} {\bibfnamefont {M.}~\bibnamefont
  {Freedman}},\ and\ \bibinfo {author} {\bibfnamefont {C.}~\bibnamefont
  {Nayak}},\ }\bibfield  {title} {\bibinfo {title} {Topologically protected
  qubits from a possible non-{A}belian fractional quantum {H}all state},\
  }\href {https://doi.org/10.1103/PhysRevLett.94.166802} {\bibfield  {journal}
  {\bibinfo  {journal} {Phys. Rev. Lett.}\ }\textbf {\bibinfo {volume} {94}},\
  \bibinfo {pages} {166802} (\bibinfo {year} {2005})}\BibitemShut {NoStop}%
\bibitem [{\citenamefont {Stern}\ and\ \citenamefont
  {Halperin}(2006)}]{Stern2006}%
  \BibitemOpen
  \bibfield  {author} {\bibinfo {author} {\bibfnamefont {A.}~\bibnamefont
  {Stern}}\ and\ \bibinfo {author} {\bibfnamefont {B.~I.}\ \bibnamefont
  {Halperin}},\ }\bibfield  {title} {\bibinfo {title} {Proposed experiments to
  probe the non-{A}belian $\ensuremath{\nu}=5/2$ quantum {H}all state},\ }\href
  {https://doi.org/10.1103/PhysRevLett.96.016802} {\bibfield  {journal}
  {\bibinfo  {journal} {Phys. Rev. Lett.}\ }\textbf {\bibinfo {volume} {96}},\
  \bibinfo {pages} {016802} (\bibinfo {year} {2006})}\BibitemShut {NoStop}%
\bibitem [{\citenamefont {Bonderson}\ \emph
  {et~al.}(2006{\natexlab{a}})\citenamefont {Bonderson}, \citenamefont
  {Kitaev},\ and\ \citenamefont {Shtengel}}]{Bonderson2006}%
  \BibitemOpen
  \bibfield  {author} {\bibinfo {author} {\bibfnamefont {P.}~\bibnamefont
  {Bonderson}}, \bibinfo {author} {\bibfnamefont {A.}~\bibnamefont {Kitaev}},\
  and\ \bibinfo {author} {\bibfnamefont {K.}~\bibnamefont {Shtengel}},\
  }\bibfield  {title} {\bibinfo {title} {Detecting non-{A}belian statistics in
  the $\ensuremath{\nu}=5/2$ fractional quantum {H}all state},\ }\href
  {https://doi.org/10.1103/PhysRevLett.96.016803} {\bibfield  {journal}
  {\bibinfo  {journal} {Phys. Rev. Lett.}\ }\textbf {\bibinfo {volume} {96}},\
  \bibinfo {pages} {016803} (\bibinfo {year} {2006}{\natexlab{a}})}\BibitemShut
  {NoStop}%
\bibitem [{\citenamefont {Bonderson}\ \emph
  {et~al.}(2006{\natexlab{b}})\citenamefont {Bonderson}, \citenamefont
  {Shtengel},\ and\ \citenamefont {Slingerland}}]{Bonderson2006a}%
  \BibitemOpen
  \bibfield  {author} {\bibinfo {author} {\bibfnamefont {P.}~\bibnamefont
  {Bonderson}}, \bibinfo {author} {\bibfnamefont {K.}~\bibnamefont
  {Shtengel}},\ and\ \bibinfo {author} {\bibfnamefont {J.~K.}\ \bibnamefont
  {Slingerland}},\ }\bibfield  {title} {\bibinfo {title} {Probing non-{A}belian
  statistics with quasiparticle interferometry},\ }\href
  {https://doi.org/10.1103/PhysRevLett.97.016401} {\bibfield  {journal}
  {\bibinfo  {journal} {Phys. Rev. Lett.}\ }\textbf {\bibinfo {volume} {97}},\
  \bibinfo {pages} {016401} (\bibinfo {year} {2006}{\natexlab{b}})}\BibitemShut
  {NoStop}%
\bibitem [{\citenamefont {Kim}(2006)}]{Kim_2006}%
  \BibitemOpen
  \bibfield  {author} {\bibinfo {author} {\bibfnamefont {E.-A.}\ \bibnamefont
  {Kim}},\ }\bibfield  {title} {\bibinfo {title} {Aharanov-{B}ohm interference
  and fractional statistics in a quantum {H}all interferometer},\ }\href
  {https://doi.org/10.1103/PhysRevLett.97.216404} {\bibfield  {journal}
  {\bibinfo  {journal} {Phys. Rev. Lett.}\ }\textbf {\bibinfo {volume} {97}},\
  \bibinfo {pages} {216404} (\bibinfo {year} {2006})}\BibitemShut {NoStop}%
\bibitem [{\citenamefont {Rosenow}\ and\ \citenamefont
  {Halperin}(2007)}]{Rosenow_2007}%
  \BibitemOpen
  \bibfield  {author} {\bibinfo {author} {\bibfnamefont {B.}~\bibnamefont
  {Rosenow}}\ and\ \bibinfo {author} {\bibfnamefont {B.~I.}\ \bibnamefont
  {Halperin}},\ }\bibfield  {title} {\bibinfo {title} {Influence of
  interactions on flux and back-gate period of quantum {H}all
  interferometers},\ }\href {https://doi.org/10.1103/PhysRevLett.98.106801}
  {\bibfield  {journal} {\bibinfo  {journal} {Phys. Rev. Lett.}\ }\textbf
  {\bibinfo {volume} {98}},\ \bibinfo {pages} {106801} (\bibinfo {year}
  {2007})}\BibitemShut {NoStop}%
\bibitem [{\citenamefont {Halperin}\ \emph {et~al.}(2011)\citenamefont
  {Halperin}, \citenamefont {Stern}, \citenamefont {Neder},\ and\ \citenamefont
  {Rosenow}}]{Halperin_2011}%
  \BibitemOpen
  \bibfield  {author} {\bibinfo {author} {\bibfnamefont {B.~I.}\ \bibnamefont
  {Halperin}}, \bibinfo {author} {\bibfnamefont {A.}~\bibnamefont {Stern}},
  \bibinfo {author} {\bibfnamefont {I.}~\bibnamefont {Neder}},\ and\ \bibinfo
  {author} {\bibfnamefont {B.}~\bibnamefont {Rosenow}},\ }\bibfield  {title}
  {\bibinfo {title} {Theory of the {Fabry-P\'erot} quantum {H}all
  interferometer},\ }\href {https://doi.org/10.1103/PhysRevB.83.155440}
  {\bibfield  {journal} {\bibinfo  {journal} {Phys. Rev. B}\ }\textbf {\bibinfo
  {volume} {83}},\ \bibinfo {pages} {155440} (\bibinfo {year}
  {2011})}\BibitemShut {NoStop}%
\bibitem [{\citenamefont {Rosenow}\ and\ \citenamefont
  {Simon}(2012)}]{Rosenow_2012}%
  \BibitemOpen
  \bibfield  {author} {\bibinfo {author} {\bibfnamefont {B.}~\bibnamefont
  {Rosenow}}\ and\ \bibinfo {author} {\bibfnamefont {S.~H.}\ \bibnamefont
  {Simon}},\ }\bibfield  {title} {\bibinfo {title} {Telegraph noise and the
  {Fabry-Perot} quantum {H}all interferometer},\ }\href
  {https://doi.org/10.1103/PhysRevB.85.201302} {\bibfield  {journal} {\bibinfo
  {journal} {Phys. Rev. B}\ }\textbf {\bibinfo {volume} {85}},\ \bibinfo
  {pages} {201302} (\bibinfo {year} {2012})}\BibitemShut {NoStop}%
\bibitem [{\citenamefont {Nakamura}\ \emph {et~al.}(2019)\citenamefont
  {Nakamura}, \citenamefont {Fallahi}, \citenamefont {Sahasrabudhe},
  \citenamefont {Rahman}, \citenamefont {Liang}, \citenamefont {Gardner},\ and\
  \citenamefont {Manfra}}]{Nakamura2019}%
  \BibitemOpen
  \bibfield  {author} {\bibinfo {author} {\bibfnamefont {J.}~\bibnamefont
  {Nakamura}}, \bibinfo {author} {\bibfnamefont {S.}~\bibnamefont {Fallahi}},
  \bibinfo {author} {\bibfnamefont {H.}~\bibnamefont {Sahasrabudhe}}, \bibinfo
  {author} {\bibfnamefont {R.}~\bibnamefont {Rahman}}, \bibinfo {author}
  {\bibfnamefont {S.}~\bibnamefont {Liang}}, \bibinfo {author} {\bibfnamefont
  {G.~C.}\ \bibnamefont {Gardner}},\ and\ \bibinfo {author} {\bibfnamefont
  {M.~J.}\ \bibnamefont {Manfra}},\ }\bibfield  {title} {\bibinfo {title}
  {Aharonov-{B}ohm interference of fractional quantum {H}all edge modes},\
  }\href {https://doi.org/10.1038/s41567-019-0441-8} {\bibfield  {journal}
  {\bibinfo  {journal} {Nature Physics}\ }\textbf {\bibinfo {volume} {15}},\
  \bibinfo {pages} {563} (\bibinfo {year} {2019})}\BibitemShut {NoStop}%
\bibitem [{\citenamefont {Willett}\ \emph {et~al.}(2019)\citenamefont
  {Willett}, \citenamefont {Shtengel}, \citenamefont {Nayak}, \citenamefont
  {Pfeiffer}, \citenamefont {Chung}, \citenamefont {Peabody}, \citenamefont
  {Baldwin},\ and\ \citenamefont {West}}]{Willett2019}%
  \BibitemOpen
  \bibfield  {author} {\bibinfo {author} {\bibfnamefont {R.~L.}\ \bibnamefont
  {Willett}}, \bibinfo {author} {\bibfnamefont {K.}~\bibnamefont {Shtengel}},
  \bibinfo {author} {\bibfnamefont {C.}~\bibnamefont {Nayak}}, \bibinfo
  {author} {\bibfnamefont {L.~N.}\ \bibnamefont {Pfeiffer}}, \bibinfo {author}
  {\bibfnamefont {Y.~J.}\ \bibnamefont {Chung}}, \bibinfo {author}
  {\bibfnamefont {M.~L.}\ \bibnamefont {Peabody}}, \bibinfo {author}
  {\bibfnamefont {K.~W.}\ \bibnamefont {Baldwin}},\ and\ \bibinfo {author}
  {\bibfnamefont {K.~W.}\ \bibnamefont {West}},\ }\bibfield  {title} {\bibinfo
  {title} {Interference measurements of non-{Abelian} $e/4$ and {Abelian} $e/2$
  quasiparticle braiding},\ }\Eprint {https://arxiv.org/abs/1905.10248}
  {arXiv:1905.10248}  (\bibinfo {year} {2019}),\ \bibinfo {note}
  {unpublished}\BibitemShut {NoStop}%
\bibitem [{\citenamefont {Nakamura}\ \emph {et~al.}(2020)\citenamefont
  {Nakamura}, \citenamefont {Liang}, \citenamefont {Gardner},\ and\
  \citenamefont {Manfra}}]{manfra_2020}%
  \BibitemOpen
  \bibfield  {author} {\bibinfo {author} {\bibfnamefont {J.}~\bibnamefont
  {Nakamura}}, \bibinfo {author} {\bibfnamefont {S.}~\bibnamefont {Liang}},
  \bibinfo {author} {\bibfnamefont {G.~C.}\ \bibnamefont {Gardner}},\ and\
  \bibinfo {author} {\bibfnamefont {M.~J.}\ \bibnamefont {Manfra}},\
  }\href@noop {} {\bibinfo {title} {Direct observation of anyonic braiding
  statistics at the $\nu$=1/3 fractional quantum {H}all state}} (\bibinfo
  {year} {2020}),\ \Eprint {https://arxiv.org/abs/2006.14115} {arXiv:2006.14115
  [cond-mat.mes-hall]} \BibitemShut {NoStop}%
\bibitem [{\citenamefont {Aasen}\ \emph {et~al.}(2020)\citenamefont {Aasen},
  \citenamefont {Mong}, \citenamefont {Hunt}, \citenamefont {Mandrus},\ and\
  \citenamefont {Alicea}}]{Aasen_2020}%
  \BibitemOpen
  \bibfield  {author} {\bibinfo {author} {\bibfnamefont {D.}~\bibnamefont
  {Aasen}}, \bibinfo {author} {\bibfnamefont {R.~S.~K.}\ \bibnamefont {Mong}},
  \bibinfo {author} {\bibfnamefont {B.~M.}\ \bibnamefont {Hunt}}, \bibinfo
  {author} {\bibfnamefont {D.}~\bibnamefont {Mandrus}},\ and\ \bibinfo {author}
  {\bibfnamefont {J.}~\bibnamefont {Alicea}},\ }\bibfield  {title} {\bibinfo
  {title} {Electrical probes of the non-{A}belian spin liquid in {K}itaev
  materials},\ }\href {https://doi.org/10.1103/PhysRevX.10.031014} {\bibfield
  {journal} {\bibinfo  {journal} {Phys. Rev. X}\ }\textbf {\bibinfo {volume}
  {10}},\ \bibinfo {pages} {031014} (\bibinfo {year} {2020})}\BibitemShut
  {NoStop}%
\bibitem [{\citenamefont {Klocke}\ \emph {et~al.}(2021)\citenamefont {Klocke},
  \citenamefont {Aasen}, \citenamefont {Mong}, \citenamefont {Demler},\ and\
  \citenamefont {Alicea}}]{Klocke_2020}%
  \BibitemOpen
  \bibfield  {author} {\bibinfo {author} {\bibfnamefont {K.}~\bibnamefont
  {Klocke}}, \bibinfo {author} {\bibfnamefont {D.}~\bibnamefont {Aasen}},
  \bibinfo {author} {\bibfnamefont {R.~S.~K.}\ \bibnamefont {Mong}}, \bibinfo
  {author} {\bibfnamefont {E.~A.}\ \bibnamefont {Demler}},\ and\ \bibinfo
  {author} {\bibfnamefont {J.}~\bibnamefont {Alicea}},\ }\bibfield  {title}
  {\bibinfo {title} {Time-domain anyon interferometry in {K}itaev honeycomb
  spin liquids and beyond},\ }\href
  {https://doi.org/10.1103/PhysRevLett.126.177204} {\bibfield  {journal}
  {\bibinfo  {journal} {Phys. Rev. Lett.}\ }\textbf {\bibinfo {volume} {126}},\
  \bibinfo {pages} {177204} (\bibinfo {year} {2021})}\BibitemShut {NoStop}%
\bibitem [{\citenamefont {Feldmeier}\ \emph {et~al.}(2020)\citenamefont
  {Feldmeier}, \citenamefont {Natori}, \citenamefont {Knap},\ and\
  \citenamefont {Knolle}}]{Feldmeier_2020}%
  \BibitemOpen
  \bibfield  {author} {\bibinfo {author} {\bibfnamefont {J.}~\bibnamefont
  {Feldmeier}}, \bibinfo {author} {\bibfnamefont {W.}~\bibnamefont {Natori}},
  \bibinfo {author} {\bibfnamefont {M.}~\bibnamefont {Knap}},\ and\ \bibinfo
  {author} {\bibfnamefont {J.}~\bibnamefont {Knolle}},\ }\bibfield  {title}
  {\bibinfo {title} {Local probes for charge-neutral edge states in
  two-dimensional quantum magnets},\ }\href
  {https://doi.org/10.1103/PhysRevB.102.134423} {\bibfield  {journal} {\bibinfo
   {journal} {Phys. Rev. B}\ }\textbf {\bibinfo {volume} {102}},\ \bibinfo
  {pages} {134423} (\bibinfo {year} {2020})}\BibitemShut {NoStop}%
\bibitem [{\citenamefont {Pereira}\ and\ \citenamefont
  {Egger}(2020)}]{Pereira_2020}%
  \BibitemOpen
  \bibfield  {author} {\bibinfo {author} {\bibfnamefont {R.~G.}\ \bibnamefont
  {Pereira}}\ and\ \bibinfo {author} {\bibfnamefont {R.}~\bibnamefont
  {Egger}},\ }\bibfield  {title} {\bibinfo {title} {Electrical access to
  {I}sing anyons in {K}itaev spin liquids},\ }\href
  {https://doi.org/10.1103/PhysRevLett.125.227202} {\bibfield  {journal}
  {\bibinfo  {journal} {Phys. Rev. Lett.}\ }\textbf {\bibinfo {volume} {125}},\
  \bibinfo {pages} {227202} (\bibinfo {year} {2020})}\BibitemShut {NoStop}%
\bibitem [{\citenamefont {Udagawa}\ \emph {et~al.}(2021)\citenamefont
  {Udagawa}, \citenamefont {Takayoshi},\ and\ \citenamefont
  {Oka}}]{Udagawa_2021}%
  \BibitemOpen
  \bibfield  {author} {\bibinfo {author} {\bibfnamefont {M.}~\bibnamefont
  {Udagawa}}, \bibinfo {author} {\bibfnamefont {S.}~\bibnamefont {Takayoshi}},\
  and\ \bibinfo {author} {\bibfnamefont {T.}~\bibnamefont {Oka}},\ }\bibfield
  {title} {\bibinfo {title} {Scanning tunneling microscopy as a single majorana
  detector of {K}itaev's chiral spin liquid},\ }\href
  {https://doi.org/10.1103/PhysRevLett.126.127201} {\bibfield  {journal}
  {\bibinfo  {journal} {Phys. Rev. Lett.}\ }\textbf {\bibinfo {volume} {126}},\
  \bibinfo {pages} {127201} (\bibinfo {year} {2021})}\BibitemShut {NoStop}%
\bibitem [{\citenamefont {Bonderson}\ \emph {et~al.}(2013)\citenamefont
  {Bonderson}, \citenamefont {Fidkowski}, \citenamefont {Freedman},\ and\
  \citenamefont {Walker}}]{Bonderson_2013}%
  \BibitemOpen
  \bibfield  {author} {\bibinfo {author} {\bibfnamefont {P.}~\bibnamefont
  {Bonderson}}, \bibinfo {author} {\bibfnamefont {L.}~\bibnamefont
  {Fidkowski}}, \bibinfo {author} {\bibfnamefont {M.}~\bibnamefont
  {Freedman}},\ and\ \bibinfo {author} {\bibfnamefont {K.}~\bibnamefont
  {Walker}},\ }\href@noop {} {\bibinfo {title} {Twisted interferometry}}
  (\bibinfo {year} {2013}),\ \Eprint {https://arxiv.org/abs/1306.2379}
  {arXiv:1306.2379 [quant-ph]} \BibitemShut {NoStop}%
\bibitem [{Note2()}]{Note2}%
  \BibitemOpen
  \bibinfo {note} {We note that $\kappa _{\protect \mathrm {b}}$ is the
  \protect \emph {two-dimensional} thermal conductivity which is the product of
  the usual three-dimensional thermal conductivity and the layer
  thickness.}\BibitemShut {Stop}%
\bibitem [{Note3()}]{Note3}%
  \BibitemOpen
  \bibinfo {note} {Here we have set $\hbar = k_B = 1$.}\BibitemShut {Stop}%
\bibitem [{Note4()}]{Note4}%
  \BibitemOpen
  \bibinfo {note} {We emphasize that this thermal Hall conductivity corresponds
  to a \protect \emph {thermally isolated} edge mode and could only be \protect
  \emph {directly} measured by using thermal leads and temperature sensors that
  couple to the edge mode rather than the bulk phonons.}\BibitemShut {Stop}%
\bibitem [{foo()}]{footnote}%
  \BibitemOpen
  \href@noop {} {}\bibinfo {note} {We use the symbol $\kappa$ for both thermal
  conductivities and thermal conductances. In two dimensions, these quantities
  have the same dimensions and are directly comparable to each
  other.}\BibitemShut {Stop}%
\bibitem [{Note5()}]{Note5}%
  \BibitemOpen
  \bibinfo {note} {We implicitly assume that the outgoing edge thermalizes with
  itself and reaches thermal equilibrium by the time it leaves the central
  region}\BibitemShut {NoStop}%
\bibitem [{\citenamefont {Fendley}\ \emph {et~al.}(2009)\citenamefont
  {Fendley}, \citenamefont {Fisher},\ and\ \citenamefont
  {Nayak}}]{Fendley2009}%
  \BibitemOpen
  \bibfield  {author} {\bibinfo {author} {\bibfnamefont {P.}~\bibnamefont
  {Fendley}}, \bibinfo {author} {\bibfnamefont {M.~P.}\ \bibnamefont
  {Fisher}},\ and\ \bibinfo {author} {\bibfnamefont {C.}~\bibnamefont
  {Nayak}},\ }\bibfield  {title} {\bibinfo {title} {Boundary conformal field
  theory and tunneling of edge quasiparticles in non-{A}belian topological
  states},\ }\href {https://doi.org/10.1016/j.aop.2009.03.005} {\bibfield
  {journal} {\bibinfo  {journal} {Annals of Physics}\ }\textbf {\bibinfo
  {volume} {324}},\ \bibinfo {pages} {1547–1572} (\bibinfo {year}
  {2009})}\BibitemShut {NoStop}%
\bibitem [{\citenamefont {Fendley}\ \emph {et~al.}(2007)\citenamefont
  {Fendley}, \citenamefont {Fisher},\ and\ \citenamefont
  {Nayak}}]{Fendley2007}%
  \BibitemOpen
  \bibfield  {author} {\bibinfo {author} {\bibfnamefont {P.}~\bibnamefont
  {Fendley}}, \bibinfo {author} {\bibfnamefont {M.~P.~A.}\ \bibnamefont
  {Fisher}},\ and\ \bibinfo {author} {\bibfnamefont {C.}~\bibnamefont
  {Nayak}},\ }\bibfield  {title} {\bibinfo {title} {Edge states and tunneling
  of non-{A}belian quasiparticles in the $5/2$ quantum {H}all state and $p+ip$
  superconductors},\ }\href {https://doi.org/10.1103/PhysRevB.75.045317}
  {\bibfield  {journal} {\bibinfo  {journal} {Phys. Rev. B}\ }\textbf {\bibinfo
  {volume} {75}},\ \bibinfo {pages} {045317} (\bibinfo {year}
  {2007})}\BibitemShut {NoStop}%
\bibitem [{\citenamefont {Fendley}\ \emph {et~al.}(2006)\citenamefont
  {Fendley}, \citenamefont {Fisher},\ and\ \citenamefont
  {Nayak}}]{Fendley_2006}%
  \BibitemOpen
  \bibfield  {author} {\bibinfo {author} {\bibfnamefont {P.}~\bibnamefont
  {Fendley}}, \bibinfo {author} {\bibfnamefont {M.~P.~A.}\ \bibnamefont
  {Fisher}},\ and\ \bibinfo {author} {\bibfnamefont {C.}~\bibnamefont
  {Nayak}},\ }\bibfield  {title} {\bibinfo {title} {Dynamical disentanglement
  across a point contact in a non-{A}belian quantum {H}all state},\ }\href
  {https://doi.org/10.1103/PhysRevLett.97.036801} {\bibfield  {journal}
  {\bibinfo  {journal} {Phys. Rev. Lett.}\ }\textbf {\bibinfo {volume} {97}},\
  \bibinfo {pages} {036801} (\bibinfo {year} {2006})}\BibitemShut {NoStop}%
\bibitem [{\citenamefont {Nilsson}\ and\ \citenamefont
  {Akhmerov}(2010)}]{Nilsson_2010}%
  \BibitemOpen
  \bibfield  {author} {\bibinfo {author} {\bibfnamefont {J.}~\bibnamefont
  {Nilsson}}\ and\ \bibinfo {author} {\bibfnamefont {A.~R.}\ \bibnamefont
  {Akhmerov}},\ }\bibfield  {title} {\bibinfo {title} {Theory of non-abelian
  {F}abry-{P}erot interferometry in topological insulators},\ }\href
  {https://doi.org/10.1103/PhysRevB.81.205110} {\bibfield  {journal} {\bibinfo
  {journal} {Phys. Rev. B}\ }\textbf {\bibinfo {volume} {81}},\ \bibinfo
  {pages} {205110} (\bibinfo {year} {2010})}\BibitemShut {NoStop}%
\bibitem [{Note6()}]{Note6}%
  \BibitemOpen
  \bibinfo {note} {We have restored appropriate factors of $\hbar $, $k_B$, and
  $v$.}\BibitemShut {Stop}%
\bibitem [{Note7()}]{Note7}%
  \BibitemOpen
  \bibinfo {note} {At the same time, we also assume that the temperature is
  high enough that the tunneling is still perturbative.}\BibitemShut {Stop}%
\bibitem [{\citenamefont {Willans}\ \emph {et~al.}(2010)\citenamefont
  {Willans}, \citenamefont {Chalker},\ and\ \citenamefont
  {Moessner}}]{Willans_2010}%
  \BibitemOpen
  \bibfield  {author} {\bibinfo {author} {\bibfnamefont {A.~J.}\ \bibnamefont
  {Willans}}, \bibinfo {author} {\bibfnamefont {J.~T.}\ \bibnamefont
  {Chalker}},\ and\ \bibinfo {author} {\bibfnamefont {R.}~\bibnamefont
  {Moessner}},\ }\bibfield  {title} {\bibinfo {title} {Disorder in a quantum
  spin liquid: Flux binding and local moment formation},\ }\href
  {https://doi.org/10.1103/PhysRevLett.104.237203} {\bibfield  {journal}
  {\bibinfo  {journal} {Phys. Rev. Lett.}\ }\textbf {\bibinfo {volume} {104}},\
  \bibinfo {pages} {237203} (\bibinfo {year} {2010})}\BibitemShut {NoStop}%
\bibitem [{\citenamefont {Willans}\ \emph {et~al.}(2011)\citenamefont
  {Willans}, \citenamefont {Chalker},\ and\ \citenamefont
  {Moessner}}]{Willans_2011}%
  \BibitemOpen
  \bibfield  {author} {\bibinfo {author} {\bibfnamefont {A.~J.}\ \bibnamefont
  {Willans}}, \bibinfo {author} {\bibfnamefont {J.~T.}\ \bibnamefont
  {Chalker}},\ and\ \bibinfo {author} {\bibfnamefont {R.}~\bibnamefont
  {Moessner}},\ }\bibfield  {title} {\bibinfo {title} {Site dilution in the
  {K}itaev honeycomb model},\ }\href
  {https://doi.org/10.1103/PhysRevB.84.115146} {\bibfield  {journal} {\bibinfo
  {journal} {Phys. Rev. B}\ }\textbf {\bibinfo {volume} {84}},\ \bibinfo
  {pages} {115146} (\bibinfo {year} {2011})}\BibitemShut {NoStop}%
\bibitem [{\citenamefont {Kao}\ \emph {et~al.}(2021)\citenamefont {Kao},
  \citenamefont {Knolle}, \citenamefont {Hal\'asz}, \citenamefont {Moessner},\
  and\ \citenamefont {Perkins}}]{Kao_2021}%
  \BibitemOpen
  \bibfield  {author} {\bibinfo {author} {\bibfnamefont {W.-H.}\ \bibnamefont
  {Kao}}, \bibinfo {author} {\bibfnamefont {J.}~\bibnamefont {Knolle}},
  \bibinfo {author} {\bibfnamefont {G.~B.}\ \bibnamefont {Hal\'asz}}, \bibinfo
  {author} {\bibfnamefont {R.}~\bibnamefont {Moessner}},\ and\ \bibinfo
  {author} {\bibfnamefont {N.~B.}\ \bibnamefont {Perkins}},\ }\bibfield
  {title} {\bibinfo {title} {Vacancy-induced low-energy density of states in
  the {K}itaev spin liquid},\ }\href
  {https://doi.org/10.1103/PhysRevX.11.011034} {\bibfield  {journal} {\bibinfo
  {journal} {Phys. Rev. X}\ }\textbf {\bibinfo {volume} {11}},\ \bibinfo
  {pages} {011034} (\bibinfo {year} {2021})}\BibitemShut {NoStop}%
\bibitem [{\citenamefont {Goldstein}(1951)}]{Goldstein_1951}%
  \BibitemOpen
  \bibfield  {author} {\bibinfo {author} {\bibfnamefont {S.}~\bibnamefont
  {Goldstein}},\ }\bibfield  {title} {\bibinfo {title} {On diffusion by
  discontinuous movements and on the telegraph equation},\ }\href
  {https://doi.org/10.1093/qjmam/4.2.129} {\bibfield  {journal} {\bibinfo
  {journal} {J. Mech. Appl. Math.}\ }\textbf {\bibinfo {volume} {4}},\ \bibinfo
  {pages} {129} (\bibinfo {year} {1951})}\BibitemShut {NoStop}%
\bibitem [{\citenamefont {Kac}(1974)}]{Kac_1974}%
  \BibitemOpen
  \bibfield  {author} {\bibinfo {author} {\bibfnamefont {M.}~\bibnamefont
  {Kac}},\ }\bibfield  {title} {\bibinfo {title} {A stochastic model related to
  the telegrapher's equation},\ }\href@noop {} {\bibfield  {journal} {\bibinfo
  {journal} {Rocky Mountain Journal of Mathematics}\ }\textbf {\bibinfo
  {volume} {4}},\ \bibinfo {pages} {497} (\bibinfo {year} {1974})}\BibitemShut
  {NoStop}%
\bibitem [{\citenamefont {Wei}\ \emph {et~al.}(2021)\citenamefont {Wei},
  \citenamefont {Mitrović},\ and\ \citenamefont {Feldman}}]{Wei2021}%
  \BibitemOpen
  \bibfield  {author} {\bibinfo {author} {\bibfnamefont {Z.}~\bibnamefont
  {Wei}}, \bibinfo {author} {\bibfnamefont {V.~F.}\ \bibnamefont {Mitrović}},\
  and\ \bibinfo {author} {\bibfnamefont {D.~E.}\ \bibnamefont {Feldman}},\
  }\href@noop {} {\bibinfo {title} {Thermal interferometry of anyons in spin
  liquids}} (\bibinfo {year} {2021}),\ \Eprint
  {https://arxiv.org/abs/2105.06873} {arXiv:2105.06873 [cond-mat.mes-hall]}
  \BibitemShut {NoStop}%
\end{thebibliography}%

\appendix
\begin{widetext}

\section{Edge and Bulk Temperatures}\label{app:phonon_appendix}

Here we elaborate on the perturbative approach from the main text and provide derivations of the first-order edge and bulk temperatures in Eqs.~\eqref{eq:T-e} and \eqref{eq:phonon_correction}.
The unperturbed (zeroth-order) temperatures correspond to vanishing edge-bulk coupling $\lambda$.
As we describe in the main text, since the hot and cold leads couple to the bulk phonons rather than the edge mode, the zeroth-order edge temperature vanishes: $T_{\mathrm{e}}^{(0)}(r,\vartheta) = 0$.
Furthermore, the zeroth-order bulk temperature within the right lobe is imposed by the boundary conditions to be $T_{\mathrm{b}}^{(0)}(r,\vartheta) = A + B \ln(r/r_0)$, where $A = \delta T [1 + \alpha \ln (R/r_0)]^{-1}$ and $B = \alpha A$ [see Eq.~\eqref{eq:T-0}].
We now seek the first-order perturbative corrections to these edge and bulk temperatures within the right lobe as a result of finite edge-bulk coupling $\lambda$.

Let us first focus on the first-order correction to the edge temperature, $T_{\mathrm{e}}^{(1)}(r,\vartheta)$.
By considering Eqs.~\eqref{eq:edge_current_variation-n} and \eqref{eq:interface-edge-n} with $n=1$, this first-order correction must satisfy the ordinary differential equations
\be
\begin{aligned}
	\frac{\partial T_{\mathrm{e}}^{(1)} (R,\vartheta)}{\partial\vartheta} &= \frac{R}{\ell} \left[T_{\mathrm{b}}^{(0)} (R,\vartheta) - T_{\mathrm{e}}^{(1)} (R,\vartheta)\right],\\
	\frac{\partial T_{\mathrm{e}}^{(1)} (r,\pm \vartheta_0)}{\partial r} &= \mp \frac{1}{\ell} \left[T_{\mathrm{b}}^{(0)} (r,\pm\vartheta_0) - T_{\mathrm{e}}^{(1)} (r,\pm\vartheta_0)\right],
\end{aligned} \label{eq:edge_temp_correction}
\ee
along with the corresponding boundary condition
\be
	T_{\mathrm{e}}^{(1)} (r_0,-\vartheta_0) = (2\beta-1) \, T_{\mathrm{e}}^{(1)} (r_0,\vartheta_0). \label{eq:edge_temp_bc}
\ee
As described in the main text, we assume that the edge-bulk thermalization length $\ell$ is much larger than the central region, $\ell \gg r_0$, but much smaller than the lobe width, $\ell \ll R \vartheta_0 \lesssim R$.
In this regime, the edge temperature thermalizes well along the outer edge of the right lobe, and we can thus write $T_{\mathrm{e}}^{(1)} (R,\vartheta_0) = \delta T$.
Solving Eq.~\eqref{eq:edge_temp_correction} for the top edge of the right lobe ($\vartheta = \vartheta_0$), the edge temperature along this edge is then given by
\be
	T_{\mathrm{e}}^{(1)}(r,\vartheta_0) = A + B\ln(r/r_0) + Be^{r/\ell}\left[\ei(-R/\ell) - \ei(-r/\ell)\right],\label{eq:edge_temp_n1}
\ee
where $\ei(x)$ is the exponential integral function with asymptotic forms
\be
	\ei(-x) = \begin{cases}
		\ln x + \gamma & (x \ll 1),\\
		-\frac{1}{x}e^{-x} & (x \gg 1).
	\end{cases}
\ee
We note that, as a result of edge-bulk thermalization, this edge temperature is independent of $\beta$.
Next, if we drop exponentially small terms from Eq.~\eqref{eq:edge_temp_n1}, and employ Eq.~\eqref{eq:edge_temp_bc}, the edge temperatures of the top and bottom edges at the interface with the central region ($r = r_0$) are found to be
\be
\begin{aligned}
	& T_{\mathrm{e}}^{(1)}(r_0, \vartheta_0) = A + Be^{r_0/\ell} \left[\ln (\ell / r_0) - \gamma\right] \approx A + B\left[\ln(\ell / r_0) - \gamma\right],\\
	& T_{\mathrm{e}}^{(1)}(r_0, -\vartheta_0) \approx (2\beta - 1)\left\{A + B\left[\ln(\ell / r_0) - \gamma\right]\right\}.
\end{aligned}
\ee
Solving Eq.~\eqref{eq:edge_temp_correction} for the bottom edge of the right lobe ($\vartheta = -\vartheta_0$), the edge temperature along this edge is then
\be
	T_{\mathrm{e}}^{(1)}(r,-\vartheta_0) \approx A + B\ln(r/r_0) + Be^{-r/\ell}\left[\ei(r_0/\ell) - \ei(r/\ell)\right] - \left[A + (1 - 2\beta)\left\{A + B\left[\ln(\ell/r_0) - \gamma\right]\right\}\right]e^{-r/\ell}.
\ee
This edge temperature has both a $\beta$-independent and a $\beta$-linear part. Differentiating with respect to $\beta$, and recalling the definitions of $A$ and $B$, we immediately recover the result in Eq.~\eqref{eq:T-e} of the main text.

Having found the first-order correction to the edge temperature, we may now consider the corresponding correction to the bulk (phonon) temperature, $T_{\mathrm{b}}^{(1)}(r,\vartheta)$.
This first-order correction must satisfy Laplace's equation [see Eq.~\eqref{eq:laplace-n}],
\be
	\frac{\partial^2 T_{\mathrm{b}}^{(1)}(r,\vartheta)}{(\partial \ln r)^2} + \frac{\partial^2 T_{\mathrm{b}}^{(1)}(r,\vartheta)}{\partial\vartheta^2} = 0,
\ee
subject to the corresponding boundary conditions [see Eqs.~\eqref{eq:exchange-n}-\eqref{eq:interface-bulk-n}],
\be
	\begin{aligned}
	T_{\mathrm{b}}^{(1)}(R,\vartheta) &= 0,\\
	\frac{\partial T_{\mathrm{b}}^{(1)}(r=r_0,\vartheta)}{\partial \ln r} &= \alpha T_{\mathrm{b}}^{(1)}(r_0, \vartheta),\\
	\frac{\partial T_{\mathrm{b}}^{(1)}(r,\vartheta=\vartheta_0)}{\partial\vartheta} &= -\frac{\lambda r}{\kappa_{\mathrm{b}}}\left[ T_{\mathrm{b}}^{(0)}(r,\vartheta_0) - T_{\mathrm{e}}^{(1)}(r,\vartheta_0) \right]\\
	&= \frac{\lambda r}{\kappa_{\mathrm{b}}} B e^{r/\ell} \left[ \ei (-R/\ell)  - \ei(-r/\ell)\right],\\
	\frac{\partial T_{\mathrm{b}}^{(1)}(r,\vartheta=-\vartheta_0)}{\partial\vartheta} &= \frac{\lambda r}{\kappa_{\mathrm{b}}} \left[ T_{\mathrm{b}}^{(0)}(r,-\vartheta_0) - T_{\mathrm{e}}^{(1)}(r, -\vartheta_0) \right]\\
	&= -\frac{\lambda r}{\kappa_{\mathrm{b}}} \left\{ Be^{-r/\ell} \left[ \ei(r_0 / \ell) - \ei(r / \ell) \right] - \left[A + (1 - 2\beta)\left\{A + B\left[\ln(\ell/r_0) - \gamma \right]\right\}\right] e^{-r/\ell}\right\}.
	\end{aligned}
\ee
Let us briefly recall the origin of these boundary conditions from the main text.
The first boundary condition reflects that the bulk temperature along the outer edge is fixed due to contact with the thermal lead and, thus, all corrections to the bulk temperature must vanish there.
The second boundary condition relates the bulk heat current through the narrow channel in the central region and the bulk heat current through the right lobe.
The remaining two boundary conditions enforce that the perpendicular heat current along the boundary of the bulk is consistent with the edge-bulk heat exchange which, in turn, depends on the edge temperatures previously found.

Given that Laplace's equation is linear and all inhomogeneities come from $T_{\mathrm{b}}^{(0)}(r,\pm\vartheta_0)$ and $T_{\mathrm{e}}^{(1)}(r,\pm\vartheta_0)$ in the last two boundary conditions, we may decompose the bulk temperature into a $\beta$-independent and a $\beta$-linear component, $T_{\mathrm{b}}^{(1)}(r,\vartheta) = \bar T_{\mathrm{b}}^{(1)}(r,\vartheta) + \beta \hat T_{\mathrm{b}}^{(1)}(r,\vartheta)$, where $\bar T_{\mathrm{b}}^{(1)}(r,\vartheta)$ and $\hat T_{\mathrm{b}}^{(1)}(r,\vartheta)$ carry no $\beta$ dependence.
Since we are interested in the sensitivity of the bulk temperature to tunneling processes in the central region, we focus on the $\beta$ derivative $\hat{T}_{\mathrm{b}}^{(1)} (r,\vartheta) = \partial T_{\mathrm{b}}^{(1)} (r,\vartheta) / \partial \beta$.
By linearity, this temperature sensitivity must satisfy the same Laplace's equation,
\be
	\frac{\partial^2 \hat T_{\mathrm{b}}^{(1)}(r,\vartheta)}{\left(\partial \ln r\right)^2} + \frac{\partial^2 \hat T_{\mathrm{b}}^{(1)}(r,\vartheta)}{\partial \vartheta^2} = 0,
\ee
but the relevant boundary conditions are now given by
\be
	\begin{aligned}
	\hat T_{\mathrm{b}}^{(1)}(R,\vartheta) &= 0,\\
	\frac{\partial \hat T_{\mathrm{b}}^{(1)}(r=r_0,\vartheta)}{\partial \ln r} &= \alpha \hat T_{\mathrm{b}}^{(1)}(r_0, \vartheta),\\
	\frac{\partial \hat T_{\mathrm{b}}^{(1)}(r,\vartheta=\vartheta_0)}{\partial\vartheta} &=  0,\\
	\frac{\partial \hat T_{\mathrm{b}}^{(1)}(r,\vartheta=-\vartheta_0)}{\partial\vartheta} &= -\frac{2\lambda}{\kappa_{\mathrm{b}}} \left\{A + B\left[\ln(\ell / r_0) - \gamma\right]  \right\}re^{-r/\ell}.
	\end{aligned}\label{eq:Tb1_hat_BCs}
\ee
Expanding in a basis of orthogonal eigenfunctions, we seek a separable solution of the general form
\be
	\hat T_{\mathrm{b}}^{(1)}(r,\vartheta) = \sum_{n=0}^\infty \sin\left[\mu_n\ln(R/r)\right]\left\{ C_n\cosh[\mu_n\vartheta] + S_n\sinh[\mu_n\vartheta]\right\},\label{eq:Tb_basis_expansion}
\ee
where $\mu_n$ is the $(n+1)$-th smallest positive $[(n+\tfrac12)\pi < \mu_n \ln(R/r_0) < (n+1)\pi]$ solution of the transcendental equation
\be
	\mu_n = -\alpha\tan[\mu_n\ln(R/r_0)]. \label{eq:mu}
\ee
The expression in Eq.~\eqref{eq:Tb_basis_expansion} automatically satisfies the first two boundary conditions ($r=r_0$ and $r=R$), while it may satisfy the remaining boundary conditions ($\vartheta = \pm \vartheta_0$) for appropriately chosen coefficients $C_n$ and $S_n$.
To find these coefficients, we consider the angular derivatives,
\be
	\frac{\partial \hat T_{\mathrm{b}}^{(1)}(r,\vartheta=\pm\vartheta_0)}{\partial\vartheta} = \sum_{n=0}^\infty \mu_n \sin[\mu_n\ln(R/r)]\{S_n\cosh[\mu_n\vartheta_0] \pm C_n\sinh[\mu_n\vartheta_0]\},
\ee
as well as the orthogonality relations between the radial eigenfunctions,
\be
	\int_{r_0}^R\frac{\dif r}{r} \sin[\mu_n\ln(R/r)]\sin[\mu_{n'}\ln(R/r)] = \frac{\delta_{n,n'}}{2}\left\{1 - \frac{\sin[2\mu_n\ln(R/r_0)]}{2\mu_n\ln(R/r_0)}\right\}\ln(R/r_0) \equiv \delta_{n,n'}N_n\ln(R/r_0),
\ee
where $N_n$ are appropriate normalization constants for these eigenfunctions.
By comparing these expressions with the boundary conditions in Eq.~\eqref{eq:Tb1_hat_BCs}, the coefficients $C_n$ and $S_n$ are then found to satisfy
\be
    C_n\sinh[\mu_n \vartheta_0] = -S_n\cosh[\mu_n\vartheta_0] = \frac{\lambda\{A + B[\ln(\ell / r_0) - \gamma]\}}{\kappa_{\mathrm{b}} \mu_n N_n\ln(R/r_0)}\int_{r_0}^R \dif r \sin[\mu_n\ln(R/r)]e^{-r/\ell} \equiv \frac{I_n}{\mu_n}.\label{eq:Tb1_coefficients}
\ee
Next, we calculate the angular average of the temperature sensitivity in the first-order approximation:
\be
	\langle \hat T_{\mathrm{b}}^{(1)}(r) \rangle \equiv \frac{1}{2\vartheta_0}\int_{-\vartheta_0}^{\vartheta_0} \dif \vartheta \, \hat T_{\mathrm{b}}^{(1)}(r,\vartheta).
\ee
This angular average is a lower bound to the maximal temperature sensitivity at the given radius, $\hat T_{\mathrm{b}}^{(1)}(r,-\vartheta_0)$, which corresponds to the bottom edge ($\vartheta = -\vartheta_0$), and becomes an accurate lower bound in the limit of $\vartheta_0 \ll 1$.
Employing the compact notation from Eq.~\eqref{eq:Tb1_coefficients}, the average temperature sensitivity may then be written as
\be
	\langle \hat T_{\mathrm{b}}^{(1)}(r) \rangle = \sum_{n=0}^\infty \frac{C_n\sinh[\mu_n\vartheta_0]\sin[\mu_n\ln(R/r)]}{\mu_n\vartheta_0} = \sum_{n=0}^\infty \frac{I_n\sin[\mu_n \ln (R/r)]}{\mu_n^2\vartheta_0}.
\ee
Since the magnitude of $I_n\sin[\mu_n\ln(R/r)]$ is bounded from above, it follows from $\mu_n > (n+\tfrac12)\pi / \ln(R/r_0)$ that the summand decays as $1/n^2$.
Thus, the sum in $n$ is convergent, and we may approximate it by its leading ($n=0$) term:
\be
	\langle \hat T_{\mathrm{b}}^{(1)}(r) \rangle \approx \frac{I_0\sin[\mu_0\ln(R/r)]}{\mu_0^2 \vartheta_0} = \frac{\sin[\mu_0\ln(R/r)]}{\mu_0^2 \vartheta_0} \frac{\lambda \{A + B[\ln(\ell / r_0) - \gamma]\}}{\kappa_{\mathrm{b}} N_0 \ln(R/r_0)}\int_{r_0}^R \dif r \sin[\mu_0\ln(R/r)]e^{-r/\ell}.
\ee
Recalling the assumption $r_0 \ll \ell \ll R$, we may then approximate the integrand as
\be
	\int_{r_0}^R \dif r \sin[\mu_0 \ln(R/r)]e^{-r/\ell} \approx \ell \sin[\mu_0 \ln (R/\ell)],
\ee
and obtain the following general expression for the average temperature sensitivity:
\be
	\langle \hat T_{\mathrm{b}}^{(1)}(r) \rangle \approx \frac{\kappa_{\mathrm{e}}\{A + B[\ln(\ell/r_0) - \gamma] \}\sin[\mu_0\ln(R/\ell)]\sin[\mu_0\ln(R/r)]}{\kappa_{\mathrm{b}}\vartheta_0\mu_0^2 N_0\ln(R/r_0)}.
\ee
Finally, we take the limits of large and small $\alpha = \kappa_{\mathrm{c}} / (\kappa_{\mathrm{b}} \vartheta_0)$.
In the first limit ($\alpha \gg 1$), the smallest positive solution of Eq.~\eqref{eq:mu} is $\mu_0 \ln (R / r_0) \approx \pi$, and the normalization constant $N_0$ is approximately $1/2$.
Since $A \approx \delta T [\alpha \ln (R / r_0)]^{-1}$ and $B \approx \delta T [\ln (R / r_0)]^{-1} \gg A$, the average temperature sensitivity becomes
\be
	\langle \hat T_{\mathrm{b}}^{(1)}(r) \rangle \approx \frac{2}{\pi^2} \left[\ln(\ell / r_0) - \gamma\right] \frac{\kappa_{\mathrm{e}} \delta T}{\kappa_{\mathrm{b}} \vartheta_0} \sin\left[\frac{\pi \ln(R/\ell)}{\ln(R/r_0)}\right]\sin\left[\frac{\pi \ln(R/r)}{\ln(R/r_0)}\right].
\ee
In the second limit ($\alpha \ll 1$), the smallest positive solution of Eq.~\eqref{eq:mu} is $\mu_0 \ln (R / r_0) \approx \pi / 2$, and the normalization constant $N_0$ is again approximately $1/2$. Since $A \approx \delta T$ and $B \approx \alpha \delta T \ll A$, the average temperature sensitivity reads
\be
	\langle \hat T_{\mathrm{b}}^{(1)}(r) \rangle \approx \frac{8\ln(R/r_0)}{\pi^2} \frac{\kappa_{\mathrm{e}} \delta T}{\kappa_{\mathrm{b}} \vartheta_0} \sin\left[\frac{\pi \ln(R/\ell)}{2\ln(R/r_0)}\right]\sin\left[\frac{\pi\ln(R/r)}{2\ln(R/r_0)}\right].
\ee
These final expressions for large and small $\alpha$ are identical to those in Eq.~\eqref{eq:phonon_correction} of the main text.

\section{Ising Anyon Tunneling}\label{app:cft_appendix}

In this appendix we show that the second order correction from Ising anyon tunneling in a single-pinch geometry diverges at zero temperature and show the appropriate power-scaling for a finite temperature effective tunneling coefficient.
Throughout this appendix we will work in units where $v=1$.
The appropriate factors of $v$ can be restored by dimensional analysis.

At second order the correction to the fermion Green's function takes the form
\be
    - \int_{-\infty}^\infty\dif s_1 \int_{-\infty}^{s_1}\dif s_2 \langle \gamma_{\mathrm{top}}(t,L) \biggl[ H_\text{tun}(s_1)H_\text{tun}(s_2) - \langle H_\text{tun}(s_1)H_\text{tun}(s_2)\rangle \biggr] \gamma_{\mathrm{top}}(0,-L)\rangle,
\ee
where we now adopt the convention that the heat current is measured at coordinate $L$ along the top edge and fermions are thermally excited by the heating due to a bath connected on the left edge (near $x=-L$).
Since we only care about the transmission across the constriction, we may freely take $L\rightarrow 0^+$.
Recall that the Ising anyon tunneling Hamiltonian for a single pinch takes the form
\be
    H_\text{tun} = t_\sigma e^{-i\pi/16}\sigma_{\mathrm{top}}(0)\sigma_{\mathrm{bot}}(0),
\ee
where $x=0$ is the coordinate of the pinch.
Evaluating the correction to the Green's function then requires that we evaluate a correlator which, after cluster decomposition between the top and bottom edges, takes the form
\be
    \langle \sigma(0,s_2)\sigma(0,s_1)\rangle_{\mathrm{bot}} \biggl[ \langle \gamma(L,t)\sigma(0,s_2)\sigma(0,s_1)\gamma(-L,0)\rangle_{\mathrm{top}} - \langle \sigma(0,s_2)\sigma(0,s_1)\rangle_{\mathrm{top}} \langle \gamma(L,t)\gamma(-L,0)\rangle_{\mathrm{top}} \biggr].
\ee
For brevity We moved the subscripts to the $\langle \cdot \rangle$ rather than the fields.
Closely related calculations have been carried out in detail in several other works~\cite{Nilsson_2010, Aasen_2020, Klocke_2020}.
In particular, Aasen \emph{et al.}~\cite{Aasen_2020} consider the first order Ising anyon tunneling correction in a loop interferometer, while Nilsson and Akhmerov~\cite{Nilsson_2010} calculate the backscattered component for the same two pinch geometry we consider in the main text, but working explicitly with the finite temperature correlations for the primary fields.

Let us then see that at zero-temperature this expression diverges.
Consider the more general scenario where we have
\be
\langle \sigma(y_1)\sigma(y_2)\rangle_{\mathrm{bot}} \left[ \langle \gamma(z_1)\sigma(\eta_1)\sigma(\eta_2)\gamma(z_2)\rangle_{\mathrm{top}} - \langle \sigma(\eta_1)\sigma(\eta_2)\rangle_{\mathrm{top}} \langle \gamma(z_1)\gamma(z_2)\rangle_{\mathrm{top}} \right].
\ee
This correlator is given explicitly by
\be
\frac{1}{2y_{12}^{1/8}\eta_{12}^{1/8}z_{12}}\left[-2 + \sqrt{\frac{(z_1-\eta_1)(z_2-\eta_2)}{(z_1-\eta_2)(z_2-\eta_1)}} + \sqrt{\frac{(z_1-\eta_2)(z_2-\eta_1)}{(z_1-\eta_1)(z_2-\eta_2)}}\right].
\label{eq:ising2_corr}
\ee
In this generic formulation we can treat both the single pinch and double pinch geometries.
For concrete purposes though, let us focus our attention on the single pinch case, fixing $y_{1,2} = \eta_{1,2} = i s_{1,2} \mp \epsilon$, $z_1 = i(t-L)$, and $z_2 = iL$.
Let us also define $s_1 = s$ and $s_2 = s - \Delta_s$ where $\Delta_s \geq 0$.
Here $\epsilon$ regulates divergences in the correlator.
A similar regulator can be taken in the coordinates $z_{1,2}$, but it should be taken smaller than $\epsilon$ to preserve the placement of a branch cut in later steps.

Since we are interested in the frequency-space transmission amplitude, let us Fourier transform this expression and insert our definitions for the coordinates into Eq.~\eqref{eq:ising2_corr}.
The expression we must evaluate then becomes
\be
    - e^{-i\pi / 8} t_\sigma^2 \int_{0}^{\infty} \dif t \int_{-\infty}^\infty \dif s \int_0^\infty \dif \Delta_s  \frac{e^{i\omega t}}{2i(i\Delta_s - 2\epsilon)^{1/4}(t-2L)}\left[-2+ \sqrt{\frac{(\Delta_s + L - s +i\epsilon)(L+s-t+i\epsilon)}{(L-s-i\epsilon)(L+s-t-\Delta_s - i\epsilon)}} + \cdots\right]
\ee
where the $\cdots$ is the inverse of the second bracketed term and $\omega = vk$.
Let us then take $L\rightarrow 0^+$ so that the expression simplifies to
\be
    - e^{-i\pi / 8} t_\sigma^2 \int_{0}^{\infty} \dif t \int_{-\infty}^\infty \dif s \int_0^\infty \dif \Delta_s  \frac{e^{i\omega t}}{2it(i\Delta_s - 2\epsilon)^{1/4}}\left[-2+ \sqrt{\frac{(s - \Delta_s - i\epsilon)(s-t+i\epsilon)}{(s+i\epsilon)(s-t-\Delta_s - i\epsilon)}} + \cdots\right].
\ee
If one then makes a shift of variables $s \rightarrow s + (\Delta_s + t)/2$, the integrand takes a particularly simple form,
\be
- e^{-i\pi / 8} t_\sigma^2 \int_{0}^{\infty} \dif t \int_{-\infty}^\infty \dif s \int_0^\infty \dif \Delta_s  \frac{e^{i\omega t}}{2it(i\Delta_s -2\epsilon)^{1/4}}\left[-2+ \sqrt{\frac{s^2 + a^2}{s^2 + b^2}} + \sqrt{\frac{s^2 + b^2}{s^2 + a^2}} \right],
\ee
where $a \equiv \epsilon + i(t - \Delta_s)/2$ and $b \equiv \epsilon - i(t + \Delta_s)/2$.
For $\epsilon > 0$, we see that $\RePart a, \, \RePart b > 0$.
Then we may invoke an integral identity to evaluate the integral with respect to $s$, giving
\be
    - e^{-i\pi / 8} t_\sigma^2 \int_{0}^{\infty} \dif t \int_0^\infty \dif \Delta_s  \frac{e^{i\omega t}}{2it(i\Delta_s - 2\epsilon)^{1/4}}4(a+b)\left[ K\left(\frac{a-b}{a+b}\right) - E\left(\frac{a-b}{a+b}\right) \right],
\ee
where $K$ and $E$ are the complete elliptic integrals of the first and second kind.
Now substituting the definitions of $a$ and $b$, one finds that the expression becomes
\be
    - e^{-i\pi / 8} t_\sigma^2 \int_{0}^{\infty} \dif t \int_0^\infty \dif \Delta_s  \frac{e^{i\omega t}}{2it(i\Delta_s - 2\epsilon)^{1/4}}4(-i\Delta_s + 2\epsilon)\left[ K\left(-\frac{t}{\Delta_s + 2i\epsilon}\right) - E\left(-\frac{t}{\Delta_s+2i\epsilon}\right) \right].
\ee
Taking $\epsilon\rightarrow 0^+$, one can then straightforwardly integrate with respect to $\Delta_s$ and obtain
\be
    e^{-i 7\pi/8} t_\sigma^2 \frac{21}{128}\frac{\Gamma\left(-\frac78\right)\Gamma\left(\frac38\right)^2}{\Gamma\left(\frac{15}{8}\right)} \int_0^\infty \dif t e^{i\omega t} t^{3/4}.
\ee
Now integrating with respect to $t$ just gives
\be
-e^{-i 3\pi/4} \frac{21 t_\sigma^2}{128} \frac{\Gamma\left(-\frac78\right)\Gamma\left(\frac38\right)^2\Gamma\left(\frac74\right)}{\Gamma\left(\frac{15}{8}\right)\omega^{7/4}} \approx 7.65 e^{-i 3\pi /4}t_\sigma^2 \omega^{-7/4}
\ee
The above corresponds to the correction to $A(k)$ where $\omega = v k$.
The correction to $\abs{A(k)}^2$ then is given by
\be
\frac{21}{64\sqrt{2}} \frac{t_\sigma^2}{\omega^{7/4}} \frac{\Gamma\left(-\frac78\right)\Gamma\left(\frac38\right)^2\Gamma\left(\frac74\right)}{\Gamma\left(\frac{15}{8}\right)} \approx -10.8 t_\sigma^2 \omega^{-7/4} \label{eq:ising_tunneling_norm}
\ee

Consider then calculating the correction to the thermal conductance associated with this term.
Temporarily neglecting the constant prefactor, we calculate the heat current
\be
	I_\sigma(T) = v \int_0^\infty \frac{\dif k}{2\pi} \varepsilon(k) n(k,T) (vk)^{-7/4} = -4\left(-1 + 2^{3/4}\right) T^{1/4} \Gamma\left(\frac54\right)\zeta\left(\frac14\right)
\ee
Then differentiating with respect to temperature we get the correction to the conductance
\be
    \Delta \kappa_\sigma = -\frac{\left(-1 + 2^{3/4}\right)  \Gamma\left(\frac54\right)\zeta\left(\frac14\right)}{T^{3/4}} = -\kappa_{\mathrm{e}} \frac{\left(-1 + 2^{3/4}\right)\Gamma\left(\frac54\right)\zeta\left(\frac14\right)}{\pi T^{7/4}} \approx \kappa_{\mathrm{e}} \frac{1.92}{T^{7/4}}.
\ee
Restoring the appropriate prefactors from Eq.~\eqref{eq:ising_tunneling_norm}, this gives $\Delta\kappa_\sigma \approx -21 t_\sigma^2 \kappa_{\mathrm{e}} T^{-7/4}$.
Indeed then we have recovered the Ising anyon tunneling correction to thermal transport, $\Delta \kappa_\sigma \propto \kappa_{\mathrm{e}} T^{-7/4}$, as reported in the main text.
Given this result it is convenient to define a renormalized tunneling amplitude $\tilde t_\sigma \equiv  4.55 t_\sigma T^{-7/8}$ so that $\Delta \kappa_\sigma = -\tilde t_\sigma^2 \kappa_{\mathrm{e}}$.

\end{widetext}

\end{document}